\documentclass[twocolumn,pra,superscriptaddress]{revtex4-1}
\usepackage{graphicx}
\usepackage{amssymb,amsmath}
\usepackage{bm}
\usepackage{dcolumn}
\usepackage{subfigure}
\usepackage{float}
\usepackage[OT1]{fontenc} 
\usepackage{url}
\usepackage{slashed}
\usepackage{color}
\usepackage{verbatim}
\usepackage{enumitem}
\usepackage{txfonts}
\usepackage{soul,physics}
\usepackage[driverfallback=dvipdfm]{hyperref}
\hypersetup{pdfpagemode=FullScreen,colorlinks=true,breaklinks,urlcolor=blue,linkcolor=blue,citecolor=blue}
\usepackage{natbib}

\usepackage{subfigure}
\usepackage{comment}
\usepackage{hyperref}
\usepackage{amssymb}
\usepackage{bm}
\usepackage{graphicx}
\usepackage{amsmath,amssymb}
\usepackage{tikz,fp}
\usepackage{tikz-cd}
\usetikzlibrary{arrows}
\usetikzlibrary{intersections}
\usetikzlibrary{shapes.geometric}
\usetikzlibrary{decorations.pathmorphing, patterns,shapes,fixedpointarithmetic}
\usetikzlibrary{decorations.markings}

\pgfdeclarepatternformonly{south west lines}{\pgfqpoint{-0pt}{-0pt}}{\pgfqpoint{3pt}{3pt}}{\pgfqpoint{3pt}{3pt}}{
	\pgfsetlinewidth{0.4pt}
	\pgfpathmoveto{\pgfqpoint{0pt}{0pt}}
	\pgfpathlineto{\pgfqpoint{3pt}{3pt}}
	\pgfpathmoveto{\pgfqpoint{2.8pt}{-.2pt}}
	\pgfpathlineto{\pgfqpoint{3.2pt}{.2pt}}
	\pgfpathmoveto{\pgfqpoint{-.2pt}{2.8pt}}
	\pgfpathlineto{\pgfqpoint{.2pt}{3.2pt}}
	\pgfusepath{stroke}}

\tikzset{
	mid arrow/.style={postaction={decorate,decoration={
				markings,
				mark=at position .575 with {\arrow{stealth}}
	}}},
	near arrow/.style={postaction={decorate,decoration={
				markings,
				mark=at position .275 with {\arrow{stealth}}
	}}},
	far arrow/.style={postaction={decorate,decoration={
				markings,
				mark=at position .800 with {\arrow{stealth}}
	}}},
	snake arrow/.style={fixed point arithmetic, decorate, decoration={snake,amplitude=2pt, segment length=11pt},postaction={decoration={markings,mark=at position 0.625 with {\arrow{stealth}}},decorate}},
}

\begin{document}
	
	\title{Quantum Entanglement in the Sachdev-Ye-Kitaev Model and its Generalizations}
	
	\author{Pengfei Zhang}
	\affiliation{Institute for Quantum Information and Matter and Walter Burke Institute for Theoretical Physics, California Institute of Technology, Pasadena, CA 91125, USA}
	\date{\today}
	
	\begin{abstract}
    Entanglement is one of the most important concepts in quantum physics. We review recent progress in understanding the quantum entanglement in many-body systems using large-$N$ solvable models: the Sachdev-Ye-Kitaev (SYK) model and its generalizations. We present the study of entanglement entropy in the original SYK Model using three different approaches: the exact diagonalization, the eigenstate thermalization hypothesis, and the path-integral representation. For coupled SYK models, the entanglement entropy shows linear growth and saturation at the thermal value. The saturation is related to replica wormholes in gravity. Finally, we consider the steady-state entanglement entropy of quantum many-body systems under repeated measurements. The traditional symmetry breaking in the enlarged replica space leads to the measurement-induced entanglement phase transition.
	\end{abstract}
	
	\maketitle
	\tableofcontents

	\section{Introduction}
Quantum entanglement, a fascinating property of quantum states, has been widely studied in quantum information, condensed matter, and high-energy physics. For example, in quantum information, entanglement is crucial for pure-state quantum algorithms to offer a speed-up over classical computation~\cite{jozsa1997entanglement,jozsa2003role,ding2007review}. In condensed matter physics, entanglement entropy can be used to distinguish thermalized/localized systems~\cite{pal2010many,nandkishore2015many,abanin2019colloquium,deutsch1991quantum,srednicki1994chaos}, as well as to characterize topologically ordered phases~\cite{kitaev2006topological,levin2006detecting}. In high energy physics, the discovery of the Ryu-Takayanagi (RT) formula~\cite{Ryu:2006bv,Ryu:2006ef,Lewkowycz:2013nqa,Hubeny:2007xt,Faulkner:2013ana,Engelhardt:2014gca}, which relates the fine-grained entropy of a subsystem to geometric properties, has considerably deepened our understanding of gravity, and recently led to recent progress in the information paradox~\cite{penington2019entanglement,almheiri2019entropy,almheiri2019page,almheiri2019islands,almheiri2019replica,penington2019replica}. However, the entanglement entropy is notoriously difficult to calculate due to the lack of efficient algorithms, in particular, for strongly interacting many-body systems. Most previous studies focus on systems with conformal symmetry~\cite{calabrese2009entanglement}, holographic duality~\cite{rangamani2017holographic}, or well-defined large-$N$ expansions \cite{metlitski2009entanglement,whitsitt2017entanglement,donnelly2019entanglement}. 

The Sachdev-Ye-Kitaev (SYK) model, proposed by Kitaev~\cite{kitaev2014hidden} and related to the early work by Sachdev and Ye~\cite{sachdev1993gapless}, describes $N$ Majorana fermions with random interactions. The SYK model is solvable under large-$N$ expansion~\cite{maldacena2016remarks}. For example, to the leading order of $1/N$, the Schwinger-Dyson equation for the Green's function can be derived in a closed form, the solution of which also determines thermodynamics. In the low-temperature limit, the SYK model shows emergent conformal symmetry. Its low-energy dynamics have a holographic dual representation by the 2D Jackiw-Teitelboim gravity~\cite{maldacena2016conformal,kitaev2018soft}, which implies the system is maximally chaotic~\cite{maldacena2016bound}. There are also studies on quantum dynamics of the SYK model, which reveals it exhibits quantum thermalization \cite{eberlein2017quantum}. Later, different generalizations of the SYK model have been proposed to study various physical problems \cite{chowdhury2021sachdev}, including models with charge conservation~\cite{davison2017thermoelectric,gu2019notes,chaturvedi2018note,bulycheva2017note}, Brownian interaction~\cite{saad2018semiclassical,sunderhauf2019quantum}, or multiple sites~\cite{gu2017local,gu2017energy,banerjee2017solvable,chen2017competition,song2017strongly,jian2017solvable,chen2017tunable,zhang2017dispersive,bi2017instability,narayan2017syk}. 

In this review, we summarize recent progress in the quantum entanglement in the SYK model and its generalizations. In the original SYK model, the entanglement entropy has been studied using three different approaches~\cite{liu2018quantum,fu2016numerical,huang2019eigenstate,zhang2020subsystem,zhang2020entanglement,haldar2020renyi,kudler2021information}: the exact diagonalization (ED), the eigenstate thermalization hypothesis (ETH), and the path-integral representation. Using the ED approach~\cite{liu2018quantum,fu2016numerical}, existing numerical analysis give results of the ground state entanglement of the SYK model, up a system size $N=44$~\cite{liu2018quantum}. On the other hand, the ETH states that in thermalized systems, small subsystems can be approximated by thermal ensembles with an effective temperature determined by the energy density~\cite{deutsch1991quantum,srednicki1994chaos}. In particular, the entanglement entropy is equal to the corresponding thermal entropy. For systems with local interaction, the boundary effect is small, and the effective temperature is independent of subsystem size. Whereas for systems with non-local interaction, the effective temperature depends on the subsystem size~\cite{huang2019eigenstate}. One can also express the $n$-th R\'enyi entropy as a path integral and perform the large-$N$ expansion as for thermodynamics~\cite{zhang2020subsystem,zhang2020entanglement,haldar2020renyi}. This leads to results exact in the large-$N$ limit. Results from these three approaches match to good precision.

For generalized SYK models that contain multiple sites, with some of the sites described by the SYK model, there is no simple approach to employ the ETH analysis, and the ED is limited to small system size $N$. Consequently, previous studies rely on the path-integral approach~\cite{gu2017spread,penington2019replica,sohal2022thermalization,chen2020replica,chen2021entropy,dadras2021perturbative,jian2021note,jian2021chaos,kaixiang2021page,nedel2021time,chen2019entanglement,pouria2022disentangling}. We are interested in the entanglement dynamics of the system prepared in some short-range entangled state. In the simplest case, we consider two SYK sites with $N_1$ and $N_2$ Majorana modes, coupled by random pair hopping~\cite{gu2017spread,penington2019replica,chen2020replica,sohal2022thermalization}. In the early-time regime, the entanglement entropy shows linear growth in time~\cite{gu2017spread,penington2019replica,chen2020replica,chen2021entropy,dadras2021perturbative,sohal2022thermalization}. In the long-time limit, when $N_1=N_2$, it is found that in the canonical ensemble the R\'enyi entanglement entropy does not saturate to the thermal value due to the existence of soft modes~\cite{gu2017spread,penington2019replica,sohal2022thermalization}. On the contrary, if $N_2\gg N_1$, the second site serves as a heat bath, and the R\'enyi entanglement entropy of the first site always matches the thermal entropy~\cite{chen2020replica,sohal2022thermalization}. When the entanglement entropy approaches the thermal value, a first-order transition occurs. This is a direct analog of the appearance of replica wormholes~\cite{almheiri2019replica,penington2019replica} in the entanglement calculations of systems with gravity, which is an important breakthrough in high-energy physics that resolves the information paradox~\cite{hawking1976breakdown}. Alternatively, one may instead consider replacing the static SYK interaction with the Brownian counterpart~\cite{jian2021note}. Analytical results can then be derived under certain assumptions. There are also attempts on understanding the von Neumann entanglement entropy~\cite{dadras2021perturbative}.

In addition to unitary evolutions where thermalization generally takes place, the evolution of quantum many-body systems can also become non-unitary when measurements are performed~\cite{nielsen2002quantum}. Recent studies show that if we follow the quantum trajectories of systems with repeated measurements, the steady-state can exhibit volume law, logarithmic, or area law entanglement~\cite{li2018quantum,10.21468/SciPostPhys.7.2.024,li2019measurement,skinner2019measurement,chan2019unitary,bao2020theory,choi2020quantum,gullans2020dynamical,gullans2020scalable,jian2020measurement,szyniszewski2019entanglement,zabalo2020critical,tang2020measurement,zhang2020nonuniversal,goto2020measurement,jian2020criticality,bao2021symmetry,alberton2020trajectory,Chen_2020,Nahum_2020,liu2021non,zhang2021emergent,jian2021measurement,zhang2021universal,sahu2021entanglement,jian2021quantum,jian2021phase}. This mechanism works for either projective or general measurements, regardless of whether the outcome is post-selected or not. SYK-like models are ideal platforms for revealing the underlying mechanism of these transitions~\cite{liu2021non,zhang2021emergent,jian2021measurement,jian2021phase,zhang2021universal,sahu2021entanglement,jian2021quantum}. In the last part of this review, we summarize recent progress in understanding the steady-state entanglement entropy of non-unitary SYK models. For small measurement rates, when the system is non-interacting, an emergent replica conformal symmetry shows up due to the existence of Goldstone modes after introducing multiple replicas~\cite{zhang2021emergent,jian2021measurement}. This leads to logarithmic entanglement entropy. After adding interactions, the Goldstone modes become massive~\cite{jian2021measurement}, and the system becomes volume law entangled. For large measurement rates, the steady-state of the system may become area law entangled, and the transition is a traditional symmetry-breaking transition on the enlarged replica space~\cite{jian2021measurement,jian2021phase}. 

This review is organized as follows: In section \ref{sec_introduceSYK}, we give a brief introduction to the SYK model and its various generalizations. Then, in section \ref{sec_singleSYK}, we summarize existing results for the entanglement entropy of the original SYK model. The entanglement entropy for coupled SYK models and non-unitary SYK models are discussed separately in sections \ref{sec_SYKbath} and \ref{sec_nonunitarySYK}. Finally, we conclude our review by a few outlooks in section \ref{sec_outlook}.

\section{The SYK model and its generalizations}\label{sec_introduceSYK}

\begin{figure}[t]
\center
\includegraphics[width=0.6\linewidth]{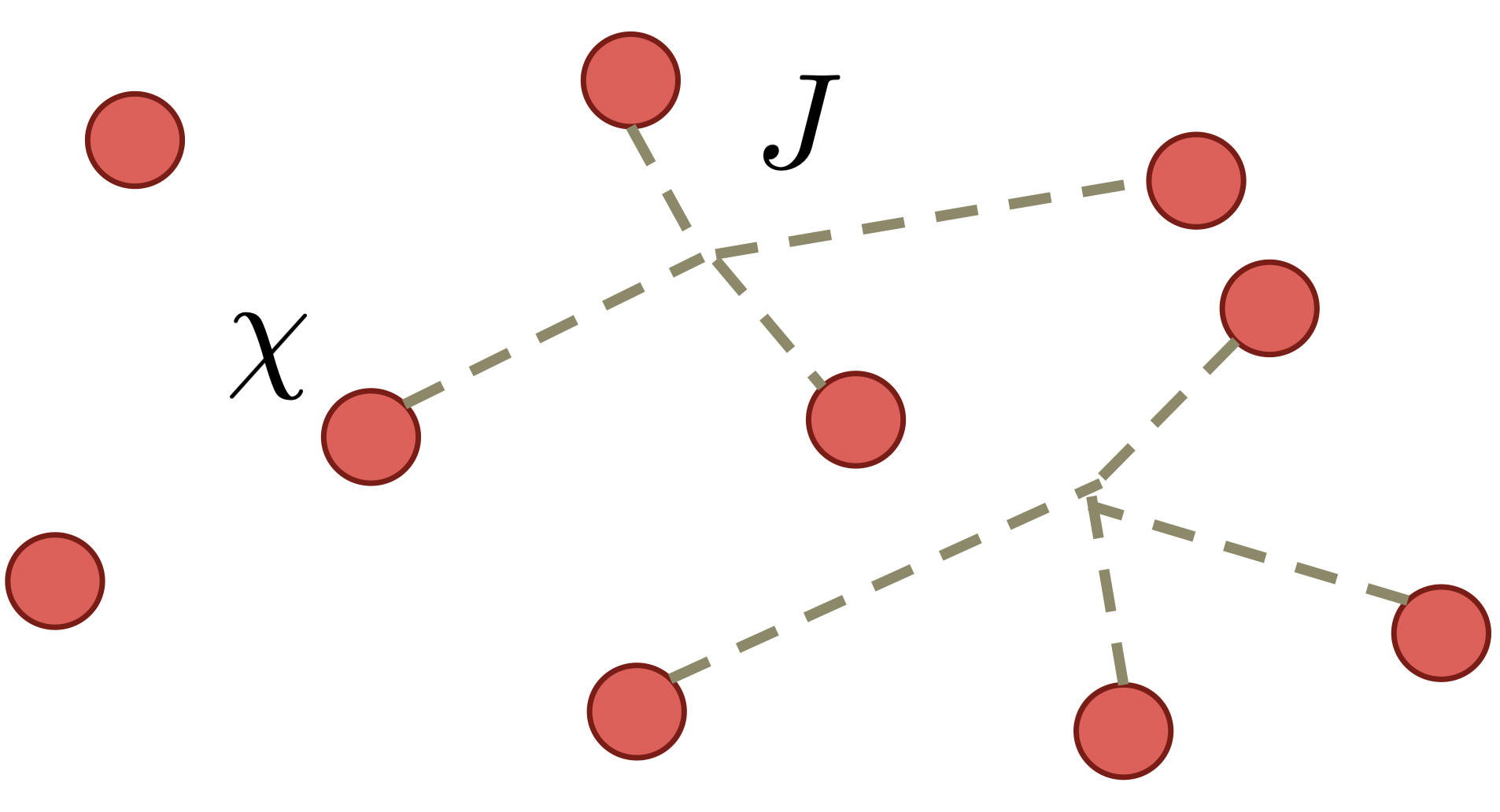}
\caption{ Schematics of the SYK$_4$ model, which describes Majorana fermions with random 4-body interactions.}\label{fig_schematic}
\end{figure}

In this section, we give a brief introduction to the original SYK model~\cite{kitaev2014hidden,maldacena2016remarks}. The Hamiltonian of the SYK$_q$ model reads 
\begin{equation}\label{H0}
H_{\text{SYK}_q}=\sum_{1\leq i_1< i_2< ...< i_q\leq N}i^{q/2}J_{i_1i_2...i_q}\chi_{i_1}\chi_{i_2}...\chi_{i_q}.
\end{equation}
Here $q$ is an even integer. It describes $q$-body random interactions between $N$ Majorana modes $\chi_i$. We choose the convention that $\{\chi_i,\chi_j\}=\delta_{ij}$. $J_{i_1i_2...i_q}$ with different set of indices are independent Gaussian variables, with expectations and variances given by
\begin{equation}
\overline{J_{i_1i_2...i_q}}=0,\ \ \ \ \ \ \overline{J_{i_1i_2...i_q}^2}=\frac{(q-1)!J^2}{N^{q-1}}=\frac{2^{q-1}(q-1)!\mathcal{J}^2}{qN^{q-1}}.
\end{equation}
Here the scaling of $N$ is chosen such that the energy is extensive. $\mathcal{J}$ is introduced for convenience in the large-$q$ analysis \cite{maldacena2016remarks}, which should be fixed as a constant as $q\rightarrow \infty$. The model describes Majorana fermions with random hopping for $q=2$, and becomes randomly interacting for $q\geq 4$.

\subsection{Large-\texorpdfstring{$N$}{TEXT} expansion} \label{subsec_SYKlargeN}
The large-$N$ expansion is a standard approach for analyzing quantum field theories that contain large number of fields~\cite{coleman1988aspects}. Here we explain the large-$N$ expansion in the SYK model by considering the partition function $Z=\text{tr}~e^{-\beta H_{\text{SYK}_q}}$. The path-integral representation takes the form of $Z=\int \mathcal{D}\chi \exp\left(-S[\chi,J]\right)$, with
\begin{equation}\label{eq_SYKaction}
S[\chi,J]=\int d\tau~\left[\sum_i\frac{1}{2}\chi_i\partial_\tau \chi_i+H_{\text{SYK}_q}[\chi]\right].
\end{equation}

By definition, the free-energy after the disorder average is given by $F(\beta)=-\beta^{-1}\overline{\log Z}$. For general systems, the computation of $F(\beta)$ then requires introducing $n$ disorder replicas and using $\log Z=\text{lim}_{n\rightarrow 0}(Z^n-1)/n$~\cite{altland2010condensed}. However, for the SYK model, it is known that replica off-diagonal solutions are sub-leading in $N$~\cite{kitaev2018soft}, and one can approximate $\overline{\log Z}\approx\log \overline{Z}$. Physically, this indicates the thermodynamical properties of the SYK model are self-averaging. Performing the Gaussian integral over $J_{i_1i_2...i_q}$, we have 
\begin{equation}\label{eq_Zbar}
\overline{Z}=\int \mathcal{D}\chi~e^{-\frac{1}{2}\int d\tau~\sum_i\chi_i(\tau)\partial_\tau \chi_i(\tau)+\frac{J^2}{2qN^{q-1}}\int d\tau d\tau'~[\sum_i\chi_i(\tau)\chi_i(\tau')]^q}.
\end{equation}
The combination $\sum_i\chi_i(\tau)\chi_i(\tau')/N$ takes the form of averaging over multiple variables. By the central limit theorem~\cite{grinstead1997introduction}, its variance should vanish in the large-$N$ limit. This motivates the introduction of bilocal fields $\tilde G(\tau,\tau')=\sum_i\chi_i(\tau)\chi_i(\tau')/N$, which should be imposed by an additional Lagrangian multiplier field $\tilde \Sigma(\tau,\tau')$
\begin{equation}\label{eq_delta}
1=\int \mathcal{D}\tilde G\mathcal{D}\tilde \Sigma~e^{\frac{1}{2}\int d\tau d\tau'~\tilde \Sigma(\tau,\tau')\left(\sum_i\chi_i(\tau)\chi_i(\tau')-N\tilde G(\tau,\tau')\right)}.
\end{equation}
Here the integral over $\Sigma$ is along the imaginary axis. Inserting \eqref{eq_delta} into \eqref{eq_Zbar} and replacing $\sum_i\chi_i(\tau)\chi_i(\tau')/N$ with $G(\tau,\tau')$, we find 
\begin{equation}
\overline{Z}=\int \mathcal{D}\chi\mathcal{D}\tilde G\mathcal{D}\tilde \Sigma~e^{-\int d\tau d\tau'[\frac{1}{2}\sum_i\chi_i(\partial_\tau-\tilde \Sigma)\chi_i+\frac{N}{2}(\tilde \Sigma \tilde G-\frac{J^2}{q}\tilde G^q)]}.
\end{equation}
Now the fermion fields become bilinear and thus can be integrated out straightforwardly
\begin{equation}\label{eq_GSigma}
\begin{aligned}
\overline{Z}&=\int \mathcal{D}\tilde G\mathcal{D}\tilde \Sigma~\exp(-S_{G\Sigma}[\tilde G,\tilde \Sigma]).\\
S_{G\Sigma}&=-\frac{N}{2}\log \det (\partial_\tau-\tilde \Sigma)+\frac{N}{2}\int d\tau d\tau'~(\tilde \Sigma \tilde G-\frac{J^2}{q}\tilde G^q).
\end{aligned}
\end{equation}
$S_{G\Sigma}$ is known as the $G$-$\Sigma$ action for the SYK model. Importantly, $S_{G\Sigma}$ is proportional to $N$. As a result, we can apply the saddle-point approximation for the integration over $\tilde{G}$ and $\tilde{\Sigma}$. This is consistent with our previous arguments using the central limit theorem. The saddle-point equations read
\begin{equation}\label{eq_thermalSD}
\begin{aligned}
\delta(\tau)&=\partial_\tau G(\tau)-\int d\tau'~\Sigma(\tau-\tau')G(\tau'),\\
\Sigma(\tau)&=J^2G^{q-1}(\tau)=\begin{tikzpicture}[baseline={([yshift=-6pt]current bounding box.center)}, scale=1.2]
\draw[thick] (-24pt,0pt) -- (-15pt,0pt);
\draw[thick,dashed] (-15pt,0pt)..controls (-8pt,16pt) and (8pt,16pt)..(15pt,0pt);
\draw[thick] (-15pt,0pt)..controls (-8pt,10pt) and (8pt,10pt)..(15pt,0pt);
\draw[thick] (-15pt,0pt)..controls (-8pt,-10pt) and (8pt,-10pt)..(15pt,0pt);
\draw[thick] (15pt,0pt) -- (-15pt,0pt);
\draw[thick] (15pt,0pt) -- (24pt,0pt);
\end{tikzpicture}.
\end{aligned}
\end{equation}
Here $(G,\Sigma)$ is the saddle-point value of $(\tilde{G},\tilde{\Sigma})$, which is only a function of the time difference due to the time translation symmetry. The first equation can also be written in short as $(\partial_\tau-\Sigma)\circ G=I$. This is equivalent to the Schwinger-Dyson equation of Majorana fermions with melon diagrams. In the diagram, we take $q=4$ as an example. The solid lines are Green's functions for Majorana fermions, and the dashed line represents the disorder average. The numerical algorithm for solving \eqref{eq_thermalSD} can be found in \cite{maldacena2016remarks}. After obtaining the solution, we can determine the free energy as $\beta F=I=S_{G\Sigma}[G,\Sigma]$, or explicitly
\begin{equation}
\frac{2\beta F}{N}=-\sum_n\log(-i\omega_n-\Sigma(i\omega_n))+\frac{(q-1)\beta}{q} \int d\tau ~\Sigma G,
\end{equation}
where $\omega_n=2\pi(n+1/2)/\beta$ with $n\in \mathbb{Z}$ is the Matsubara frequency. 

In the large-$q$ limit, the Green's function can be expanded in terms of $1/q$ and the free energy can be obtained in the closed form \cite{maldacena2016remarks}
\begin{equation}
-\frac{\beta F}{N}=\frac{\log 2}{2}+\frac{\pi v}{q^2}\left[\tan \frac{\pi v}{2}-\frac{\pi v}{4}\right],
\end{equation}
with $\beta \mathcal{J}=\pi v/\cos(\pi v/2)$. Under the Legendre transformation, this gives the thermal entropy
\begin{equation}\label{eq_SE}
\frac{S_\text{th}(E)}{N}=\frac{\log 2}{2}-\frac{1}{q^2}\arcsin^2\left(\frac{E}{E_0}\right).
\end{equation} 
We have defined the ground state energy $E_0=-\mathcal{J}/q^2$. In the zero-temperature limit, there is a residue entropy $S_0/N={\log 2}/{2}-{\pi^2}/{4q^2}$. Such a finite zero-temperature entropy is a general phenomenon for the SYK model. It is known that for finite $q$~\cite{georges2001quantum,maldacena2016remarks,gu2019notes},
\begin{equation}
\frac{S_0}{N}=\int_{1/q}^{1/2}dx~\pi\left(\frac{1}{2}-x\right)\tan \pi x.
\end{equation}
This indicates the existence of exponentially many low-energy states. Although \eqref{eq_SE} is derived in the large-$q$ limit, it turns out to be a good approximation for finite $q$ away from $|{E}/{E_0}|=1$. It has also been derived using a different approach in \cite{garcia2017analytical}.

One can go further by considering the fluctuation around saddle-point solutions. The quadratic order expansion determines the fluctuation $\left<(\tilde{G}-G)(\tilde{G}-G)\right>$, which corresponds to a connected four-point function of Majorana fermions. Detailed analysis can be found in \cite{maldacena2016remarks}.

\subsection{Low-energy limit and holography}\label{subsec_conformallimitAdS}
The SYK model shows emergent conformal symmetry and holographic duality in the low-temperature limit $\beta \mathcal{J}\gg 1$. Let us first perform the scaling analysis for the kinetic term $\chi \partial \chi$ and the interaction $J\chi^q$ in \eqref{eq_SYKaction}. If we assume the kinetic term is marginal, this gives the scaling dimension $[\chi]=0$, and the interaction term is relevant. This is not a stable fixed point. Consequently, we should have assumed the interaction term is marginal, which gives $[\chi]\equiv \Delta=1/q$. This implies in the low-energy limit $\tau \mathcal{J} \gg1$, we can neglect the $\partial_\tau$ term in the Schwinger-Dyson equation
\begin{equation}
-J^2\int d\tau'~G^{q-1}(\tau-\tau')G(\tau')=\delta(\tau),
\end{equation}
 and the solution reads
\begin{equation}\label{eq_SYKconformalG}
G(\tau)=b\left(\frac{\pi}{\beta \sin \frac{\pi\tau}{\beta}}\right)^{2\Delta},\ \ \ \ \ \ J^2b^q\pi=\left(\frac{1}{2}-\Delta\right)\tan\pi \Delta.
\end{equation}
This is the standard form of Green's functions in the conformal field theory (CFT), which indicates the emergence of conformal symmetry in the SYK model. Indeed, after neglecting the kinetic term, the $G$-$\Sigma$ action \eqref{eq_GSigma} is invariant under the reparametrization (or conformal transformation in 1D) $\tau \rightarrow f(\tau)$, with   
\begin{equation}
\tilde G(\tau_1,\tau_2)=(f(\tau_1)'f(\tau_2)')^\Delta \tilde G(f(\tau_1),f(\tau_2)).
\end{equation}
In particular, this indicates that fluctuations $$\tilde G_f(\tau_1,\tau_2)= (f(\tau_1)'f(\tau_2)')^\Delta G(f(\tau_1)-f(\tau_2))$$ cost no energy. These are called reparametrization modes, which governs the SYK low-energy dynamics. After adding back contributions from the kinetic term $\partial_\tau$, a non-vanishing action for these reparametrization modes can be derived \cite{maldacena2016remarks,kitaev2018soft}, which is known as the Schwarzian action
\begin{equation}\label{eq_Schwarzian}
S_\text{Sch}=-\frac{N\alpha_S}{\mathcal{J}}\int d\tau~\left\{\tan \frac{\pi f(\tau)}{\beta},\tau\right\}.
\end{equation}
Here $\{f,\tau\}=f'''/f'-3 (f''/f')^2/2$ is the Schwarzian derivative and $\alpha_S$ is some numerical factor. There are extensive discussions of the exact solution of the Schwarzian action from different perspectives \cite{bagrets2016sachdev,stanford2017fermionic,mertens2017solving,yang2019quantum}. 

The holography dual of the SYK low-energy sector refers to the fact that the Schwarzian action can also be obtained from an independent gravity calculation \cite{maldacena2016conformal}. The gravity theory is the Jackiw-Teitelboim (JT) gravity in two-dimensional Anti-de-Sitter (AdS$_2$) space. The action reads
\begin{equation}\label{JT}
S_\text{JT}=-\frac{1}{16\pi G}\left[\int dx^2\sqrt{g}\phi(R+2)+2\int_\text{bdy}\phi_b K\right].
\end{equation}
The first/second term is a bulk/boundary term. $\phi$ is the dilaton field, which can be integrated out as a Lagrangian multiplier to impose $R=-2$. In 2D, this fixes all degrees of freedom in bulk. We choose the bulk to be a Poincar\'e disk, with metric
\begin{equation}
ds^2=d\rho^2+\sinh ^2\rho ~d\theta^2.
\end{equation}
The only remaining dynamical degrees of freedom is the fluctuation of the boundary curve. We fix the boundary condition that $g_{\tau\tau}=1/\epsilon^2 \gg 1$ with boundary time $\tau$, this gives
\begin{equation}
\sinh \rho(\tau)~{\theta'(\tau)} \approx{\epsilon}^{-1}. 
\end{equation}
As a result, the shape of the boundary curve can be parametrized by a single function $\theta(\tau)$, and the energy cost is determined by the extrinsic curvature term \eqref{JT}. Explicit calculation shows the result is exactly the Schwarzian action \eqref{eq_Schwarzian}, with $2\pi f(\tau)=\beta\theta(\tau)$. In particular, the saddle-point solution $f(\tau)=\tau$ corresponds to a circle on the Poincar\'e disk. This completes the discussion of the holographic dual of the SYK low-energy sector.

\subsection{Generalizations of the SYK model}
We have introduced the original SYK model in previous subsections. Here we turn to its generalizations. The simplest generalization includes adding SYK terms with different $q$ into a single Hamiltonian $H=\sum_q c_q H_{\text{SYK}_q}$, with $c_q \neq 0$ for some $q$. In addition, there are mainly three routes for generalizing the SYK model: adding symmetry, making coupling time-dependent, and introducing multiple sites. In this subsection, we illustrate these ideas using a few examples. \\

\textbf{Adding Symmetry} Symmetry plays an important role in modern physics. The original SYK model only exhibits energy conservation, and there are several attempts to add symmetries for the SYK model. The simplest generalization is to replace Majorana fermions with complex fermions~\cite{davison2017thermoelectric,gu2019notes,chaturvedi2018note,bulycheva2017note} and impose the $U(1)$ charge conservation. This leads to the complex SYK (or cSYK) model. The Hamiltonian of the cSYK$_q$ model reads
\begin{equation}
H_{\text{cSYK}_q}=\sum_{\substack{i_1< ...< i_{q/2} \\  j_1< ...< j_{q/2}}}J_{i_1i_2...i_{q/2}j_1j_2...j_{q/2}}c^\dagger_{i_1}...c^\dagger_{i_{q/2}}c_{j_1}...c_{j_{q/2}},
\end{equation}
where the Gaussian distributions of random couplings $J_{i_1i_2...i_{q/2}j_1j_2...j_{q/2}}$ satisfy 
\begin{equation}
\overline{J_{i_1i_2...i_{q/2}j_1j_2...j_{q/2}}}=0,\ \ \ \ \ \ \overline{J_{i_1i_2...i_{q/2}j_1j_2...j_{q/2}}^2}=\frac{\frac{q}{2}!(\frac{q}{2}-1)!J^2}{N^{q-1}}.
\end{equation}
In the grand canonical ensemble, an chemical potential $\mu$ can be added to tune the charge filling $Q=\sum_i c^\dagger_i c_i$.  

The large-$N$ expansion of the cSYK model is similar to that of the Majorana SYK model. For example, the Schwinger-Dyson equation can be written as~\cite{davison2017thermoelectric,sachdev1993gapless}
\begin{equation}\label{eq_thermalSDcomplex}
\begin{aligned}
\delta(\tau)&=(\partial_\tau-\mu) G(\tau)-\int d\tau'~\Sigma(\tau-\tau')G(\tau'),\\
\Sigma(\tau)&=J^2G^{q/2}(\tau)[-G(-\tau)]^{q/2-1}=\begin{tikzpicture}[baseline={([yshift=-6pt]current bounding box.center)}, scale=1.2]
\draw[thick,mid arrow] (-24pt,0pt) -- (-15pt,0pt);
\draw[thick,dashed] (-15pt,0pt)..controls (-8pt,16pt) and (8pt,16pt)..(15pt,0pt);
\draw[thick,mid arrow] (-15pt,0pt)..controls (-8pt,10pt) and (8pt,10pt)..(15pt,0pt);
\draw[thick,mid arrow] (-15pt,0pt)..controls (-8pt,-10pt) and (8pt,-10pt)..(15pt,0pt);
\draw[thick,mid arrow] (15pt,0pt) -- (-15pt,0pt);
\draw[thick,mid arrow] (15pt,0pt) -- (24pt,0pt);
\end{tikzpicture},
\end{aligned}
\end{equation}
and the free energy can also be expressed as a functional of $G$ and $\Sigma$. In the low-energy limit, where is an additional low-energy parameter $\theta$, which is related to the charge filling $Q$ and characterizes the asymmetry of the spectral function between particles and holes~\cite{davison2017thermoelectric}. There is also an additional soft mode $\lambda(\tau)$, which is associated with the phase fluctuation~\cite{davison2017thermoelectric,gu2019notes}. \\

\textbf{Time-dependent Coupling} 
The Hamiltonian of the original SYK model is static. Later, there are studies in which couplings between Majorana fermions are also random in time. This is known as the Brownian SYK (or BSYK) model~\cite{saad2018semiclassical,sunderhauf2019quantum}. The Hamiltonian of the BSYK model is time-dependent:
\begin{equation}
H_{\text{BSYK}_q}(t)=\sum_{1\leq i_1< i_2< ...< i_q\leq N}i^{q/2}J_{i_1i_2...i_q}(t)\chi_{i_1}\chi_{i_2}...\chi_{i_q},
\end{equation}
$J_{i_1i_2...i_q}(t)$ are Gaussian variables with zero expectations and
\begin{equation}
\begin{aligned}
&\overline{J_{i_1i_2...i_q}(t)J_{i_1i_2...i_q}(t')}=\frac{(q-1)!J}{N^{q-1}}\delta(t-t').
\end{aligned}
\end{equation}

Since there is no energy conservation, only the infinite temperature ensemble is well-defined. Instead of the imaginary-time path-integral, we consider the Keldysh contour which contains one forward and one backward evolution \cite{kamenev2011field}. The partition function $Z=\text{tr}~e^{iHT}e^{-iHT}$ with $T\rightarrow \infty$ reads
\begin{equation}
Z=\int \mathcal{D}\chi^+\mathcal{D}\chi^- e^{-\int_{-\infty}^{\infty} dt~\sum_\pm\left[\sum_{i}\pm\frac{1}{2}\chi_i^\pm\partial_t \chi_i^\pm \pm iH_{\text{BSYK}_q}[\chi^\pm]\right]}.
\end{equation}
$\chi_i^\pm$ are Majorana fields on the forward/backward evolution contour. One can follow derivation in section \ref{subsec_SYKlargeN} to integrate over the random interactions, introduce the $(\tilde{G},\tilde{\Sigma})$ fields, and derive the saddle-point equations \cite{zhang2021obstacle}. The result reads 
\begin{equation}
\begin{aligned}
\left[\begin{pmatrix}
\partial_t&0\\
0&-\partial_t
\end{pmatrix}-\begin{pmatrix}
\Sigma^{++}&\Sigma^{+-}\\
\Sigma^{-+}&\Sigma^{--}
\end{pmatrix}\right]\circ \begin{pmatrix}
G^{++}&G^{+-}\\
G^{-+}&G^{--}
\end{pmatrix}=I,
\end{aligned}
\end{equation}
with self-energy $\Sigma^{ss'}(t,t')=-Jss'\delta(t-t')G^{ss'}(t,t')^{q-1}$, which is non-zero only when two time variables coincide. This is the consequence of having Brownian couplings. \\

\textbf{Multiple Sites} By introducing different copies of the SYK model and adding interactions between them, we can construct SYK-like models with more than one site~\cite{gu2017local,gu2017energy,banerjee2017solvable,chen2017competition,song2017strongly,jian2017solvable,chen2017tunable,zhang2017dispersive,bi2017instability,narayan2017syk}. As an example, we consider coupling two SYK$_4$ models using SYK$_4$ random pair hopping:
\begin{equation}\label{eq_coupledSYK44}
\begin{aligned}
H=&\sum_{\{i_n\}}J^\chi_{i_1 i_2 i_3 i_4} \chi_{i_1}\chi_{i_2}\chi_{i_3}\chi_{i_4}+\sum_{\{a_n\}}J^\psi_{a_1 a_2 a_3 a_4} \psi_{a_1}\psi_{a_2}\psi_{a_3}\psi_{a_4}\\
&+\sum_{i_1< i_2, a_1< a_2} V_{i_1 i_2 a_1 a_2}\chi_{i_1}\chi_{i_2}\psi_{a_1}\psi_{a_2}.
\end{aligned}
\end{equation}
Here we have $N_1$ $\chi$ fermions ($1\leq i_1<...< i_4\leq N_1$) and $N_2$ $\psi$ fermions ($1\leq a_1<...< a_4\leq N_2$). We take both $N_1,N_2\rightarrow \infty$ with their ratio $r=N_2/N_1$ fixed. We consider that both $J^\chi_{i_1 i_2 i_3 i_4}$ and $J^\psi_{a_1 a_2 a_3 a_4}$ are SYK random interactions with strength $J$. The random pair hopping strengths satisfy
\begin{equation}
\overline{V_{i_1 i_2 a_1 a_2}}=0,\ \ \ \ \ \ \overline{V_{i_1 i_2 a_1 a_2}^2}=\frac{2V^2}{N_1 N_2^2}.
\end{equation}
Now the self-energies become
\begin{equation}
\begin{aligned}
\Sigma_\chi(\tau)&=J^2G_\chi(\tau)^3+V^2G_\psi(\tau)^2G_\chi(\tau),\\
\Sigma_\psi(\tau)&=J^2G_\psi(\tau)^3+V^2\frac{N_1}{N_2}G_\psi(\tau)G_\chi(\tau)^2.
\end{aligned}
\end{equation}
In the low-energy limit, the emergent conformal symmetry and the soft reparametrization mode $f(\tau)$ still exist, with a Schwarzian effective action. Following the same spirit, one can introduce a chain of SYK sites with random pair hopping. This leads to higher-dimensional SYK lattices \cite{gu2017local}, in which the low-energy butterfly velocity and the energy diffusion constant has been studied.

\section{Entanglement in the SYK model}\label{sec_singleSYK}

\begin{figure}[t]
\centering
\includegraphics[width=0.5\textwidth]{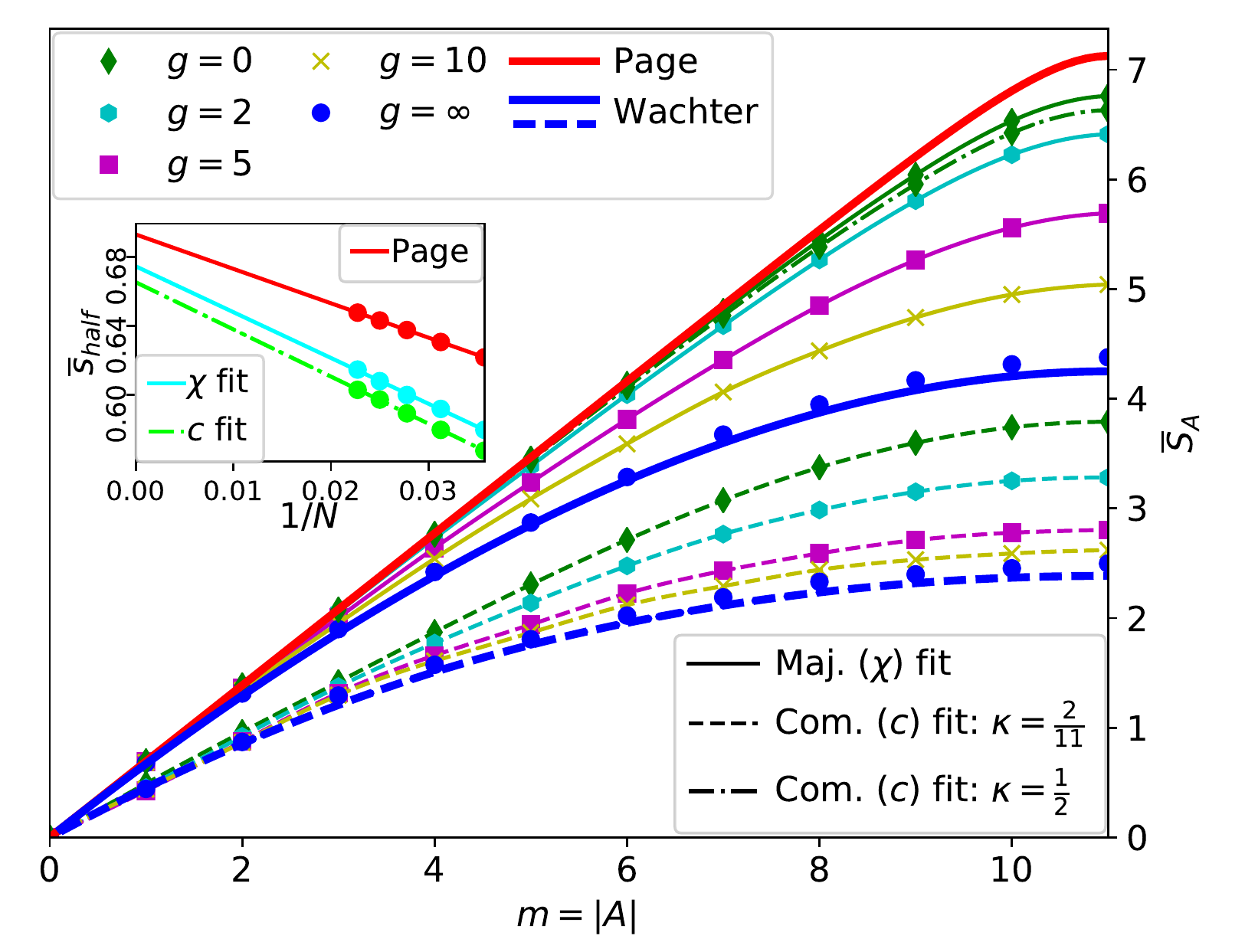}
\caption{The von Neumann entanglement entropy for the ground state of the SYK$_4$+SYK$_2$ model using ED. The result is averaged over 10 disorder realizations. $M=m$ is the subsystem size of the cSYK model, while for the Majorana SYK the subsystem size is $M=2m$. The system contains $N=22$ complex fermions with filling fraction $\kappa=Q/N$ or $N=44$ Majorana fermions. Adapted from \cite{liu2018quantum}. }\label{fig_SYKED}
\end{figure}

After introducing the SYK model in section \ref{sec_introduceSYK}, we now review existing results for its entanglement entropy. We divide the system into $A$ and $B$, where $A$ contains $M$ Majorana modes and $B$ contains $N-M$ Majorana modes. We fix $M/N=\lambda$ when taking the large-$N$ limit. The definition of $A$ and $B$ subsystems in the cSYK model is similar. Given the density matrix of the total system $\rho$, the reduced density matrix of $A$ can be computed as $\rho_A=\text{tr}_B~\rho$. The $n$-th R\'enyi entropy $S^{(n)}_A$ and the von Neumann entropy $S_A$ are defined as 
\begin{equation}
S^{(n)}_A=-\frac{1}{n-1}\log~\text{tr}_A~\rho_A^n,\ \ \ \ \ \ S_A=-\text{tr}_A~\rho_A \log \rho_A.
\end{equation}
It is straightforward to show that the von Neumann entropy is a special limit of the R\'enyi entropy $S_A=S^{(1)}_A$. 

There are three different approaches for studying either $S^{(n)}_A$ or $S_A$ in the SYK model: the ED, the ETH assumption, and the path-integral representation. All of them have some advantages and disadvantages. The ED can give exact numerical results for both $S^{(n)}_A$ and $S_A$ of any state, but is limited to small system size $N$ and small $q$. The ETH analysis can lead to large-$N$ results for $S_A$, but it requires further justification and knowledge about thermodynamics. The path-integral approach works in the large-$N$ limit and can be easily extended to other models, but in most cases is restricted to $S^{(n)}_A$ with small integers $n\geq 2$. In the following subsections, we summarize results obtained using these approaches separately, and show their consistency. 

\subsection{The exact diagonalization}
We first review results obtained from the exact diagonalization in~\cite{liu2018quantum}. In this work, authors numerically study the ground-state von Neumann entropy $S_A$ for a complex fermion model with Hamiltonian
\begin{equation}
H_\text{c}=H_{\text{cSYK}_4}+\frac{g}{N}H_{\text{cSYK}_2},
\end{equation}
or a Majorana fermion model with $H_{\text{cSYK}_q}$ replaced by its Majorana counterpart $H_{\text{SYK}_q}$. Using the sparse matrix method, authors study systems with $N=22$ complex fermions or $N=44$ Majorana fermions. For complex fermions, one can further fix the charge filling $\kappa=Q/N$. The result is shown in Figure \ref{fig_SYKED}. 

For $M\ll N$, the entanglement entropy is maximal for any $g$. For the Majorana fermion model, this corresponds to $S_A=M\log 2/2$. For the complex fermion model, it depends on the filling fraction $\kappa$:
\begin{equation}\label{eq_maximal}
S_A=-M\left(\kappa \log\kappa+(1-\kappa) \log(1-\kappa) \right).
\end{equation} 
For $M/N\sim O(1)$, the entanglement entropy is lowered when $g$ is increased. For $g\rightarrow \infty$, the entanglement entropy matches the SYK$_2$ result, as will be discussed in section \eqref{subsec_SYK2EE}. The authors also compare the numerical results with the Page curve~\cite{page1993average}, which is the von Neumann entanglement entropy of a random pure state 
\begin{equation}
S_A^\text{Page}=M\log 2-2^{2M-N-1}.
\end{equation}
Here we take the complex fermion model as an example. For the Majorana model, one should make the replacement of $(N,M)\rightarrow (N,M)/2$. The first term is the maximal entropy, and the second term is a small correction that vanishes in the large-$N$ limit. As shown in Figure \ref{fig_SYKED}, the numerical result of the SYK$_4$ model is smaller than the Page value. The difference is expected to become smaller when we consider models with larger $q$.

\subsection{The eigenstate thermalization hypothesis}

\begin{figure}[t]
\centering
\includegraphics[width=0.85\linewidth]{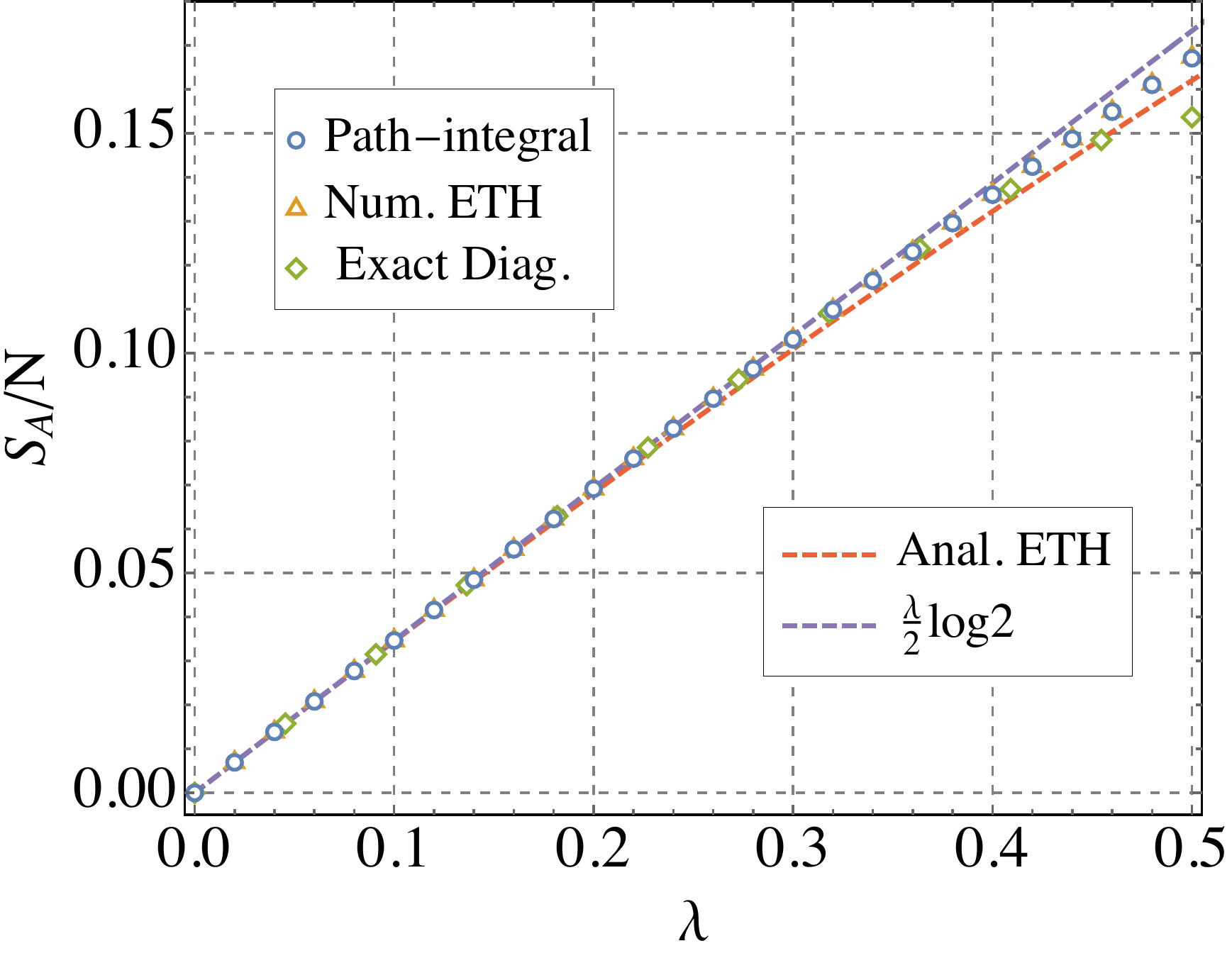}
\caption{A comparison between different approaches for the entangelement entropy for the SYK$_4$ model. In the path-integral approach, we consider $S^{(2)}_A$ set $\beta J=50$~\cite{zhang2020subsystem,zhang2020entanglement}. The ED data is copied from Figure \eqref{fig_SYKED}. The analytical ETH result is plotted according to \eqref{eq_anaETH} and the numerical ETH result is adapted from~\cite{zhang2020subsystem}. }\label{fig_SYKEEcompare}
\end{figure}

The entanglement entropy of the SYK$_q$ model can also be studied using the ETH. The ETH states that an isolated quantum system can serve as its own bath. More precisely, let us consider a eigenstate $|\psi\rangle$ of some local chaotic Hamiltonian $H$ with energy $E=\langle\psi|H|\psi\rangle$. If we measure some local observable $O$ in the subsystem $A$, the result will match the prediction from a thermal ensemble
\begin{equation}
\langle\psi|O|\psi\rangle=\frac{\text{tr}~O e^{-\beta H}}{\text{tr}~e^{-\beta H}}.
\end{equation}
Here $\beta$ is fixed by matching the energy with $O=H$. This can be understood as proposing the density matrix of the subsystem $A$ is almost thermal. Consequently, the entanglement entropy of the subsystem $A$ matches the corresponding thermal entropy $S_A=M S_\text{th}(E)/N$. 

The SYK model describes Majorana fermions with non-local interactions. As a result, this analysis should be modified following the discussions in~\cite{huang2019eigenstate}. In this subsection, we outline the basic idea. For simplicity, we choose the subsystem $A$ to contain Majorana modes $\chi_i$ with $i=1,2,...,M$. To understand the reduced density matrix of subsystem $A$, we first define its Hamiltonian $H_A$ by restricting the summation over indices in \eqref{H0}:
\begin{equation}
H_{A}=\sum_{1\leq i_1<i_2<...<i_q\leq M}i^{q/2}J_{i_1i_2...i_q}\chi_{i_1}\chi_{i_2}...\chi_{i_q}.
\end{equation}
Comparing to the original Hamiltonian, which contains $\sim N^q$ terms, here the summation is restricted to $\sim M^q$ terms. As a result, to the leading order of $1/N$, we have
\begin{equation}
E_A=\langle\psi|H_A|\psi\rangle=\frac{M^q}{N^q}\langle\psi|H|\psi\rangle=\lambda^qE.
\end{equation}
Here we have assumed the state is symmetric under the permutation of indices. This is very different from the case with local Hamiltonians, where the energy density will not change. 

In the spirit of ETH, we assume the reduced density matrix $\rho_A$ is a thermal density matrix of $H_A$ for $\lambda<1/2$, and the entanglement entropy $S_A$ matches thermal entropy. To determine the entropy density $s=S_A/M$, we need additional knowledge about $H_A$. Assume the ground state energy of the SYK model with $N$ Majorana fermions with coupling $J$ is $E_0=N c_0 J$, the ground state energy $E_{A,0}$ of $H_A$ is determined by realizing 
\begin{equation}
\overline{J_{i_1i_2...i_q}^2}=\frac{(q-1)!J^2}{N^{q-1}}=\frac{(q-1)!(\lambda^{(q-1)/2} J)^2}{M^{q-1}}.
\end{equation}
As a result, $H_A$ is an SYK model that contains $M$ Majorana fermions with coupling strength $J\lambda^{(q-1)/2}$. This gives $E_{A,0}=M c_0 \lambda^{(q-1)/2}=\lambda^{{(q+1)}/{2}}E_0$. We have
\begin{equation}
\frac{E_{A}}{E_{A,0}}=\lambda^{\frac{q-1}{2}}\frac{E}{E_0},
\end{equation}
Physically, this means that for non-local Hamiltonians, the effective temperature changes when varying the subsystem size. The entanglement entropy is then determined by the thermal entropy with the corresponding energy density. In the limit of $\lambda \rightarrow 0$, the energy density vanishes, and the thermal ensemble is at infinite temperature. This explains the observed maximal entropy behavior for small subsystem size in ED. 

A simple analytical approximation of $S_A$ can be derived using \eqref{eq_SE}:
\begin{equation}\label{eq_anaETH}
\frac{S_A}{N}=\lambda\left(\frac{\log 2}{2}-\frac{1}{q^2}\arcsin^2\left(\lambda^{\frac{q-1}{2}}\frac{E}{E_0}\right)\right).
\end{equation}
In particular, for the ground state of the SYK$_4$ model ($E/E_0=1$), this gives $S_A=0.338M$ at $\lambda=1/2$. On the other hand, one can also perform the numerical study to derive the equation of state as explained in \eqref{subsec_SYKlargeN}. The numerical results are shown in Figure \ref{fig_SYKEEcompare}, which match results of the path-integral approach (explained in the next subsection) to good precision.

\subsection{The path-integral approach}\label{subsec_singleSYKpathintegral}
The R\'enyi entropy in the SYK model can also be studied using the large-$N$ expansion. This is only possible for certain states where the path-integral takes a simple form. In this section, we consider the SYK model prepared in either a thermal ensemble or a Kourkoulou-Maldacena (KM) pure state~\cite{kourkoulou2017pure}. We focus on the second R\'enyi entropy $S_A^{(2)}$, while a generalization to higher orders is straightforward. When the total system is in a thermal ensemble, the result contains both entanglement contributions and thermal contributions. However, as we will show later, the subsystem entropy of a thermal ensemble matches the steady-state entanglement entropy of the KM pure-state with the same energy for $\lambda<1/2$, and can be directly compared to the ED and the ETH results.

\textbf{Thermal Ensemble} We first consider the subsystem entropy $S^{(2)}_A$ of a thermal ensemble $\rho=e^{-\beta H}/Z$~\cite{zhang2020subsystem}. A pictorial representation of the density matrix $\rho$ and the reduced density matrix $\rho_A$ read
\begin{equation}
Z\rho=
\begin{tikzpicture}[scale=0.7,baseline={([yshift=0pt]current bounding box.center)}]
   \draw (1,-0.2) arc(0:-180:1 and 1);
   \draw (1.3,-0.2) arc(0:-180:1.3 and 1.3);

       \draw[dotted,thick] (0.866,-0.725) -- (1.126,-0.875);
        \draw[dotted,thick] (0.5,-1.091) -- (0.65,-1.35);
              \draw[dotted,thick] (0,-1.225) -- (0,-1.525);
        \draw[dotted,thick] (-0.5,-1.091) -- (-0.65,-1.35);
           \draw[dotted,thick] (-0.866,-0.725) -- (-1.126,-0.875);

      \draw (0.4,-0.8) node{$B$};
       \draw (0.9,-1.4) node{$A$};
       \draw (-1.5,-0.2) node{$\beta$};

       \draw (1.5,-0.2) node{$0$};
\end{tikzpicture},\ \ \ \ Z\rho_A=
\begin{tikzpicture}[scale=0.7,baseline={([yshift=0pt]current bounding box.center)}]
   \draw (0.65,-0.65) arc(0:-360:0.65 and 0.65);
   \draw (1.3,-0.2) arc(0:-180:1.3 and 1.3);

       \draw[dotted,thick] (0.6,-0.4) -- (1.126,-0.875);
        \draw[dotted,thick] (0.5,-1.091) -- (0.65,-1.35);
              \draw[dotted,thick] (0,-1.35) -- (0,-1.475);
        \draw[dotted,thick] (-0.5,-1.091) -- (-0.65,-1.35);
           \draw[dotted,thick] (-0.6,-0.4) -- (-1.126,-0.875);

      \draw (0.,-0.6) node{$B$};
       \draw (0.9,-1.4) node{$A$};

       \draw (-1.5,-0.2) node{$\beta$};
       \draw (1.5,-0.2) node{$0$};
       \filldraw  (0pt,0.15pt) circle (1.5pt) node[left]{\scriptsize $ $};
\end{tikzpicture}.
\end{equation}
Here the solid lines represent the imaginary-time evolutions, and the dashed lines represent the interaction between subsystems. The black dot indicates the anti-periodic boundary condition for fermion fields, and the free ends can be assigned with quantum states to give matrix elements of density matrices. To compute $S^{(2)}_A$, we sew two $\rho_A$ together:
\begin{equation}
\exp(-S_A^{(2)})=\text{tr}~\rho_A^2=\frac{1}{Z^2}\times \left[\begin{tikzpicture}[scale=0.7,baseline={([yshift=0pt]current bounding box.center)}]
   \draw (0.6,-0.85) arc(0:-360:0.6 and 0.6);
   \draw (0.6,0.45) arc(0:-360:0.6 and 0.6);
   \draw (1.3,-0.2) arc(0:-360:1.3 and 1.3);

        \draw[dotted,thick] (0.4,-0.4) -- (1.126,-0.875);

        \draw[dotted,thick] (0.5,-1.191) -- (0.65,-1.35);

        \draw[dotted,thick] (0,-1.44) -- (0,-1.475);

        \draw[dotted,thick] (-0.5,-1.191) -- (-0.65,-1.35);

        \draw[dotted,thick] (-0.4,-0.4) -- (-1.126,-0.875);

        \draw[dotted,thick] (0.4,0) -- (1.126,0.475);

        \draw[dotted,thick] (0.5,0.791) -- (0.65,0.95);

        \draw[dotted,thick] (0,1.04) -- (0,1.075);

        \draw[dotted,thick] (-0.5,0.791) -- (-0.65,0.95);

        \draw[dotted,thick] (-0.4,0) -- (-1.126,0.475);

      \draw (0.,-0.85) node{$B_1$};
      \draw (0.,0.45) node{$B_2$};
       \draw (0.9,-1.4) node{$A$};
       \draw (-1.5,-0.2) node{$\beta$};
       \draw (1.5,0.2) node{$2\beta$};
       \draw (1.5,-0.6) node{$0$};
       \filldraw  (0pt,-7pt) circle (1pt) node[left]{\scriptsize $ $};
       \filldraw  (0pt,-4pt) circle (1pt) node[left]{\scriptsize $ $};
       \filldraw  (37pt,-5.5pt) circle (1pt) node[left]{\scriptsize $ $};
\end{tikzpicture}\right].
\end{equation}
The contribution in the square bracket is very similar to the path-integral of a thermal ensemble with inverse temperature $2\beta$, which reads
\begin{equation}
\begin{aligned}
Z^{(2)}=\int_{\text{b.c.}} \mathcal{D}\chi~e^{-\int_0^{2\beta} d\tau~\left[\sum_i\frac{1}{2}\chi_i\partial_\tau \chi_i+H_{\text{SYK}_q}[\chi]\right]}.
\end{aligned}
\end{equation}
The only difference is the boundary condition. Here we should impose 
\begin{equation}
\begin{aligned}
&\chi_{i}(0^+)=-\chi_{i}(2\beta^-),\ \ \ \ \ \chi_{i}(\beta^+)=\chi_{i}(\beta^-),\ \ \ \ \ \ \text{if} \ \ \ i\in A;\\
&\chi_{i}(0^+)=-\chi_{i}(\beta^-),\ \ \ \ \ \chi_{i}(\beta^+)=-\chi_{i}(2\beta^-),\ \ \ \ \text{if} \ \ \ i\in B.\label{eq_bc}
\end{aligned}
\end{equation}
These boundary conditions break the time translation symmetry and the Green's functions depend on both time variables. In \cite{haldar2020renyi}, this boundary condition is expressed in an equivalent form with an integral over auxiliary fermion modes.

To proceed, we need to average over the disorder realizations as $\overline{S_A^{(2)}}=-\overline{\log Z^{(2)}/Z^2}$. Similar to the calculation of the free energy, here we assume the disorder replica is diagonal, and thus $\overline{S_A^{(2)}}=-\log \overline{Z^{(2)}}/\overline{Z}^2$ with
\begin{equation}
\overline{Z^{(2)}}=\int dJ P(J)\int_{\text{b.c.}} \mathcal{D}\chi~e^{-\int_0^{2\beta} d\tau~\left[\sum_i\frac{1}{2}\chi_i\partial_\tau \chi_i+H_{\text{SYK}_q}[\chi]\right]}.
\end{equation}

The calculation of $\overline{Z^{(2)}}$ is parallel to that of $\overline{Z}$. One can integrate over disorder realizations, introduce bilocal fields 
\begin{equation}
\begin{aligned}
\tilde G_A(\tau,\tau')&=\frac{1}{M}\sum_{i\in A}\chi_i(\tau)\chi_i(\tau'),\\\tilde G_B(\tau,\tau')&=\frac{1}{N-M}\sum_{i\in B}\chi_i(\tau)\chi_i(\tau'),
\end{aligned}
\end{equation}
 with Lagrangian multipliers $\tilde{\Sigma}_{A/B}$, and make the saddle-point approximation after integrating out Majorana modes. Skipping the details, the saddle-point equations read
\begin{equation}
\begin{aligned}
G_A&=(\partial_\tau-\Sigma_A)^{-1}_A,\ \ \ \ \ \ G_B=(\partial_\tau-\Sigma_B)^{-1}_B,\\\Sigma_A&=\Sigma_B=J^2(\lambda G_A+(1-\lambda)G_B)^{q-1}. \label{eq_SDee}
\end{aligned}
\end{equation}
These self-energies can again be understood by melon diagrams, with each internal line being $G_A$ or $G_B$ with probability $\lambda$ or $1-\lambda$. We add subscripts $A/B$ as a reminder that the inversion should be performed under the corresponding boundary condition \eqref{eq_bc}. As a result, although the self-energy for fermions in $A/B$ systems are the same, Green's functions are different. Generalizations to the cSYK model or with multiple flavors are straightforward. After solving \eqref{eq_SDee}, the on-shell action $I^{(2)}=-\log \overline{Z^{(2)}}$ reads
\begin{equation}
\begin{aligned}\label{eq_SYKthermalentropy}
\frac{I^{(2)}}{N}=&\frac{\lambda}{2}\log \det G_A+\frac{\lambda(q-1)}{2q}\int ds ds'G_A\Sigma_A\\
&+ (\lambda\rightarrow 1-\lambda,\ A\rightarrow B).
\end{aligned}
\end{equation}
Recall that we have $\overline{Z}=e^{-I}$, the subsystem R\'enyi entropy is given by $\overline{S^{(2)}_A}=I^{(2)}-2I$. Generalizations to the cSYK model and coupled SYK models are straightforward. In particular, for cSYK model at half-fililng, the large-$N$ entropy is only different from the Majorana case by a factor of $2$~\cite{zhang2020subsystem}. 

For $\lambda=1$, the calculation is reduced to the thermal R\'enyi entropy. For $\lambda \ll 1$, one can relate the partition function to the normalization of KM pure states and prove that the subsystem entropy is maximal~\cite{zhang2020subsystem}, consistent with previous discussions. In more general cases, one needs to perform a numerical study by solving the saddle-point equation \eqref{eq_SDee}. Compared to the thermal equilibrium calculation, this requires additional efforts since Green's function depends on two-time variables. Fortunately, Green's function of the SYK model is smooth, and a small discretization is enough. Then we can store each Green's functions as matrices and directly perform iterations as explained in~\cite{zhang2020subsystem}. As an example, we show results with $q=4$ and $\beta J=50$ in Figure \ref{fig_SYKEEcompare} for $\lambda<1/2$. It clearly shows the subsystem entropy is smaller than the maximal value for $\lambda \sim O(1)$. Interestingly, this subsystem entropy matches the entanglement entropy of thermalized pure states. This will be explained below. 

\textbf{KM Pure States} Since the entanglement entropy is messed up with the thermal entropy if the full system is in a mixed state, it is natural to ask whether some pure-state entanglement entropy can be computed using the path-integral approach for the SYK model. As discussed in~\cite{zhang2020entanglement}, such a calculation is possible for KM pure states~\cite{kourkoulou2017pure}. 

The KM pure states are defined similarly to the thermofield double (TFD) state~\cite{israel1976thermo} (also see section \ref{sec_SYKbath}). We first prepare a maximally entangled state between Majorana modes. Without loss of generality, we choose to entangle fermions with even and odd indices. Defining $c_j=\frac{\chi_{2j-1}+i\chi_{2j}}{\sqrt{2}}$, we introduce the (unnormalized) KM pure states
\begin{equation}
|\text{KM}(\{s\},\beta)\rangle=e^{-\frac{\beta H}{2}}|\{s\}\rangle,\ \ \ \ \ \ (1-2n_j)|\{s\}\rangle=s_j|\{s\}\rangle.
\end{equation}
Here $n_j=c^\dagger_j c_j$ and $s_j=\pm1$. Under transformation $\chi_{2j}\rightarrow - \chi_{2j}$, $s_j$ changes sign. As a result, we are free to average over all $\{s_j=\pm 1\}$ when computing the energy. Due to the completeness of $|\{s\}\rangle$, we find
\begin{equation}
\frac{\langle\text{KM}(\{s\},\beta)|H|\text{KM}(\{s\},\beta)\rangle}{\langle\text{KM}(\{s\},\beta)|\text{KM}(\{s\},\beta)\rangle}=\frac{1}{Z}\text{tr}~H e^{-\beta H},
\end{equation}
and the energy density matches the corresponding thermal ensemble. A holographic interpretation for KM states has also been proposed in \cite{kourkoulou2017pure}. The KM pure states are generally non-thermal. In particular, they are product states when $\beta=0$. To make the states thermal, we add additional real-time evolutions, and the states become
\begin{equation}
|\text{KM}(\{s\},\beta,t)\rangle=e^{-iHt}e^{-\frac{\beta H}{2}}|\{s\}\rangle,.
\end{equation}
\begin{figure}[t]
\centering
\includegraphics[width=0.85\linewidth]{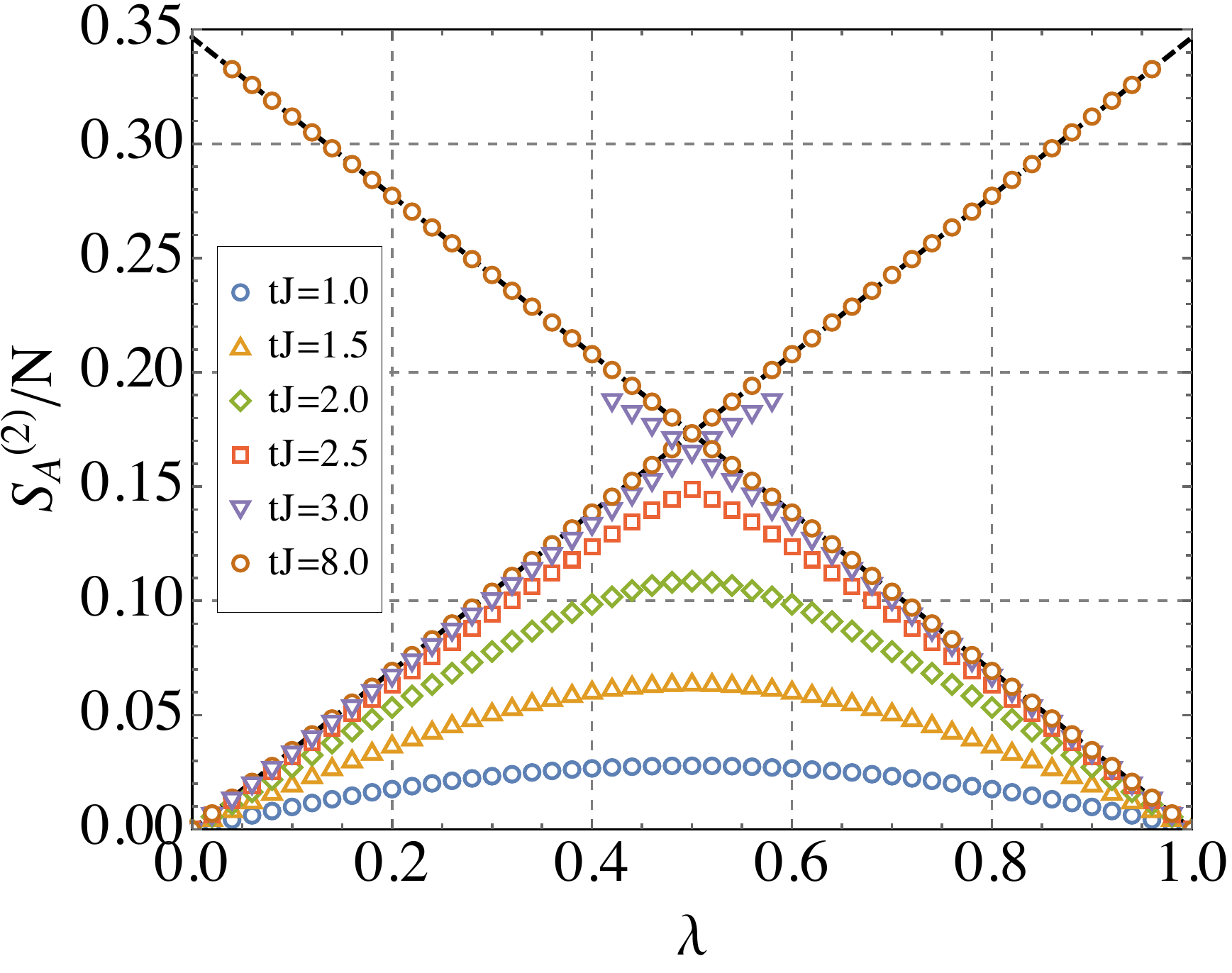}
\caption{The R\'enyi entanglement entropy $\mathcal{S}_A^{(2)}(t)$ on KM pure state as a function of subsystem size $\lambda$ for $q=4$ and $\beta=0$. The black dashed lines are $\lambda \log(2)/2$ and $(1-\lambda)\log(2)/2$. Adapted from \cite{zhang2020entanglement}. }\label{fig_SYKKM}
\end{figure}

The path-integral representation of KM pure states can be pictorially drawn as
\begin{equation}
|\text{KM}(\{s\},\beta)\rangle=e^{-\frac{\beta H}{2}}|\{1\}\rangle=\begin{tikzpicture}[thick,scale = 0.4,baseline={([yshift=0pt]current bounding box.center)}]
 \draw (0.7,0.8) arc(0:-180:0.5 and 0.5);

 \draw  (0.7,0.8)  --  (0.7,2.8);
 \draw  (-0.3,0.8)  --  (-0.3,2.8);

 \draw[dotted]  (-0.3,0.8)  --  (0.7,0.8);
 \draw[dotted]  (-0.3,1.4)  --  (0.7,1.4);
 \draw[dotted]  (-0.3,2)  --  (0.7,2);
 \draw[dotted]  (-0.3,2.6)  --  (0.7,2.6);

 \filldraw  (0.2,0.3) circle (1.5pt) node[left]{\scriptsize $ $};

 \draw (1,-0.1) node{\scriptsize $|\{s\}\rangle$};
 \draw (1,0.6) node{\scriptsize $0$};

 \draw (1,2.8) node{\scriptsize $\frac{\beta}{2}$};
 \draw (-0.6,0.6) node{\scriptsize $0$};

 \draw (-0.6,2.8) node{\scriptsize $\frac{\beta}{2}$};

 \draw (1.4,1.7) node{\scriptsize $\chi_2$};
 \draw (-1,1.7) node{\scriptsize $\chi_1$};
\end{tikzpicture}
\end{equation}
Here $\chi_1/\chi_2$ represents Majorana modes with odd/even indices. The solid line represents an imaginary-time evolution $\tau\in[0,\beta/2]$. The black dot represents the boundary condition that $(\chi_{2j-1}+is_j\chi_{2j})|\{s\}\rangle=0$, which can be solved by introducing a single $\chi_j(s)$ field with $s\in [0,\beta]$, where
\begin{equation}
\chi_j(s)=\begin{cases}
\chi_{2j-1}(\beta/2-s)\ \ \ \ \ \ \ \ \ \ \ \ \ \text{for}\ s\in[0,\beta/2),\\
-is_j\chi_{2j}(s-\beta/2)\ \ \ \ \ \ \ \ \ \ \text{for}\ s\in[\beta/2,\beta).
\end{cases}
\end{equation}
This reduces the original problem for $N$ Majorana fields with boundary conditions to $N/2$ Majorana fields with the smooth condition and doubled the contour length. Choosing the $A$ subsystem containing $M/2$ pair of Majorana fermions, we can express $S^{(2)}_A(t)$ as 
\vspace{-10pt}
\begin{equation}
e^{-S^{(2)}_A(t)}=\frac{1}{Z_{\text{KM}}^2}\times\begin{tikzpicture}[thick,scale = 0.35,baseline={([yshift=-4pt]current bounding box.center)}]
 \draw[red] (-2,-1.5) arc(0:-180:0.5 and 0.5);
 \draw[red] (-2,1.5) arc(0:180:0.5 and 0.5);
 \draw[red] (2,-1.5) arc(-180:0:0.5 and 0.5);
 \draw[red] (2,1.5) arc(180:0:0.5 and 0.5);
 \draw[blue] (-1.5,2) arc(-90:-270:0.5 and 0.5);
 \draw[blue] (-1.5,-2) arc(90:270:0.5 and 0.5);
 \draw[blue] (1.5,2) arc(-90:90:0.5 and 0.5);
 \draw[blue] (1.5,-2) arc(90:-90:0.5 and 0.5);

\filldraw  (-2.5,2) circle (1.5pt) node[left]{\scriptsize $ $};
\filldraw  (-2.5,-2) circle (1.5pt) node[left]{\scriptsize $ $};
\filldraw  (2.5,2) circle (1.5pt) node[left]{\scriptsize $ $};
\filldraw  (2.5,-2) circle (1.5pt) node[left]{\scriptsize $ $};

\filldraw  (-2,2.5) circle (1.5pt) node[left]{\scriptsize $ $};
\filldraw  (-2,-2.5) circle (1.5pt) node[left]{\scriptsize $ $};
\filldraw  (2,2.5) circle (1.5pt) node[left]{\scriptsize $ $};
\filldraw  (2,-2.5) circle (1.5pt) node[left]{\scriptsize $ $};

 \draw[red]  (-2,-1.5)  --  (-2,1.5);
 \draw[red]  (-3,-1.5)  --  (-3,1.5);
 \draw[red]  (2,-1.5)  --  (2,1.5);
 \draw[red]  (3,-1.5)  --  (3,1.5);
 
 \draw[blue]  (-1.5,-2)  --  (1.5,-2);
 \draw[blue]  (-1.5,-3)  --  (1.5,-3);
 \draw[blue]  (-1.5,2)  --  (1.5,2);
 \draw[blue]  (-1.5,3)  --  (1.5,3);

 \draw[red] (-2.5,0) node{$A$};
 \draw[blue] (0,2.5) node{$B$};
 \draw[red] (2.5,0) node{$A$};
 \draw[blue] (0,-2.5) node{$B$};

  \draw[dotted]  (-3,1.5)  --  (-2,1.5);
  \draw[dotted]  (-3,0.9)  --  (-2,0.9);
  \draw[dotted]  (-3,0.3)  --  (-2,0.3);
  \draw[dotted]  (-3,-1.5)  --  (-2,-1.5);
  \draw[dotted]  (-3,-0.9)  --  (-2,-0.9);
  \draw[dotted]  (-3,-0.3)  --  (-2,-0.3);

  \draw[dotted]  (3,1.5)  --  (2,1.5);
  \draw[dotted]  (3,0.9)  --  (2,0.9);
  \draw[dotted]  (3,0.3)  --  (2,0.3);
  \draw[dotted]  (3,-1.5)  --  (2,-1.5);
  \draw[dotted]  (3,-0.9)  --  (2,-0.9);
  \draw[dotted]  (3,-0.3)  --  (2,-0.3);

  \draw[dotted]  (1.5,3)  --  (1.5,2);
  \draw[dotted]  (0.9,3)  --  (0.9,2);
  \draw[dotted]  (0.3,3)  --  (0.3,2);
  \draw[dotted]  (-1.5,3)  --  (-1.5,2);
  \draw[dotted]  (-0.9,3)  --  (-0.9,2);
  \draw[dotted]  (-0.3,3)  --  (-0.3,2);

  \draw[dotted]  (1.5,-3)  --  (1.5,-2);
  \draw[dotted]  (0.9,-3)  --  (0.9,-2);
  \draw[dotted]  (0.3,-3)  --  (0.3,-2);
  \draw[dotted]  (-1.5,-3)  --  (-1.5,-2);
  \draw[dotted]  (-0.9,-3)  --  (-0.9,-2);
  \draw[dotted]  (-0.3,-3)  --  (-0.3,-2);

  \draw[dotted]  (-2,1.5)  --  (-1.5,2);
  \draw[dotted]  (-2,0.9)  --  (-0.9,2);
  \draw[dotted]  (-2,0.3)  --  (-0.3,2);

  \draw[dotted]  (2,1.5)  --  (1.5,2);
  \draw[dotted]  (2,0.9)  --  (0.9,2);
  \draw[dotted]  (2,0.3)  --  (0.3,2);

  \draw[dotted]  (-2,-1.5)  --  (-1.5,-2);
  \draw[dotted]  (-2,-0.9)  --  (-0.9,-2);
  \draw[dotted]  (-2,-0.3)  --  (-0.3,-2);

  \draw[dotted]  (2,-1.5)  --  (1.5,-2);
  \draw[dotted]  (2,-0.9)  --  (0.9,-2);
  \draw[dotted]  (2,-0.3)  --  (0.3,-2);

  \draw[thick]  (-2.1,0)  --  (-1.9,0);
  \draw[thick]  (-3.1,0)  --  (-2.9,0);
  \draw[thick]  (2.1,0)  --  (1.9,0);
  \draw[thick]  (3.1,0)  --  (2.9,0);

  \draw[thick]  (-2.1,0.9)  --  (-1.9,0.9);
  \draw[thick]  (-3.1,0.9)  --  (-2.9,0.9);
  \draw[thick]  (2.1,0.9)  --  (1.9,0.9);
  \draw[thick]  (3.1,0.9)  --  (2.9,0.9);

  \draw[thick]  (-2.1,-0.9)  --  (-1.9,-0.9);
  \draw[thick]  (-3.1,-0.9)  --  (-2.9,-0.9);
  \draw[thick]  (2.1,-0.9)  --  (1.9,-0.9);
  \draw[thick]  (3.1,-0.9)  --  (2.9,-0.9);

  \draw[thick]  (0,-2.1)  --  (0,-1.9);
  \draw[thick]  (0,-3.1)  --  (0,-2.9);
  \draw[thick]  (0,2.1)  --  (0,1.9);
  \draw[thick]  (0,3.1)  --  (0,2.9);

  \draw[thick]  (0.9,-2.1)  --  (0.9,-1.9);
  \draw[thick]  (0.9,-3.1)  --  (0.9,-2.9);
  \draw[thick]  (0.9,2.1)  --  (0.9,1.9);
  \draw[thick]  (0.9,3.1)  --  (0.9,2.9);

  \draw[thick]  (-0.9,-2.1)  --  (-0.9,-1.9);
  \draw[thick]  (-0.9,-3.1)  --  (-0.9,-2.9);
  \draw[thick]  (-0.9,2.1)  --  (-0.9,1.9);
  \draw[thick]  (-0.9,3.1)  --  (-0.9,2.9);

  \draw[mid arrow, red] (-3,-0.8) -- (-3,0);
  \draw[mid arrow, red] (-2,-0.8) -- (-2,0);
  \draw[mid arrow, red] (3,-0.8) -- (3,0);
  \draw[mid arrow, red] (2,-0.8) -- (2,0);

  \draw[mid arrow, red] (-3,0.8) -- (-3,0);
  \draw[mid arrow, red] (-2,0.8) -- (-2,0);
  \draw[mid arrow, red] (3,0.8) -- (3,0);
  \draw[mid arrow, red] (2,0.8) -- (2,0);

  \draw[mid arrow, blue] (-0.8,-3) -- (0,-3);
  \draw[mid arrow, blue] (-0.8,-2) -- (0,-2);
  \draw[mid arrow, blue] (-0.8,3) -- (0,3);
  \draw[mid arrow, blue] (-0.8,2) -- (0,2);

  \draw[mid arrow, blue] (0.8,-3) -- (0,-3);
  \draw[mid arrow, blue] (0.8,-2) -- (0,-2);
  \draw[mid arrow, blue] (0.8,3) -- (0,3);
  \draw[mid arrow, blue] (0.8,2) -- (0,2);

\end{tikzpicture},\label{contourrhoA2t}
\end{equation}
Here we have defined the normalization of KM states $Z_{\text{KM}}=\langle\text{KM}|\text{KM}\rangle$. The evolution contour involves both imaginary-time and real-time evolution, with a direction indicated by the arrows. Now we can parametrize the contour by a single real variable $s\in[0,4(\beta+2t)]$. The saddle-point equation can be derived as
\begin{equation}
\begin{aligned}
G_A&=(\partial_s-\Sigma_A)^{-1}_A,\ \ \ \ \ \ G_B=(\partial_s-\Sigma_B)^{-1}_B,\\
\Sigma_A&=\Sigma_B=J^2(\lambda G_A+(1-\lambda)G_B)^{q-1}\tilde P.\label{SDKMt}
\end{aligned}
\end{equation}
Comparing to the thermal ensemble case \eqref{eq_SDee}, the differences are two-folds. Firstly, the boundary condition is now different. We have two separate contours for both subsystems. Secondly, there is an additional factor $\tilde{P}(s,s')=P(s,s')f(s)f(s')$. $P(s,s')=1$ if $\chi_j(s)$ and $\chi_j(s')$ both correspond to original fields with even/odd indices and otherwise zero. This is due to the fact that random interactions for $\chi_{2j-1}$ and $\chi_{2j}$ are independent. The factor $f(s)=1$ for imaginary-time evolutions, and $f(s)=\pm i$ for forward/backward real-time evolutions. The detailed derivation can be found in \cite{zhang2020entanglement}. With the saddle-point solution, we can compute
\begin{equation}
\begin{aligned}
\frac{I^{(2)}_{\text{KM}}(\lambda)}{N}=&\frac{\lambda}{4}\log \det G_A+\frac{\lambda(q-1)}{4q}\int ds ds'G_A\Sigma_A\\
&+ (\lambda\rightarrow 1-\lambda,\ A\rightarrow B). \label{eq_SKM}
\end{aligned}
\end{equation} 
Comparing to \eqref{eq_SYKthermalentropy}, there is an additional factor of $1/2$. This is because we are now counting Majorana modes in terms of $N/2$ pairs. The R\'enyi entanglement entropy is given by $S^{(2)}_A(t)=I^{(2)}_{\text{KM}}(\lambda)-I^{(2)}_{\text{KM}}(0)$. 

As an example, we show numerical results for the $S^{(2)}_A(t)$ at $\beta=0$ for different time $tJ$. In the short-time regime, the entanglement entropy is a smooth function of $\lambda$. However, at the longer time regime, there are two saddle-point solutions near $\lambda=1/2$, with a first-order transition at $\lambda=1/2$. (These solutions take the form of ``replica wormholes'', which will be explained in more details in section \ref{sec_SYKbath}.) The final result of $S^{(2)}_A(t)$ is determined by the solution with smaller on-shell action $I^{(2)}_{\text{KM}}$. This leads to a singularity of $S^{(2)}_A$ at $\lambda=1/2$. In the long-time limit, $S^{(2)}_A$ saturates to the maximal value, while each saddle-point gives a curve with entropy $\lambda \log(2)/2$ or $(1-\lambda)\log(2)/2$. These values corresponds to the entropy of subsystem $A$ or $B$ when the total system is prepared in an thermal ensemble, which has been discussed above. This can be understood as follows~\cite{zhang2020entanglement}: Let us consider a general pure state $|\psi(t)\rangle=\sum_c c_ne^{-iE_n t}|E_n\rangle$, the entanglement entropy of $A$ can be written as
\begin{equation}
e^{-\mathcal{S}_A^{(2)}(t)}=\text{tr}\left[\hat{S}^A|\psi(t) \rangle \langle \psi(t)| \otimes |\psi(t) \rangle \langle \psi(t)|\right].
\end{equation}
In the long-time limit, the eigenstates should pair up to cancel the fast oscillation phase factor. As a result, we find
\begin{equation}
\begin{aligned}
e^{-\mathcal{S}_A^{(2)}(t)}\approx&\sum_{m,n}|c_mc_n|^2\text{tr}\left[(\hat{S}^A+\hat{S}^B)|E_m \rangle \langle E_m| \otimes |E_n \rangle \langle E_n|\right]\\
=&\text{tr}\left[(\hat{S}^A+\hat{S}^B)\rho_{\text{th}} \otimes\rho_{\text{th}}\right]=e^{-\mathcal{S}_{A,th}^{(2)}}+e^{-\mathcal{S}_{B,th}^{(2)}}.\label{eq_SYKKMdynamics}
\end{aligned}
\end{equation}
Here we have assumed the diagonal ensemble looks thermal  $\sum_m|c_m|^2|E_m \rangle \langle E_m|=\rho_\text{th}$. In the path-integral representation, each of the term comes from a separate saddle point, and there is a first-order transition in the large-$N$ limit. The equivalence between the subsystem entropy of a thermal ensemble and the long-time entanglement of KM pure state has also been directly proved in~\cite{zhang2020entanglement} using the path-integral approach.

\subsection{Exact results for SYK\texorpdfstring{$_2$}{TEXT} }\label{subsec_SYK2EE}
Before ending this section, we review exact results for the entanglement entropy in the SYK$_2$ and cSYK$_2$ random hopping models~\cite{liu2018quantum,zhang2020subsystem,lydzba2020eigenstate}. We would focus on the cSYK$_2$ model since the path-integral approach suggests the large-$N$ entanglement entropy of the SYK$_q$ model is given by one half of the cSYK$_q$ model result at half-filling. 

The idea is to use the correlation matrix approach for the entanglement entropy of the free particles~\cite{casini2009entanglement}. The ground-state correlation function of free fermion Hamiltonians satisfies the wick theorem, even if restricted to arbitrary subsystems. The validity of the wick theorem indicates the reduced density matrix takes the form of Gaussian states 
\begin{equation}
\rho_A=e^{-K_A}/\text{tr}[e^{-K_A}],\ \ \ \ \ \ K_A=\sum_{ij\in A}K_{ij}c^\dagger_ic_j.
\end{equation}
In other words, $\rho_A$ is a thermal ensemble of an auxiliary quadratic Hamiltonian $K_A$, which is known as the Modular Hamiltonian or the entanglement Hamiltonian. The parameter $K_{ij}$ can be fixed by matching the correlation matrix $C_{ij}=\langle c^\dagger_i c_j\rangle$ with $i,j\in A$. The simplest way is to transform into the diagonal basis of $C_{\eta \xi}=\sum_{ij}U_{\eta i}^*U_{\xi j}C_{ij}$, where $C_{\eta \xi}=C_{\eta}\delta_{\eta \xi}$. In the same basis, we expect $K_A$ is also diagonal. Writing $K_A=K_\eta \delta_{\eta \xi}$, we have 
\begin{equation}
C_{\eta}=\frac{e^{-K_\eta}}{1+e^{-K_\eta}}=\frac{1}{e^{K_\eta}+1}.
\end{equation}
The entanglement entropy of the system is then given by $S_A=-\text{tr}~\rho_A\log \rho_A$, which reads
\begin{equation}
S_A=-\sum_\eta\left(C_\eta \log C_\eta + (1-C_\eta) \log (1-C_\eta)\right).
\end{equation}
In the continuum limit, if we define a normalized distribution $f(x)$ of eigenvalue $x=C_\eta$, we have 
\begin{equation}\label{eq_Sfree}
\begin{aligned}
S_A&=-M\int dx~\left(x\log x+(1-x)\log(1-x)\right)f(x).\\
S_A^{(n)}&=-\frac{M}{n-1}\int dx~\log\left(x^n+(1-x)^n\right)f(x).
\end{aligned}
\end{equation}

The only remaining task is to determine $f(x)$ for the cSYK$_2$ model. Since the single-particle Hamiltonian is a random Hermitian matrix, its eigenstates form a Haar random unitary matrix $U$~\cite{liu2018quantum}. For a filling fraction $\kappa=Q/N$, first Q states are occupied. We parametrize $U$ as \cite{liu2018quantum}
\begin{equation}
U  =  \left(\begin{array}{cc} V_{Q\times M} & V'_{Q\times( N-M)}\\ W_{(N-Q)\times M} & W'_{(N-Q)\times (N-M)}\end{array}\right)^T,
\end{equation}
and the correlation matrix restricted to the subsystem $A$ is given by $C_{ij}=V^\dagger V$. It is known that $C$ then forms a $\beta$-Jacobi ensemble with $\beta=2$~\cite{forrester2006quantum}. The eigenvalue distribution $f(x)$ is known to satisfy the Wachter law \cite{forrester2010log}:
\begin{equation}
\begin{aligned}
f(x)=&\frac{1}{2\pi\lambda}\frac{\sqrt{(\lambda_+-x)(x-\lambda_-)}}{x(1-x)}\Theta\left[(\lambda_+-x)(x-\lambda_-)\right]\\
&+\left(1-\frac{\kappa}{\lambda}\right)\Theta(\lambda-\kappa) \delta(x).
\end{aligned}
\end{equation}
with $\lambda_\pm=[\sqrt{\kappa(1-\lambda)}\pm\sqrt{\lambda(1-\kappa)}]^2$. For $\lambda \rightarrow 0$, \eqref{eq_Sfree} immediately predicts the maximal entropy \eqref{eq_maximal}, consistent with all previous discussions. For general $\lambda$, the entropy can be analytically obtained for $n=1$ and all even $n$~\cite{lydzba2020eigenstate,zhang2020subsystem}, and also for finite $N$ \cite{bianchi2021page}. They show that the entanglement entropy is much smaller than the Page value, as expected for non-interacting systems.

\section{Entanglement in coupled SYK models}\label{sec_SYKbath}
Having summarized results for the SYK/cSYK model, in this section, we review studies of the entanglement dynamics in coupled SYK models. We focus on the coupled $\chi-\psi$ system with the Hamiltonian \eqref{eq_coupledSYK44} for different $r=N_2/N_1$. We copy the Hamiltonian here for convenience:
\begin{equation}\label{eq_coupledSYK44copy}
\begin{aligned}
H=&\sum_{\{i_n\}}J^\chi_{i_1 i_2 i_3 i_4} \chi_{i_1}\chi_{i_2}\chi_{i_3}\chi_{i_4}+\sum_{\{a_n\}}J^\psi_{a_1 a_2 a_3 a_4} \psi_{a_1}\psi_{a_2}\psi_{a_3}\psi_{a_4}\\
&+\sum_{i_1< i_2, a_1< a_2} V_{i_1 i_2 a_1 a_2}\chi_{i_1}\chi_{i_2}\psi_{a_1}\psi_{a_2}.
\end{aligned}
\end{equation}
Unlike the single SYK model, where many approaches can be employed, here the path-integral approach is the only efficient method. Motivated by gravity calculations~\cite{almheiri2019islands}, which would be reviewed in section \ref{subsec_SYKbathwormhole}, the coupled system is prepared in a TFD state with an auxiliary system and evolves under the coupled Hamiltonian $H$ \footnote{There are also studies on subsystem entropy of systems prepared in thermal ensembles and coupled to a bath \cite{kaixiang2021page}.}. In the early-time regime, the entanglement entropy grows linearly for both R\'enyi and von Neumann entropies. In the long-time limit, R\'enyi entropy saturates to a non-thermal value for $r=1$. On the other hand, for $r\gg 1$, the $\psi$ system contains large degrees of freedom and serves as a heat bath \cite{chen2017tunable,zhang2019evaporation,almheiri2019universal}. The coupled system is then qualitatively similar to~\cite{chen2020replica}
        \begin{equation}\label{eq_chainbath}
        H=H_{\text{SYK}_4}[\chi]+i \sum_{x,i}\frac{\Lambda}{2}\psi_{i,x}\psi_{i,x+1}+\sum_{i}iV\sqrt{\Lambda}\chi_i\psi_{i,0},
    \end{equation}
which is closely related to its gravity counterpart in AdS$_2$ spacetime~\cite{almheiri2019islands}. As a result, a replica wormhole solution appears in the long-time limit, and the entanglement entropy approaches the thermal value \cite{kaixiang2021page}. These results will be explained in details in the following subsections.

\subsection{The set-up and \texorpdfstring{$G$-$\Sigma$}{TEXT} action}
We begin with the setup of the problem. TFD states can be defined generally by replicating the system \cite{gu2017spread}, with arbitrariness in choosing the basis. Here we focus on a convenient convention for Majorana fermions. We consider a system that contains $N$ Majorana modes $\chi_i$ with Hamiltonian $H[\chi]$. We introduce $2N$ fermion modes $\chi^L_i$, $\chi^R_i$. Similar to the definition of KM pure states, we pair up the left and the right fermions by defining $c_i=\frac{\chi_{i}^L+i\chi_{i}^R}{\sqrt{2}}$. The unnormalized TFD state reads
\begin{equation}\label{eq_TFDdef}
|\text{TFD}\rangle=e^{-\frac{\beta H_L}{2}}|I\rangle,\ \ \ \ \ \ c_{i}|I\rangle=0.
\end{equation}
Here we choose a single maximal entangled state $|I\rangle$ with no filling of $c_i$, and keep the $\beta$ dependence of $|\text{TFD}\rangle$ implicit. We have defined the Hamiltonian for the left system as $H_L=H[\chi^L]$. We could also introduce the Hamiltonian for the right system as $H_R=H[-i\chi^R]$, which satisfies $H_R|I\rangle=H_L|I\rangle$, and the TFD state can be equivalently prepared by evolving $|I\rangle$ with $H_\text{tot}=H_L+H_R$ for an imaginary time $\beta/4$, or with $H_R$ for an imaginary time $\beta/2$. 

We add a few comments. Firstly, any maximally entangled state can be written as $|I\rangle=2^{-N/4}\sum_{i=1}^{2^{N/2}}|i\rangle_L|i'\rangle_R$, where both $\{|i\rangle_L\}$ and $\{|i'\rangle_R\}$ are orthonormal bases in the corresponding Hilbert space. As a result, the reduced density matrix of a single side is thermal $\rho^\alpha=e^{-\beta H^\alpha}/Z$ with $\alpha=L,R$. Secondly, for small $\beta$, $|\text{TFD}\rangle$ mainly contains entanglement between left and right copies, while the entanglement between Majorana modes in the same copy is weak. This indicates the TFD state is not thermal, similar to the KM pure states. After long-time evolution, we expect the system thermalize and $S^{(2)}_A=\text{min}~\{2S^{(2)}_{A,\text{th}},2S^{(2)}_{B,\text{th}}\}$. Here $S^{(2)}_{A/B,\text{th}}$ is the subsystem R\'enyi entropy of thermal density matrices. Thirdly, under holographic duality, the TFD states are described by the Hartle-Hawking states in the two-sided black holes~\cite{israel1976thermo}, in which a version of the information paradox exists~\cite{almheiri2019islands}. 

To study the entanglement dynamics, we further evolve the coupled system with $H_\text{tot}$. This gives 
\begin{equation}
|\text{TFD}(t)\rangle=e^{-i H_\text{tot}t}|\text{TFD}\rangle=e^{-2iH_L t}|\text{TFD}\rangle.
\end{equation}
Then one can compute the reduced density matrix $\rho_A$ and R\'enyi entanglement entropy $S^{(n)}_A$.

In this section, we focus on coupled SYK models with both $\chi$ and $\psi$ fermions. After doubling the system, we have four species of fermions $\chi^L$, $\psi^L$, $\chi^R$, and $\psi^R$. The Hamiltonian $H_\alpha$ contains interactions between $\chi^\alpha$ and $\psi^\alpha$, and there is no interaction between left and right copies. The TFD state after the real-time evolution can be represented pictorially
\begin{equation}
|\text{TFD}\rangle=
\begin{tikzpicture}[thick,scale = 0.7,baseline={([yshift=-0pt]current bounding box.center)}]
  \draw[mid arrow] (-1,-0.2) -- (-2.5,-0.2);
   \draw (1,-0.2) arc(0:-180:1 and 1);
   \draw (1.2,-0.2) arc(0:-170:1.2 and 1.2);

    \draw[mid arrow] (-1.18,-0.4) -- (-2.5,-0.4);

       \draw[dotted] (0.866,-0.7) -- (1.039,-0.8);

        \draw[dotted] (0.5,-1.066) -- (0.6,-1.239);

              \draw[dotted] (0,-1.2) -- (0,-1.4);

        \draw[dotted] (-0.5,-1.066) -- (-0.6,-1.239);

           \draw[dotted] (-0.866,-0.7) -- (-1.039,-0.8);

             \draw[dotted] (-1.4,-0.2) -- (-1.4,-0.4);

             \draw[dotted] (-1.8,-0.2) -- (-1.8,-0.4);

             \draw[dotted] (-2.2,-0.2) -- (-2.2,-0.4);
      \draw (0.4,-0.8) node{$\chi$};
       \draw (0.9,-1.4) node{$\psi$};
\end{tikzpicture}
\end{equation}
Here the horizontal solid lines represent real-time evolutions, while the arcs represent imaginary-time evolutions. We choose the subsystem $A$ as $\psi^L \cup \psi^R$, which is in a pure state when $\beta=t=0$. The purity $\text{tr}~\rho_A^2$ then takes two equivalent forms \cite{chen2020replica}
\vspace{-5pt}
\begin{equation}\label{diagram_TFD}
\begin{tikzpicture}[thick,scale = 0.6,baseline={([yshift=-0pt]current bounding box.center)}]
 \draw (1,0.2) arc(0:180:1 and 1);
 \draw (1,-0.2) arc(0:-180:1 and 1);
  \draw (-1,-0.2) -- (-2.5,-0.2);
   \draw (-1,0.2) -- (-2.5,0.2);

   \draw (1.2,0.2) arc(0:170:1.2 and 1.2);
   \draw (1.2,-0.2) arc(0:-170:1.2 and 1.2);
   \draw[dashed] (1.2,0.2) --(1.2,-0.2) ;

    \draw (-1.18,-0.4) -- (-2.5,-0.4);
     \draw (-1.18,0.4) -- (-2.5,0.4);

     \draw[dashed] (-2.5,0.4) arc (90:270:0.4 and 0.4);
     \draw[dotted] (0.866,0.7) -- (1.039,0.8);
       \draw[dotted] (0.866,-0.7) -- (1.039,-0.8);
      \draw[dotted] (0.5,1.066) -- (0.6,1.239);
        \draw[dotted] (0.5,-1.066) -- (0.6,-1.239);

           \draw[dotted] (0,1.2) -- (0,1.4);
              \draw[dotted] (0,-1.2) -- (0,-1.4);
          \draw[dotted] (-0.5,1.066) -- (-0.6,1.239);

        \draw[dotted] (-0.5,-1.066) -- (-0.6,-1.239);
         \draw[dotted] (-0.866,0.7) -- (-1.039,0.8);
           \draw[dotted] (-0.866,-0.7) -- (-1.039,-0.8);
           \draw[dotted] (-1.4,0.2) -- (-1.4,0.4);

             \draw[dotted] (-1.4,-0.2) -- (-1.4,-0.4);
                     \draw[dotted] (-1.8,0.2) -- (-1.8,0.4);
             \draw[dotted] (-1.8,-0.2) -- (-1.8,-0.4);

              \draw[dotted] (-2.2,0.2) -- (-2.2,0.4);
             \draw[dotted] (-2.2,-0.2) -- (-2.2,-0.4);
      \draw (0.4,-0.8) node{$\chi$};
       \draw (0.9,-1.4) node{$\psi$};
 
 \begin{scope}[shift={(0,3)}]
 \draw (1,0.2) arc(0:180:1 and 1);
 \draw (1,-0.2) arc(0:-180:1 and 1);
  \draw (-1,-0.2) -- (-2.5,-0.2);
   \draw (-1,0.2) -- (-2.5,0.2);

   \draw (1.2,0.2) arc(0:170:1.2 and 1.2);
   \draw (1.2,-0.2) arc(0:-170:1.2 and 1.2);
   \draw[dashed] (1.2,0.2) --(1.2,-0.2) ;
    \draw (-1.18,-0.4) -- (-2.5,-0.4);
     \draw (-1.18,0.4) -- (-2.5,0.4);

     \draw[dashed] (-2.5,0.4) arc (90:270:0.4 and 0.4);
     \draw[dotted] (0.866,0.7) -- (1.039,0.8);
       \draw[dotted] (0.866,-0.7) -- (1.039,-0.8);
      \draw[dotted] (0.5,1.066) -- (0.6,1.239);
        \draw[dotted] (0.5,-1.066) -- (0.6,-1.239);

           \draw[dotted] (0,1.2) -- (0,1.4);
              \draw[dotted] (0,-1.2) -- (0,-1.4);
          \draw[dotted] (-0.5,1.066) -- (-0.6,1.239);
        \draw[dotted] (-0.5,-1.066) -- (-0.6,-1.239);
         \draw[dotted] (-0.866,0.7) -- (-1.039,0.8);
           \draw[dotted] (-0.866,-0.7) -- (-1.039,-0.8);

           \draw[dotted] (-1.4,0.2) -- (-1.4,0.4);
             \draw[dotted] (-1.4,-0.2) -- (-1.4,-0.4);
                     \draw[dotted] (-1.8,0.2) -- (-1.8,0.4);
             \draw[dotted] (-1.8,-0.2) -- (-1.8,-0.4);

              \draw[dotted] (-2.2,0.2) -- (-2.2,0.4);
             \draw[dotted] (-2.2,-0.2) -- (-2.2,-0.4);
      \draw (0.4,-0.8) node{$\chi$};
       \draw (0.9,-1.4) node{$\psi$};
       
\end{scope}
\draw[dashed] (-2.5,3+0.2) arc (90:270:1 and 1.7);
\draw[dashed] (-2.5,3-0.2) arc (90:270:0.5 and 1.3);
\draw[dashed] (1,3+0.2) arc (90:-90:1 and 1.7);
\draw[dashed] (1,3-0.2) arc (90:-90:0.5 and 1.3);

  \draw (-2.5,3+0.6-0.05) node[circle,fill,scale = 0.1,above]{A};
 \draw[->] (-2.5, 3+0.6) -- (-1.5, 3+0.6) ;
 \draw (-2.3,3+0.9) node{\small $s=0$} ;

 \draw (-2.5,0.6-0.05) node[circle,fill,scale = 0.1,above]{A};
 \draw[->] (-2.5, 0.6) -- (-1.5, 0.6) ;
 \draw (-2.1,0.9) node{\small $s=\beta + 4t$} ;
\end{tikzpicture}\ \ \ \ 
\begin{tikzpicture}[thick,scale = 0.5,baseline={([yshift=-0pt]current bounding box.center)}]
\draw (0.1,0.3+0.1) arc (0:-180:0.1 and 0.1);
\draw (0.3+0.1,0.1) arc (90:270:0.1 and 0.1);
\draw (-0.3-0.1,-0.1) arc (-90:90:0.1 and 0.1);
\draw (0.1,-0.3-0.1) arc (0:180:0.1 and 0.1);

\draw  (0.1,0.3+0.1)  --  (0.1,0.3+0.1+1.2);
\draw  (-0.1,0.3+0.1)  --  (-0.1,0.3+0.1+1.2);  

\draw (0.3+0.1,0.1) --(0.3+0.1+1.2,0.1) ;
\draw  (0.3+0.1,-0.1) --(0.3+0.1+1.2,-0.1) ;
\draw (-0.3-0.1,0.1) --(-0.3-0.1-1.2,0.1) ;

\draw  (-0.3-0.1,-0.1) --(-0.3-0.1-1.2,-0.1) ;
\draw  (0.1,-0.3-0.1)  --  (0.1,-0.3-0.1-1.2);
\draw  (-0.1,-0.3-0.1)  --  (-0.1,-0.3-0.1-1.2);

\draw  (0.1,0.3+0.1+1.2) arc (-90+5.74:270-5.74:1 and 1);

\draw  (0.3+0.1+1.2,0.1)  arc (180-5.74:-180+5.74:1 and 1);
\draw  (-0.3-0.1-1.2,0.1)  arc (5.74:360-5.74:1 and 1);
\draw  (-0.1,-0.3-0.1-1.2) arc (90+5.74:450-5.74:1 and 1);

\draw[dotted]  (0.1,0.3+0.1+0.3) arc (90:0:0.6 and 0.6 );
\draw[dotted]   (0.3+0.1+0.3,-0.1) arc (0:-90:0.6 and 0.6 );
\draw[dotted]   (-0.3-0.1-0.3,0.1)  arc (180:90:0.6 and 0.6 );

\draw[dotted]   (-0.3-0.1-0.3,-0.1)  arc (180:270:0.6 and 0.6 );

\draw[dotted]  (0.1,0.3+0.1+0.6) arc (90:0:0.9 and 0.9 );

\draw[dotted]   (0.3+0.1+0.6,-0.1) arc (0:-90:0.9 and 0.9 );
\draw[dotted]   (-0.3-0.1-0.6,0.1)  arc (180:90:0.9 and 0.9 );
\draw[dotted]   (-0.3-0.1-0.6,-0.1)  arc (180:270:0.9 and 0.9 );

\draw[dotted]  (0.1,0.3+0.1+0.9) arc (90:0:1.2 and 1.2 );
\draw[dotted]   (0.3+0.1+0.9,-0.1) arc (0:-90:1.2 and 1.2 );

\draw[dotted]   (-0.3-0.1-0.9,0.1)  arc (180:90:1.2 and 1.2 );
\draw[dotted]   (-0.3-0.1-0.9,-0.1)  arc (180:270:1.2 and 1.2 );

\draw[dotted]  (0.5, 1.73) arc (90-16.12:16.12:1.8 and 1.8 );

\draw[dotted]  (-0.5, 1.73) arc (90+16.12:180-16.12:1.8 and 1.8 );
\draw[dotted]  (0.5, -1.73) arc (-90+16.12:-16.12:1.8 and 1.8 );
\draw[dotted]  (-0.5, -1.73) arc (270-16.12:180+16.12:1.8 and 1.8 );

\draw[dotted]  (1, 2.25) arc (90-23.96:23.96:2.46 and 2.46 );
\draw[dotted]  (-1, 2.25) arc (90+23.96:180-23.96:2.46 and 2.46 );
\draw[dotted]  (1,-2.25) arc (-90+23.96:-23.96:2.46 and 2.46 );

\draw[dotted]  (-1,-2.25) arc (270-23.96:180+23.96:2.46 and 2.46 );

\draw[dotted]  (0.6, 3.4) arc (90-10:10:3.45 and 3.45);
\draw[dotted]  (-0.6, 3.4) arc (90+10:180-10:3.45 and 3.45);
\draw[dotted]  (0.6,-3.4) arc (-90+10:-10:3.45 and 3.45);

\draw[dotted]  (-0.6, -3.4) arc (270-10:180+10:3.45 and 3.45);

\draw[dashed,gray] (0,-4) -- (0,4);
\draw[dashed,gray] (-4,0) -- (4,0);

\draw (0,3) node{$\chi^2$}; 
\draw (0,-3) node{$\chi^1$}; 
\draw (3,0) node{$\psi^2$}; 
\draw (-3,0) node{$\psi^1$}; 

\draw (-2.1,2.1) node{$C_2$}; 
\draw (2.1,-2.1) node{$C_2$}; 

\draw (2.1,2.1) node{$C_1$}; 
\draw (-2.1,-2.1) node{$C_1$}; 

\end{tikzpicture}
\end{equation}
The left graph can be understood as inserting a pair of twist operators $T_{\chi,L}/T_{\chi,R}$ to the $\chi$ system:
\begin{equation}
Z^2\text{tr}~\rho_A^2={\langle\text{TFD}(t)|^{\otimes 2}T_{\chi,L}T_{\chi,R}|\text{TFD}(t)\rangle}^{\otimes 2}.
\end{equation}
The right graph is more convenient for the path-integral approach. Following the discussions in section \ref{subsec_SYKlargeN} by introducing bilocal fields $G^{mn}_{\chi/\psi}$ and $\Sigma^{mn}_{\chi/\psi}$ with $m,n\in \{1,2\}$, one find the $G$-$\Sigma$ action reads 
\begin{equation}\label{eq_TFDGsigmaaction}
\begin{aligned}
\frac{S_{\text{TFD}}^{(2)}}{N_1}&=-\frac{1}{2}\log \det (\partial_\tau-\tilde \Sigma_\chi)-\frac{r}{2}\log \det (\partial_\tau-\tilde \Sigma_\psi)\\
&+\frac{1}{2}\int_{\tau \tau'}~\Big[\tilde \Sigma^{mn}_\chi \tilde G^{mn}_\chi-\frac{J^2}{4}(\tilde G^{mn}_\chi)^4+r\tilde \Sigma^{mn}_\psi \tilde G^{mn}_\psi\\
&-\frac{J^2r}{4}(\tilde G^{mn}_\psi)^4-\frac{V^2}{2}(\tilde G^{mn}_\chi)^2(\tilde G^{m'n'}_\psi)^2c_{mm'}c_{nn'}\Big ].
\end{aligned}
\end{equation}
Here the integrals over $\tau$ and $\tau'$ are over the contour $C_1\cup C_2$ on the complex plane. An additional phase factor appears if we use a single real parameter as we did in section \ref{subsec_singleSYKpathintegral}. We have defined $c_{mm'}(\tau)$ and $c_{nn'}(\tau')$ as a permutation between two replicas~\cite{gu2017spread,jian2021phase}:
\begin{equation}
c(\tau)= 
    \begin{cases}
      \mathbf{I} & \text{for $\tau \in C_1$,}\\
      \bm{\sigma_x} & \text{for $\tau \in C_2$.}
    \end{cases}  
\end{equation} 
The saddle-point equations of \eqref{eq_TFDGsigmaaction} read 
\begin{equation}\label{eq_TFDSD}
\begin{aligned}
&G_\chi=(\partial_\tau- \Sigma_\chi)^{-1},\ \ \ \ \ \ G_\psi=(\partial_\tau- \Sigma_\psi)^{-1},\\
&\Sigma_\chi^{mn}=J^2(G_\chi^{mn})^3+V^2c_{mm'}c_{nn'} (G_\psi^{m'n'})^2 G_\chi^{mn},\\
&\Sigma_\psi^{mn}=J^2(G_\psi^{mn})^3+V^2c_{mm'}c_{nn'} (G_\chi^{m'n'})^2 G_\psi^{mn}/r,
\end{aligned}
\end{equation}
and the on shell action $I^{(2)}_\text{TFD}=S_{\text{TFD}}^{(2)}[G,\Sigma]$. The second R\'enyi entropy is given by $S^{(2)}_A=I^{(2)}_\text{TFD}-2I$, with $I$ being the on-shell action of the thermal ensemble. Equivalently, $2I$ can be computed by setting $c(\tau)=\mathbf{I}$ in \eqref{eq_TFDGsigmaaction}.

\subsection{Early-time linear growth}\label{subsec_SYKbathlinear}
We begin to analyze the entanglement dynamics of the coupled system. An important observation~\cite{gu2017spread} is that the saddle-point equations \eqref{eq_TFDSD} admit solutions that is ``replica diagonal'':
\begin{equation}
G^{mn}_{\psi/\chi}(\tau,\tau')=\delta^{mn}G_{\psi/\chi}(\tau,\tau').
\end{equation} 
The arguments contain two steps: Without any interaction $V=0$, the solution of $G_{\psi/\chi}$ satisfies the replica diagonal condition. When we turn on a finite $V$, new terms in the self-energy is again replica diagonal due to the presence of $G_{\psi/\chi}$. As a result, the replica diagonal solution is consistent with the saddle-point equation. It turns out that this is the correct saddle-point solution in the early-time regime.

Now let's assume $V$ is small, and perform a perturbative calculation of the entanglement entropy near the saddle-point solution with $V=0$~\cite{gu2017spread,chen2021entropy,kaixiang2021page,dadras2021perturbative,penington2019replica}. We have
\begin{equation}
\frac{\delta I^{(n)}_\text{TFD}}{n N_1}=-\frac{V^2}{4}\left(\int_{\tau,\tau'\in C_1}+\int_{\tau,\tau'\in C_2}\right) G_\chi(\tau,\tau')^2G_\psi(\tau,\tau')^2.
\end{equation}
Here we consider the $n$-th R\'enyi entropy. On the other hand, there is a similar contribution for $I$:
\begin{equation}
\frac{\delta I}{N_1}=-\frac{V^2}{4}\int_{\tau,\tau'\in C_1\cup C_2} G_\chi(\tau,\tau')^2G_\psi(\tau,\tau')^2.
\end{equation}
Combining these results, we have
\begin{equation}
S^{(n)}_A=\frac{nV^2N_1}{2(n-1)}\int_{\tau\in C_1}\int_{\tau'\in C_2} G_\chi(\tau,\tau')^2G_\psi(\tau,\tau')^2.
\end{equation} 
For $V=0$, the Green's functions only depend on time differences. When the evolution time $t\geq t_{\text{dis}}$, with the dissipation time $t_{\text{dis}}$ charactering the decay of two-point functions, the main time-dependent contribution $\Delta S^{(n)}_A(t)=S^{(n)}_A(t)-S^{(n)}_A(0)$ comes from the integral over real-time evolutions. This gives
\begin{equation}
\begin{aligned}
\Delta S^{(n)}_A(t)&\approx \frac{nV^2N_1}{2(n-1)}\int_0^{2t}\int_{0}^{2t}dt_1dt_2~G^>_\chi(t_{12})^2G^>_\psi(t_{12})^2\\
&\approx \left[\frac{nV^2N_1}{n-1}\int_{-\infty}^\infty dt'~G^>_\chi(t')^2G^>_\psi(t')^2\right]t\equiv \kappa t.
\end{aligned}
\end{equation}
Here $G^>_\chi(t)=-i\left<\chi(t)\chi(0)\right>$ is the real-time Green's function~\cite{kamenev2011field}. We find at early-time the entanglement entropy grows linearly. 

Such an linear-growth behavior is valid beyond SYK-like models and the replica diagonal consistency is not necessary. As an illustration, we consider some arbitrary $\chi$ and $\psi$ systems, with a general coupling $H_\text{int}=V\sum_{i=1}^NO_\chi^iO_\psi^i$~\cite{dadras2021perturbative}. The perturbative analysis can still be carried out diagrammatically. For $n$-th R\'enyi entropy with $n\geq 2$, the leading order time-dependent contribution comes from the diagram
\begin{equation}
\begin{tikzpicture}[thick,scale = 0.3,baseline={([yshift=-2pt]current bounding box.center)}]
\draw[gray] (0.1,0.3+0.1) arc (0:-180:0.1 and 0.1);
\draw[gray] (0.3+0.1,0.1) arc (90:270:0.1 and 0.1);
\draw[gray] (-0.3-0.1,-0.1) arc (-90:90:0.1 and 0.1);
\draw[gray] (0.1,-0.3-0.1) arc (0:180:0.1 and 0.1);

\draw[gray]  (0.1,0.3+0.1)  --  (0.1,0.3+0.1+1.2);
\draw[gray]  (-0.1,0.3+0.1)  --  (-0.1,0.3+0.1+1.2);  
\draw[gray] (0.3+0.1,0.1) --(0.3+0.1+1.2,0.1) ;
\draw[gray]  (0.3+0.1,-0.1) --(0.3+0.1+1.2,-0.1) ;
\draw[gray] (-0.3-0.1,0.1) --(-0.3-0.1-1.2,0.1) ;
\draw[gray]  (-0.3-0.1,-0.1) --(-0.3-0.1-1.2,-0.1) ;
\draw[gray]  (0.1,-0.3-0.1)  --  (0.1,-0.3-0.1-1.2);
\draw[gray]  (-0.1,-0.3-0.1)  --  (-0.1,-0.3-0.1-1.2);

\draw[gray]  (0.1,0.3+0.1+1.2) arc (-90+5.74:270-5.74:1 and 1);
\draw[gray]  (0.3+0.1+1.2,0.1)  arc (180-5.74:-180+5.74:1 and 1);
\draw[gray]  (-0.3-0.1-1.2,0.1)  arc (5.74:360-5.74:1 and 1);
\draw[gray]  (-0.1,-0.3-0.1-1.2) arc (90+5.74:450-5.74:1 and 1);

\draw  (-0.1,-1.2)  --  (-1.2,-0.1);
\filldraw  (-0.1,-1.2) circle (1pt) node[left]{$ $};
\filldraw  (-1.2,-0.1) circle (1pt) node[left]{$ $};

\draw  (-0.1,-0.8)  --  (-0.8,-0.1);
\filldraw  (-0.1,-0.8) circle (1.5pt) node[left]{$ $};
\filldraw  (-0.8,-0.1) circle (1.5pt) node[left]{$ $};

\draw[gray] (0,2.6) node{\small$\chi^2$}; 
\draw[gray] (0,-2.6) node{\small$\chi^1$}; 
\draw[gray] (2.6,0) node{\small$\psi^2$}; 
\draw[gray] (-2.6,0) node{\small$\psi^1$}; 

\end{tikzpicture}=-\frac{nV^2N}{n-1}\text{tr}\left[G^>_{O_\chi}\circ G^{>,T}_{O_\psi}\right].\label{eq_Sbathlinear1}
\end{equation}
Here we treat Green's functions as matrices in the time domain. This leads to a linear growth 
\begin{equation}
\Delta S^{(n)}_A(t)\approx\left[-\frac{2nV^2N}{n-1}\int \frac{d\omega}{2\pi}G^>_{O_\chi}(\omega)G^>_{O_\psi}(-\omega)\right]t.
\end{equation}

However, above calculations do not have a well-defined limit $n\rightarrow 1$, in which the R\'enyi entropy $S_A^{(n)}$ reduces to the von Neumann entropy $S_A$. As noticed by authors~\cite{dadras2021perturbative}, this is because an additional diagram contributes for $n\rightarrow 1$:
\begin{equation}
\begin{tikzpicture}[thick,scale = 0.3,baseline={([yshift=-2pt]current bounding box.center)}]
\draw[gray] (0.1,0.3+0.1) arc (0:-180:0.1 and 0.1);
\draw[gray] (0.3+0.1,0.1) arc (90:270:0.1 and 0.1);
\draw[gray] (-0.3-0.1,-0.1) arc (-90:90:0.1 and 0.1);
\draw[gray] (0.1,-0.3-0.1) arc (0:180:0.1 and 0.1);

\draw[gray]  (0.1,0.3+0.1)  --  (0.1,0.3+0.1+1.2);
\draw[gray]  (-0.1,0.3+0.1)  --  (-0.1,0.3+0.1+1.2);  
\draw[gray] (0.3+0.1,0.1) --(0.3+0.1+1.2,0.1) ;
\draw[gray]  (0.3+0.1,-0.1) --(0.3+0.1+1.2,-0.1) ;
\draw[gray] (-0.3-0.1,0.1) --(-0.3-0.1-1.2,0.1) ;
\draw[gray]  (-0.3-0.1,-0.1) --(-0.3-0.1-1.2,-0.1) ;
\draw[gray]  (0.1,-0.3-0.1)  --  (0.1,-0.3-0.1-1.2);
\draw[gray]  (-0.1,-0.3-0.1)  --  (-0.1,-0.3-0.1-1.2);

\draw[gray]  (0.1,0.3+0.1+1.2) arc (-90+5.74:270-5.74:1 and 1);
\draw[gray]  (0.3+0.1+1.2,0.1)  arc (180-5.74:-180+5.74:1 and 1);
\draw[gray]  (-0.3-0.1-1.2,0.1)  arc (5.74:360-5.74:1 and 1);
\draw[gray]  (-0.1,-0.3-0.1-1.2) arc (90+5.74:450-5.74:1 and 1);

\draw  (-0.1,-1.2)  --  (-1.2,-0.1);
\filldraw  (-0.1,-1.2) circle (1pt) node[left]{$ $};
\filldraw  (-1.2,-0.1) circle (1pt) node[left]{$ $};

\draw  (-0.1,0.8)  --  (-0.8,0.1);
\filldraw  (-0.1,0.8) circle (1.5pt) node[left]{$ $};
\filldraw  (-0.8,0.1) circle (1.5pt) node[left]{$ $};

\draw  (0.1,-0.8)  --  (0.8,-0.1);
\filldraw  (0.1,-0.8) circle (1.5pt) node[left]{$ $};
\filldraw  (0.8,-0.1) circle (1.5pt) node[left]{$ $};

\draw  (0.1,1.2)  --  (1.2,0.1);
\filldraw  (0.1,1.2) circle (1pt) node[left]{$ $};
\filldraw  (1.2,0.1) circle (1pt) node[left]{$ $};

\draw[gray] (0,2.6) node{\small$\chi^2$}; 
\draw[gray] (0,-2.6) node{\small$\chi^1$}; 
\draw[gray] (2.6,0) node{\small$\psi^2$}; 
\draw[gray] (-2.6,0) node{\small$\psi^1$}; 

\end{tikzpicture}=-\frac{V^{2n}N}{n-1}\left[-G^>_{O_\chi}\circ G^{>,T}_{O_\psi}\right]^n.\label{eq_Sbathlinear2}
\end{equation}
Although this contribution is at a higher order of $V$ for any $n>1$, It becomes important in the limit of $n\rightarrow 1$. Summing up \eqref{eq_Sbathlinear1} and \eqref{eq_Sbathlinear2}, we get a finite expression for the von Neumann entropy, which predicts a slope
\begin{equation}
\frac{\Delta S_A}{N}\approx \left[2\int \frac{d\omega}{2\pi}g(\omega)\left(1-\log[g(\omega)]\right)\right]t,
\end{equation}
with $g(\omega)\equiv-V^2G^>_{O_\chi}(\omega)G^>_{O_\psi}(-\omega)$. Interestingly, this shows that the linear growth rate for the von Neumann entropy is proportional to $-V^2\log V $ for small $V$, parametrically faster than any $n$-th R\'enyi entropy with $n>1$.

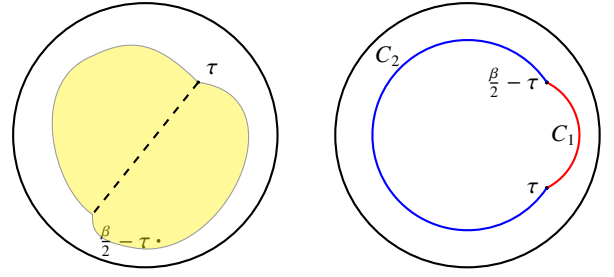
\begin{figure}[t]
\center
\begin{tikzpicture}[scale=0.5,baseline={(current bounding box.center)}]
\draw[thick] (0pt, 0pt) circle (100pt);
\filldraw (40pt,40pt) circle (1pt) node[above right]{$\tau$};
\filldraw (10pt,-80pt) circle (1pt)  node[left]{$\frac{\beta}{2}-\tau$};

\draw[fill=yellow, opacity=0.40] (40pt,40pt)  .. controls (120pt, 30pt) and (60pt,-100pt) .. (-10pt,-85pt) .. controls (-40pt,-80pt) and (-40pt, -70pt)..
(-40pt,-60pt) .. controls  (-70pt,-40pt) and (-90pt, 40pt)..(-40pt, 60pt) ..controls  (-10pt, 75pt) and (10pt, 70pt)..(40pt,40pt) ;

\draw[dashed, thick] (40pt,40pt) -- (-40pt,-60pt);
\end{tikzpicture}\ \ \ \ \ 
\begin{tikzpicture}[scale=0.5,baseline={(current bounding box.center)}]
\draw[thick] (0pt, 0pt) circle (100pt);
\filldraw (60pt,40pt) circle (1pt) node[above right]{$$} ;
\filldraw (60pt,-40pt) circle (1pt) node [below right]{$$};

\node at (-60pt,60pt) {$C_2$};
\node at (73pt,0pt) {$C_1$};

\draw[thick, red] (60pt,40pt) arc (63.43: -63.43 : 44.7pt);
\filldraw (60pt,40pt) circle (1pt) node[left]{$\frac{\beta}{2}-\tau$} ;
\filldraw (60pt,-40pt) circle (1pt) node [left]{$\tau$};

\draw[thick, blue] (60pt,40pt) arc (33.69: 326.31 : 72pt);
\end{tikzpicture}
\caption{ An illustration of the geometric problem on the Poincar\'e disk for the dynamics of R\'enyi entropy with $N_1=N_2=N$ in the low-energy limit. The saddle-point solution is given by joining two arcs. Adapted from \cite{gu2017spread}.
}
\label{fig_geo}
\end{figure}

\subsection{Geometric interpretation for \texorpdfstring{$N_2=N_1$}{TEXT}}
Now we turn to the entanglement dynamics beyond the early-time regime. Generally, this requires direct numerical studies of the saddle-point equations \eqref{eq_TFDSD}. In this subsection, we focus on the special limit with $N_1=N_2=N$, $V\ll J$, and $\beta J\rightarrow \gg 1$, where analytical results exist~\cite{gu2017spread}. The $\chi$ and $\psi$ systems now become symmetric. Following the discussions in \cite{gu2017spread}, we further assume the replica diagonal solution also works at late times~\footnote{Similar calculations has been carried out in \cite{gu2017spread} for SYK chains.}. This has been verified in numerics~\cite{chen2020replica,sohal2022thermalization}.  

Since $V \ll J$, the perturbation due to the coupling between two sites are small. As a result, one can assume the changes of Green's functions are restricted to the low-energy manifold. In other words, we can use the Schwarzian action \eqref{eq_Schwarzian} to determine the entanglement dynamics. Similar to the previous subsection, the effective action contains two parts $S_{\text{eff}}=S_\text{Sch}+ S_\text{int}$, with
\begin{equation}\label{eq_BathSch1}
\begin{aligned}
\frac{S_\text{Sch}}{n}&=-\frac{N\alpha_S}{\mathcal{J}}\int_{C_1 \cup C_2} d\tau~\left\{\tan \frac{\pi f(\tau)}{\beta},\tau\right\},\\
 \frac{S_\text{int}}{n}&=\frac{V^2}{2}\int_{C_1}d\tau_1\int_{C_2}d\tau_2~\tilde G_f(\tau_1,\tau_2)^4.
\end{aligned}
\end{equation}
 We have assumed the only relevant mode satisfies $\tilde G_\psi(\tau_1,\tau_2)=\tilde G_\chi(\tau_1,\tau_2)=\tilde G_f(\tau_1,\tau_2)$. The integral is along a contour on the complex plane. To avoid this, we perform the analytical continuation $t\rightarrow -i \tau$, and then the integral is purely on the imaginary time. We choose $C_1 \in [\tau,\beta/2-\tau)$ and $C_2\in[0,\tau)\cup[\beta/2-\tau,\beta)$. A second analytical continuation back to the real time $\tau \rightarrow it$ is performed at the end of the calculation. This gives
\begin{equation}\label{eq_BathSch2}
\begin{aligned}
 \frac{S_\text{int}}{n}=\frac{\gamma}{2}\log\frac{\beta^2\sin^2\frac{\pi}{\beta}(f(\beta/2-\tau)-f(\tau))}{\pi^2\epsilon^2f'(\beta/2-\tau)f'(\tau)}.
\end{aligned}
\end{equation}
Here $\gamma=V^2/4\pi J^2$ and $\epsilon \sim 1/\mathcal{J}$ is a cutoff.

The problem of solving \eqref{eq_BathSch1} and \eqref{eq_BathSch2} can be mapped to a geometric problem on a Poincar\'e disk thanks to the holographic duality~\cite{gu2017spread}. For a curve $(\rho(\tau),\theta(\tau))$ on a Poincar\'e disk with $\theta(\tau)=2\pi f(\tau)/\beta$ and $g_{\tau\tau}(\tau)=\mathcal J^2$, the Schwarzian action can be represented by 
\begin{equation}
\frac{S_{\text{sch}}}{n N}=\alpha_S(\beta \mathcal{J}-A-2\pi).
\end{equation}
Here $A$ is the area enclosed by the curve. This is a consequence of \eqref{JT} and the Gauss-Bonnet theorem \cite{do2016differential}, and a direct verification is given in \cite{gu2017spread}. On the other hand, $S_\text{int}$ represents an attractive force between two boundary points
\begin{equation}
    \frac{S_{\text{int}}}{n N}=\frac{\gamma}{2}\log \cosh D\left(\frac{\beta}{2}-\tau,\tau\right).
\end{equation}
Here $D(\tau_1,\tau_2)$ is the geodesic distance between two points $(\rho(\tau_1),\theta(\tau_1))$ and $(\rho(\tau_2),\theta(\tau_2))$ on the curve. Intuitively, this can be understood by realizing that the combination in \eqref{eq_BathSch2} takes the form of the two-point function, which can be approximated by $e^{-D(\tau_1,\tau_2)}$.

The entanglement dynamics is determined by minimizing $S_{\text{sch}}+S_{\text{int}}$ with respect to the shape of the curve $f(\tau)$. Away from $\tau$ and $\beta/2-\tau$, the equation of motion is the same as $V=0$, under which the solution is a circle. As a result, the solution takes the form of joining two arcs (up to an SL(2,R) transformation on the Poincar\'e disk). The minimization has been worked out in \cite{gu2017spread} by expressing both terms through the radius and open-angle of each arc. In the long-time limit, the authors find
\begin{equation}
   \frac{n-1}{nN} S^{(n)}\approx\frac{4\pi \alpha_S}{\beta \mathcal{J}}+\gamma\left(1+\log \frac{8\sqrt{2}\alpha_S}{\gamma}\right).
\end{equation}

\begin{figure}[t]
\center
\includegraphics[width=1.05\linewidth]{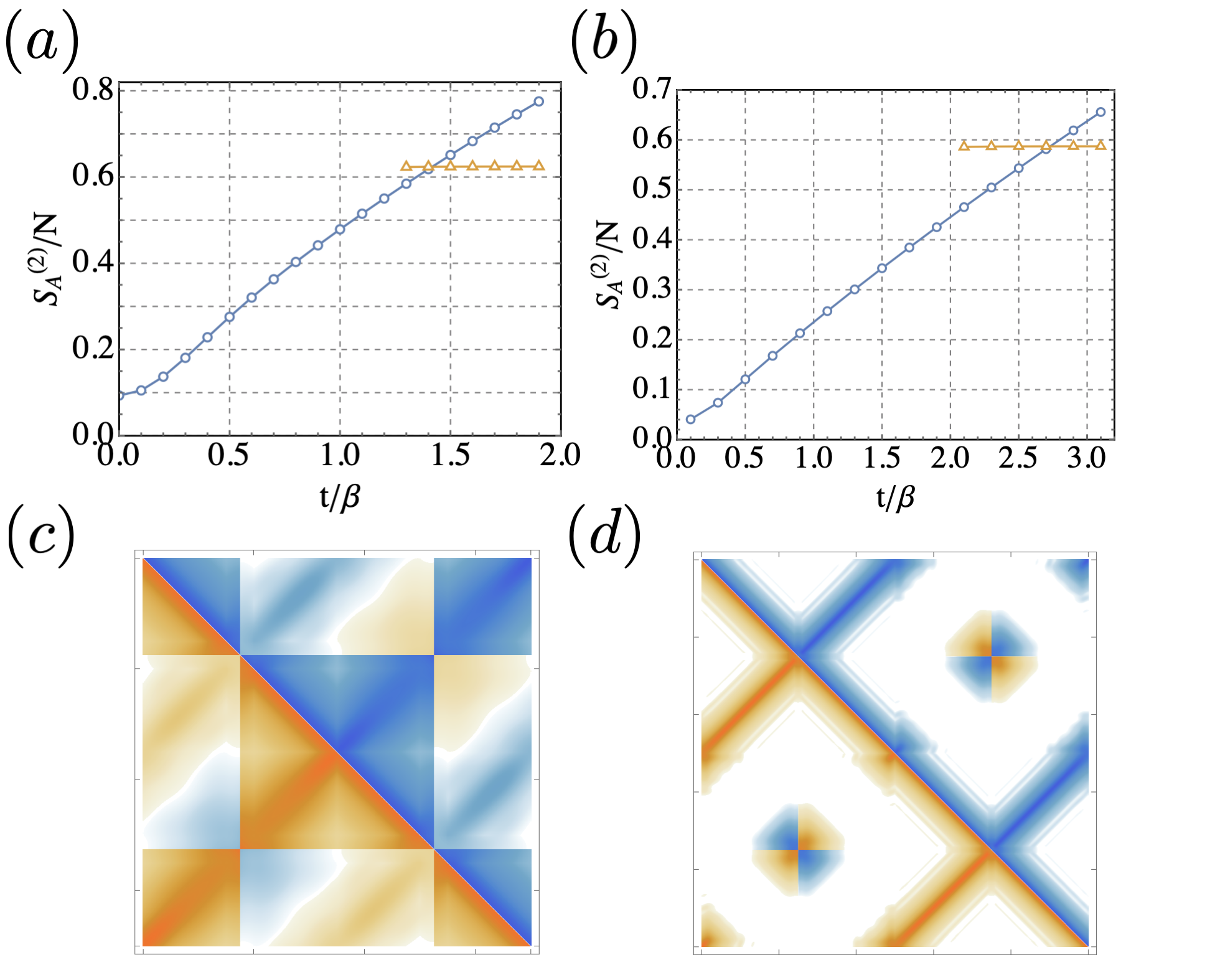}
\caption{ Numerical results for the entanglement dynamics of coupled SYK models. (a). 2-nd R\'enyi entropy for Hamiltonian \eqref{eq_coupledSYK44copy} with $\beta J=4$, $r\rightarrow \infty$, $N_1=N$ and $V/J=1$; (b). 2-nd R\'enyi entropy for Hamiltonian \eqref{eq_chainbath} with $\beta J=4$, $V^2/J=0.05$, $N_1=N$ and $\Lambda/J=5$; (c). Typical Green's function $\text{Re}~G_\chi$ in the early-time regime; (d). Typical Green's function $\text{Re}~G_\chi$ in the late-time regime. (a), (c), and (d) are adapted from \cite{chen2020replica}.}\label{fig_SYKreplicawormhole}
\end{figure}

If we take $\gamma \rightarrow 0$, this reduces to the excitation energy of the SYK model, which does not contain the zero-temperature entropy $S_0$. This may lead to the conclusion that the coupled SYK model fail to thermalize for $N_1=N_2$. However, this may be an artifact of the large-$N$ limit for R\'enyi entropies, and the von Neumann entropy may show parametrically different behaviors as we have seen in the last subsection. For example \footnote{We thank Yingfei Gu for explaining this example. }, if we consider a toy density matrix $$\rho=\frac{P}{P+Q}|E_1\rangle\langle E_1|+\frac{1}{P+Q}\mathbf{I}_{Q\times Q},$$ when $P,Q\gg 1$, We have \begin{equation}
\begin{aligned}
    S^{(n)}&=-\frac{1}{n-1}\log \frac{P^n+Q}{(P+Q)^n}\approx \frac{n }{n-1}\log\frac{P+Q}{P},\\
    S&=\frac{P}{P+Q}\log \frac{P+Q}{P}+\frac{Q}{P+Q}\log(P+Q).
    \end{aligned}
\end{equation}
Here we have assumed $P^2\gg Q$. $|E_1\rangle$ is an analog of the excited state, and $\mathbf{I}_{Q\times Q}$ is an analog of the SYK ground state ensemble. We see that although $S^{(n)}$ only contains the contribution from $|E_1\rangle$ if we take the large $P,Q$ limit, the von Neumann entropy $S$ contains both contributions. It is also known that one can get rid of the non-thermal behavior by considering the microcanonical ensemble~\cite{penington2019replica}. 

One could also ask about results at higher temperatures. In this case, the quantum dynamics is no longer restricted to the low-energy manifold, and the numerical study~\cite{sohal2022thermalization,chen2020replica} shows that the entanglement entropy saturates to the thermal value through a first-order transition. We will review similar physics in the next subsection.

\subsection{Thermalization and replica wormholes}\label{subsec_SYKbathwormhole}
Now we turn to the situation where the $\psi$ system contains much more degrees of freedom $r\gg 1$. In this case, the Green's function of $\psi$ receives no correction from $V$ \eqref{eq_TFDSD} and we can fix the $G_\psi$ by thermal two-point functions. Moreover, we can neglect all $\tilde{G}_\psi$ and $\tilde{\Sigma}_\psi$ terms in the $G$-$\Sigma$ action as
\begin{equation}\label{eq_TFDGsigmaactionrinfty}
\begin{aligned}
\frac{S_{\text{TFD}}^{(2)}}{N}=&-\frac{1}{2}\log \det (\partial_\tau-\tilde \Sigma_\chi)+\frac{1}{2}\int_{\tau \tau'}~\Big[\tilde \Sigma^{mn}_\chi \tilde G^{mn}_\chi\\
&-\frac{J^2}{4}(\tilde G^{mn}_\chi)^4-\frac{V^2}{2}(\tilde G^{mn}_\chi)^2(G^{m'n'}_\psi)^2c_{mm'}c_{nn'}\Big ].
\end{aligned}
\end{equation}
This is because they cancel out with corresponding terms in $2I$. Here we set $N_1=N$ for conciseness. \eqref{eq_TFDGsigmaactionrinfty} can be understood as firstly integrating out the $\psi$ system and focus on the response of $\chi$ fermions. For the model with a bath described by a Majorana chain \eqref{eq_chainbath}, the only difference is to replace the last term in the square bracket by \cite{chen2020replica}
\begin{equation}
-{V^2\Lambda} \tilde G^{mn}_\chi G^{m'n'}_\psi c_{mm'}c_{nn'},
\end{equation}
with $G^{m'n'}_\psi$ computed using the free Majorana chain. 

\begin{figure}[t]
\center
\includegraphics[width=1.\linewidth]{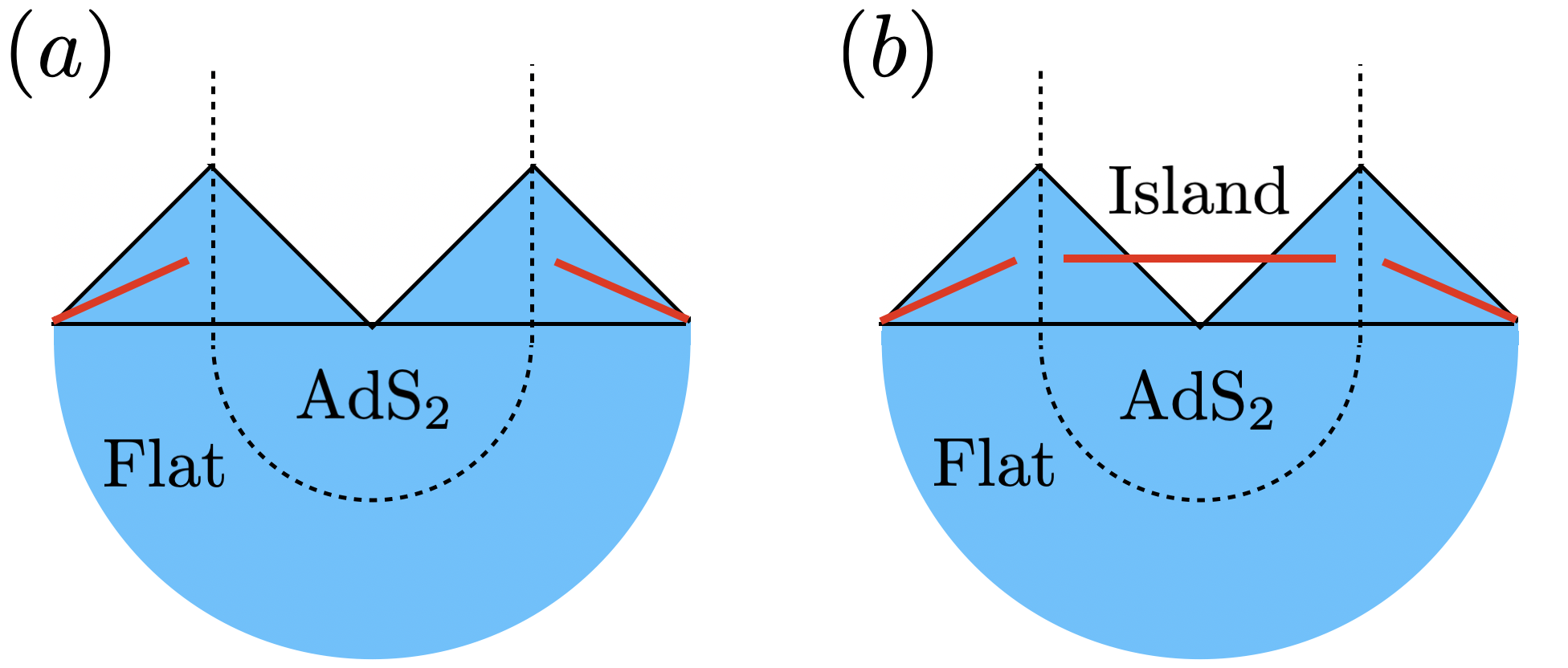}
\caption{ An illustration of the gravity calculation of the von Neumann entropy: (a). The RT surface without an island, and (b). The RT surface with an island outside the horizon.}\label{fig_AdSreplicawormhole}
\end{figure}

The similarity between $G$-$\Sigma$ actions indicates two models show qualitatively the same dynamics of the entanglement entropy. The analysis in section \eqref{subsec_SYKbathlinear} shows the entanglement entropies of both models grow linearly under perturbation. When perturbation theory fails, one should perform a numerical study of Schwinger-Dyson equations. We parametrize the contour using a single real parameter $s\in [0,4(2t+\beta/2)]$, similar to the study of KM pure states. The results are shown in Figure \ref{fig_SYKreplicawormhole} (a) for the SYK$_4$ bath and (b) for the chain bath. The results are qualitatively the same. In the early-time regime, the entanglement entropy grows following a smooth curve with a nearly constant slope. This persists to a very long time even when $S^{(2)}_A>2S^{(2)}_{A,\text{th}}$, which leads to a negative (second R\'enyi) mutual information $S_{\psi^L}^{(2)}+S_{\psi^R}^{(2)}-S_{\psi^L\cup\psi^R}^{(2)}=2S^{(2)}_{A,\text{th}}-S^{(2)}_A<0$. Such a violation of quantum information bounds is known as the information paradox~\cite{hawking1976breakdown}. Fortunately, an additional saddle-point solution appears in the intermediate time regime, which dominates the path-integral and resolves the paradox in the long-time limit by a first-order transition. Similar transitions also show up in Brownian models~\cite{jian2021note,liu2021non}. 

To understand the difference between two saddle-point solutions, we plot the typical Green's functions $G_\chi$ in Figure \ref{fig_SYKreplicawormhole} (c) and (d). Here we use the convention illustrated in \eqref{diagram_TFD}, where replica diagonal solutions of $G_\chi$ and $G_\psi$ take the form
\begin{equation}\label{eq_Gconvention}
G_\chi^{\text{diag}}:\tikz[baseline=4.5ex,x=0.2cm,y=0.2cm]{
        \draw (0,0) -- (8,0) -- (8,8) -- (0,8) -- (0,0);
        \fill[gray] (0,0) rectangle (2,2);
        \fill[gray] (2,2) rectangle (6,6);
        \fill[gray] (6,6) rectangle (8,8);
        \fill[gray] (0,6) rectangle (2,8);
        \fill[gray] (6,0) rectangle (8,2);}
        \ \ \ \ \ \ 
G_\psi^{\text{diag}}:\tikz[baseline=4.5ex,x=0.2cm,y=0.2cm]{
        \draw (0,0) -- (8,0) -- (8,8) -- (0,8) -- (0,0);
        \fill[gray] (0,4) rectangle (4,8);
        \fill[gray] (4,0) rectangle (8,4);}     
\end{equation}

Since $G_\psi$ does not depends on $V$, it is always replica diagonal and we only need to focus on $G_\chi$. On the early-time saddle, we expect the Green's functions take a similar form as $V=0$. For the SYK$_4$ bath, this suggests that $G_\psi$ is strictly replica diagonal on the early-time saddle. For the Majorana chain bath, the replica diagonal assumption is not consistent with a finite $V$. As a result, $G_\psi$ is only approximately replica diagonal, as shown in Figure \ref{fig_SYKreplicawormhole} (c). 

However, $G_\psi$ takes a complete different form on the long-time saddle point, as shown in Figure \ref{fig_SYKreplicawormhole} (d). The solution approximately takes the form of $G^{\text{diag}}_\psi$. This suggests that although the contour of $s\in[0,\beta/2+2t)$ is connected with $s\in[3(\beta/2+2t),4(\beta/2+2t))$, the solution favors the pairing between $s\in[0,\beta/2+2t)$ and $s\in[\beta/2+2t,2(\beta/2+2t))$. Due to the reasons explained below, we call this a replica wormhole solution. As we mentioned in section \ref{subsec_singleSYKpathintegral}, similar transitions also present when considering the entanglement dynamics of KM pure states. In \cite{chen2020replica}, authors explains the replica wormhole solutions can be understood as a factorization of twist operators, which gives
\begin{equation}
e^{-S^{(2)}_A}=\left<T_{\chi,L}T_{\chi,R}\right>\approx\left<T_{\chi,L}\right>\left<T_{\chi,R}\right>=e^{-2S_{A,\text{th}}^{(2)}}.
\end{equation}

The replica wormholes were introduced in the entanglement calculation of holographic systems \cite{penington2019replica,almheiri2019replica}. Towards a gravity analog of the SYK$_4$ fermions coupled to a Majorana chain, we can replace the SYK model with a JT gravity in the AdS$_2$ spacetime. The Majorana chain corresponds to a bath in flat spacetime without gravity dynamics. An SYK$_4$ fermion has a scaling dimension of $\Delta=1/4$, which is dual to a massive bulk field. However, massive fields make the analytical calculation hard. In the gravity calculation \cite{almheiri2019islands}, authors assume there is a conformal field theory in both regions, with transparent boundary conditions. The system is prepared in the TFD state, which is dual to a two-sided black hole, and evolved under the total Hamiltonian $H^\text{tot}$. An illustration of the geometry is shown in Figure \ref{fig_AdSreplicawormhole}.

The von Neumann entanglement entropy of boundary region $A$ can be computed using the RT formula~~\cite{Ryu:2006bv,Ryu:2006ef,Lewkowycz:2013nqa,Hubeny:2007xt,Faulkner:2013ana,Engelhardt:2014gca}:
\begin{equation}\label{eq_RT}
S_A=\text{min}\left[\text{ext}_{\gamma_A}\left(\frac{\text{Area}(\gamma_A)}{4G_N}+S_{\text{bulk}}\right)\right].
\end{equation} 
The first term is the geometric term, and the second term is the entanglement of bulk quantum fields. Here we should consider all possible bulk extremal surfaces $\gamma_A$ homologous to $A$, and choose the minimal one. In our case, the system is 1+1-D, and the RT surface is a set of points. The area term is replaced by the expectation of the dilaton field \cite{Lewkowycz:2013nqa}. The naive RT surface only contains points that separate gravity systems and baths, as illustrated in Figure \ref{fig_AdSreplicawormhole} (a). Since there is no gravity dynamics in the bath, only $S_{\text{bulk}}$ contributes. A CFT calculation shows the result grows linearly in the long-time limit \cite{almheiri2019islands}, due to the establish of the entanglement of bulk fields. Similar to the SYK calculation, this gives rise to an information paradox.

The resolution is to add the contribution in Figure \ref{fig_AdSreplicawormhole} (b). Here the RT surface contains additional points in the gravity region, which forms an island. Without bulk quantum fields, islands always lead to an increase of \eqref{eq_RT} because the area increases. However, in the current calculation, this considerably reduces the entanglement of bulk fields. The explicit calculation shows the long-time limit is then a constant, which approaches $2S^{(2)}_{A,\text{th}}$. The presence of the island can be dated back to the gravity path-integral with multiple replicas \cite{penington2019replica,almheiri2019replica}. For the naive saddle point, the bath is twisted while the gravity region is untwisted. This indicates the way of pairing different contours is different for the gravity system and the bath. The island can be viewed as emergent dynamical twist operators. After the island appears, most of spacetime regions are twisted, and the contours for the gravity system and the bath are paired similarly. This is consistent with the SYK result. These emergent twist operators connect gravity systems in different replicas through wormholes, which are known as ``replica wormholes'' \cite{penington2019replica,almheiri2019replica}. When the island presents, the entanglement wedge reconstruction \cite{dong2016reconstruction} implies part of the gravity region will be accessible to the bath. Different concrete protocols to extract information have been proposed in \cite{penington2019replica} and \cite{chen2020pulling}.

\subsection{Entanglement dynamics of SYK chains}\label{subsec_SYKchainunitary}
We finally extend the discussions for the coupled $\chi$-$\psi$ systems to SYK chains \cite{gu2017local}. On each site, there is an SYK$_4$ model, and they are coupled by SYK-like random pair hopping terms. The Hamiltonian reads
\begin{equation}
\begin{aligned}
H=&\sum_x\Bigg (\sum_{ i_1< i_2<i_3 < i_4}J^x_{i_1 i_2 i_3 i_4} \chi_{x,i_1}\chi_{x,i_2}\chi_{x,i_3}\chi_{x,i_4}\\
&+\sum_{i_1 < i_2, j_1< j_2} V^{x}_{i_1i_2 j_1j_2}\chi_{x,i_1}\chi_{x,i_2}\chi_{x+1,j_1}\chi_{x+1,j_2}\Bigg).
\end{aligned}
\end{equation}
Here $x\in [1,L]$ with periodic boundary condition. We prepare the system in a TFD state with an auxiliary system. Entanglement dynamics of KM states is similar. We choose the subsystem $A$ as the first $L_A\leq L/2$ sites. 

The first contribution to the R\'enyi entropy is the (approximate) replica diagonal solution. Following the discussions in section \ref{subsec_SYKbathlinear}, the entanglement entropy will grow linearly for $L_A\geq 1$, with, a slope $v_E$ independent of subsystem size. Choosing the same convention as the last subsection, the Green's functions $G_x$ look like $G_\chi^{\text{diag}}$ if $x\in A$, and $G_\psi^{\text{diag}}$ if $x\in B$. Moreover, Green's functions in the bulk of $A/B$ system approach two copies of the thermal Green's function. 

The second contribution is the replica wormhole solution, where all Green's functions $G_x$ look like $G_\psi^{\text{diag}}$. However, due to the mismatch between connectivity of the contour and the pairing indicated by the Green function, small deviations from $G_\psi^{\text{diag}}$ always exist near the insertion time of twist operators (as in Figure \ref{fig_SYKreplicawormhole} (d)), even in the bulk of subsystem $A$. As a result, the replica wormhole solution leads to a volume law phase with $S^{(2)}_A=s_0L_A$. 
Putting both contributions together, the entanglement dynamics satisfies
\begin{equation}\label{eq:scalingunitarychain}
    S^{(2)}_A (t,L_A)=\left\{
    \begin{array}{cc}
         v_E t &  {\rm if}\quad t< s_0L_A/v_E\\
         s_0 L_A &  {\rm if}\quad  t\geq s_0L_A/v_E
    \end{array}
    \right.
\end{equation}
This is consistent with numerical results in \cite{liu2021non}. 

Interestingly, this calculation can formulated as a classical spin problem. We can treat the choice of the Green's functions as a spin degree of freedom. To be concrete, we label $G_\psi^{\text{diag}}$ as $\bm{1}$ and $G_\chi^{\text{diag}}$ as $\bm{\sigma}$. The boundary condition is fixed by twist operators: \eqref{diagram_TFD} suggests we should impose $\bm{1}$ for $x\in B$ and $\bm{\sigma}$ for $x\in A$ near $s=0$ and $\beta/2+2t$. The replica diagonal and replica wormhole solution can then be illustrated as
\begin{equation}\label{eq_unitarydomain}
\tikz[baseline=4.5ex,x=0.2cm,y=0.25cm]{
        \draw (-2,-2) -- (12,-2) -- (12,6) -- (-2,6) -- (-2,-2);
        \node at (5,7) {\small \text{Diagonal}};
        \fill[gray] (3,-2) rectangle (7,6);
        \node at (5,2) {\small $\bm{\sigma}$};
        \node at (0.5,2) {\small $\bm{1}$};
        \node at (9.5,2) {\small $\bm{1}$};
        \node at (-4.2,6) {\small $s=0$};
        \node at (-4.2,-2) {\small $2t+\frac{\beta}{2}$};
        \node at (5,-2.7) {\small $A$};
        \node at (0.5,-2.7) {\small $B$};
        \node at (9.5,-2.7) {\small $B$};
        }
        \ \ \ \ \ \ \ 
\tikz[baseline=4.5ex,x=0.2cm,y=0.25cm]{
        \draw (-2,-2) -- (12,-2) -- (12,6) -- (-2,6) -- (-2,-2);
        \node at (5,7) {\small \text{Wormhole}};
        \fill[gray] (3,5) rectangle (7,6);
        \fill[gray] (3,-2) rectangle (7,-1);
        \node at (5,2) {\small $\bm{1}$};
        \node at (5,5.5) {\small $\bm{\sigma}$};
        \node at (5,-1.5) {\small $\bm{\sigma}$};
        \node at (5,-2.7) {\small $A$};
        \node at (0.5,-2.7) {\small $B$};
        \node at (9.5,-2.7) {\small $B$};
        }   
\end{equation}
The replica diagonal solution corresponds to domain walls along the time-direction, the energy cost of which grows linearly. The replica wormhole solution corresponds to having domain walls parallel to the spatial direction, which indicates a volume law entanglement entropy. Similar picture was proposed in random tensor networks \cite{hayden2016holographic} and random circuits \cite{nahum2018operator,von2018operator}. As we will see in the next subsection, such a picture gives a natural understanding of measurement induced phase transitions.

\section{Entanglement in non-unitary SYK models}\label{sec_nonunitarySYK}
Recent years, it was realized novel quantum entanglement dynamics can be achieved for ``hybrid'' quantum systems, in which evolutions involve both unitary evolutions and measurements~\cite{li2018quantum,10.21468/SciPostPhys.7.2.024,li2019measurement,skinner2019measurement,chan2019unitary,bao2020theory,choi2020quantum,gullans2020dynamical,gullans2020scalable,jian2020measurement,szyniszewski2019entanglement,zabalo2020critical,tang2020measurement,zhang2020nonuniversal,goto2020measurement,jian2020criticality,bao2021symmetry,alberton2020trajectory,Chen_2020,Nahum_2020,liu2021non,zhang2021emergent,jian2021measurement,jian2021phase,zhang2021universal,jian2021quantum,sahu2021entanglement}. When tunning the measurement rate, the steady states can show entanglement phase transitions between volume law entangled phases and area law entangled phases, separated by a critical point with logarithmic entanglement entropy if the transition is second order.

Since SYK-like models are solvable, they are ideal platforms for understanding the entanglement phase transitions \cite{liu2021non,zhang2021emergent,jian2021measurement,jian2021quantum,jian2021phase,zhang2021universal,sahu2021entanglement}. In this section, we review the studies of non-unitary SYK models, focusing on continuous forced measurements which are equivalent to non-Hermitian Hamiltonian dynamics. A general measurement is described by a set of Kraus operators $\{M_\xi\}$ that satisfies $\sum_\xi M_\xi ^\dagger M_\xi=I$. Let us assume the system is prepared in some state $|\psi_0\rangle$. After performing the measurement, the probability $p_\xi$ of getting the result $\xi$ and the corresponding post-selected state $|\psi_\xi\rangle$ read
\begin{equation}
p_\xi=\langle\psi_0|M_\xi^\dagger M_\xi|\psi_0\rangle,\ \ \ \ \ \ |\psi_\xi\rangle=\frac{1}{\sqrt{p_\xi}}M_\xi|\psi_0\rangle.
\end{equation}
Now we focus on the special case where $\xi=1,2$ with $M_1\approx I-{\epsilon} H_\text{I}$ and $M_2\approx\sqrt{2\epsilon} H_\text{I}$, with Hermitian and positive semidefinite $H_\text{I}$ and $\epsilon \ll 1$. We now post select the measurement outcome with $\xi=1$. This leads to
\begin{equation}
|\psi_\xi\rangle\propto \left(1-{\epsilon} H_\text{I}\right)|\psi_0\rangle\approx e^{-\epsilon H_\text{I}}|\psi_0\rangle.
\end{equation}
This is equivalent to an imaginary-time evolution of the initial state. Taking the continuum limit by setting $\epsilon=dt$ and adding the unitary evolution part with Hamiltonian $H_\text{R}$, we find 
\begin{equation}
|\psi(T)\rangle \propto e^{-iHT}|\psi_0\rangle, \ \ \ \ \ \ H=H_R-iH_I.
\end{equation}
We choose the subsystem $A$ as first $L_A$ sites. The $n$-th R\'enyi entropy reads
\begin{equation}
e^{-(n-1)S^{(n)}_A}=\frac{\text{tr}_A\left(\text{tr}_B~e^{-iHT}\left|\psi_0\right>\left<\psi_0\right|e^{iH^\dagger T}\right)^n}{\left(\left<\psi_0\right|e^{iH^\dagger T}e^{-iHT}\left|\psi_0\right>\right)^n}.
\end{equation}
Main previous studies focus on $n=2$ \cite{liu2021non,zhang2021emergent,jian2021measurement,jian2021quantum,zhang2021universal,sahu2021entanglement}. The von Neumann entropy limit $n\rightarrow 1$ has been studied in \cite{jian2021phase} with further assumptions.

In the following subsections, we will consider different non-Hermitian SYK chains and derive effective actions that govern the entanglement dynamics of the 2-nd R\'enyi entropy. An important ingredient is the emergence of conformal symmetry in replicated non-interacting systems for small measurement rates, which leads to logarithmic entanglement entropy. After adding interactions, the gapless modes become gapped, and the volume law entanglement replaces the logarithmic law entanglement. We can drive transitions to area law phases in certain setups for a larger measurement rate. We will also discuss the consequence of adding long-range couplings.

\subsection{Emergent replica conformal symmetry }
The non-unitary dynamics of free fermions has been studied numerically \cite{Chen_2020}, and the result shows logarithmic entanglement entropy with emergent conformal symmetry. In this subsection, we review the progress of understanding these results using SYK solvable models. We consider the non-interacting SYK$_2$ chains \cite{zhang2021emergent}. The Hamiltonian reads \footnote{Here $H_I$ is not positive semidefinite. However, we can always make it positive semidefinite by shifting a large enough constant.}
\begin{equation}\label{eq:H0non2}
\begin{aligned}
H_R&=i\sum_{x}\Big[\sum_{j,k}J_{jk}^{x}(t) \chi_{x,j}\chi_{x+1,k}+\sum_{j<k}\tilde{J}_{jk}^{x}(t) \chi_{x,j}\chi_{x,k}\Big],\\
H_I&=i\sum_{x}\Big[\sum_{j,k}V_{jk}^{x}(t) \chi_{x,j}\chi_{x+1,k}+\sum_{j<k}\tilde{V}_{jk}^{x}(t) \chi_{x,j}\chi_{x,k}\Big].
\end{aligned}
\end{equation}
For simplicity, we choose couplings to be Brownian:
\begin{equation}
\begin{aligned}
&\overline{J_{ij}^{x}(t)J_{ij}^{x}(0)}=\frac{J_1}{2N}\delta(t),\ \ \ \ \overline{\tilde{J}_{ij}^{x}(t)\tilde{J}_{ij}^{x}(0)}=\frac{J_0}{N}\delta(t),\\&\overline{V_{ij}^{x}(t)V_{ij}^{x}(0)}=\frac{V_1}{2N}\delta(t),\ \ \ \ \overline{\tilde{V}_{ij}^{x}(t)\tilde{V}_{ij}^{x}(0)}=\frac{V_0}{N}\delta(t).
\end{aligned}
\end{equation}
The static coupling case has been analyzed similarly in \cite{zhang2021emergent}.

\begin{figure}[t]
\center
\includegraphics[width=1.\linewidth]{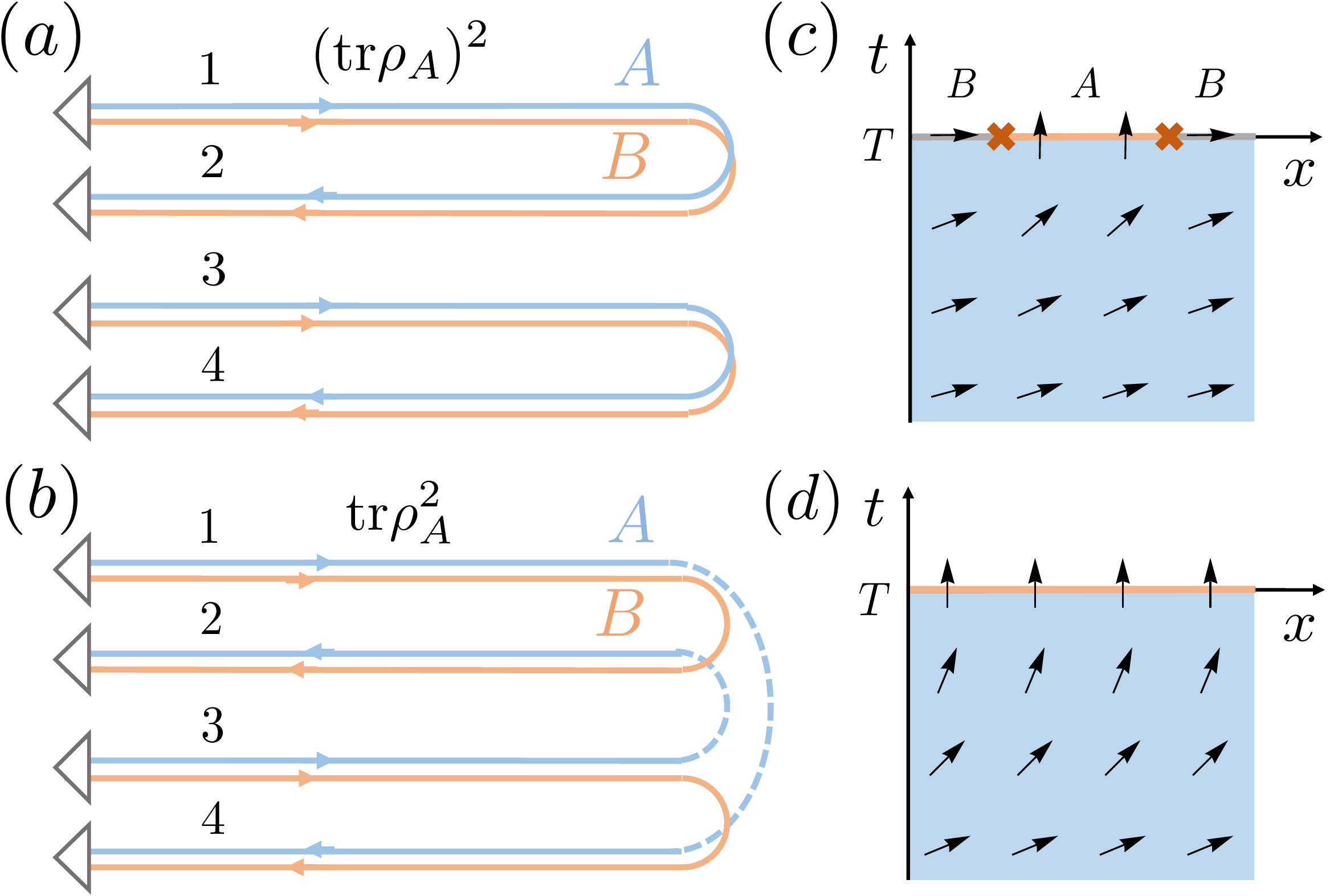}
\caption{(a-b). A sketch of the path-integral contour without/with twist operators for subsystems $A$. (c). The configuration of $\theta$ when computing the entanglement entropy in non-Hermitian SYK$_2$ chains. (d). The configuration of $\theta$ for the purification dynamics in non-Hermitian SYK$_2$ chains. (c), (d) are adapted from \cite{zhang2021universal}.}\label{fig_nonhermitianSYK2}
\end{figure}

In section \ref{subsec_SYKchainunitary}, we have seen that the entanglement entropy calculation in interacting systems with unitary evolution corresponds to the energy of domain walls. We expect similar pictures work here: the twist operator creates some defects which excite low-energy excitations, and the entanglement entropy is given by the corresponding energy. To understand the possible low-energy excitations, we start with the path-integral without any twists $\text{lim}_{T\rightarrow \infty}(\text{tr}~e^{iHT}e^{iHT})^2$. We label four evolution contours by $1,2,3,4$, where the evolution is forward/backward on $\{1,3\}$/$\{2,4\}$, as illustrated in Figure \ref{fig_nonhermitianSYK2} (a). Following the derivation in section \ref{subsec_SYKlargeN}, the saddle-point equation reads 
\begin{equation}\label{eq_saddlenonSYK21}
\begin{aligned}
&\left[(-1)^{a+1}\delta^{ac}\partial_t-\Sigma_x^{ac}\right]\circ G_x^{cb}=I^{ab},
\end{aligned}
\end{equation}
and 
\begin{equation}\label{eq_saddlenonSYK22}
\begin{aligned}
\Sigma_x^{ab}(t,t')=&\Bigg[(V_1^2-(-1)^{a+b}J_1^2)\frac{G_{x+1}^{ab}(t,t)+G_{x-1}^{ab}(t,t)}{2}\\&+(V_0^2-(-1)^{a+b}J_0^2)G_{x}^{ab}(t,t)\Bigg]\delta(t-t').
\end{aligned}
\end{equation}

Interestingly, there is a $O(2)\otimes O(2)$ symmetry: $G(t,t')\rightarrow OG(t,t')O^{T}$ with $O=\exp(-\gamma_{13}\theta_{13}-\gamma_{24}\theta_{24})$ left the equation invariant. Here $(\gamma_{cd})^{ab}=\delta_{ac}\delta_{bd}-\delta_{bc}\delta_{ad}$ is a $4\times4$ symmetry generator in the contour space. This can be understood as the promotion of the permutation symmetry $(1 \leftrightarrow 3)$, $(2 \leftrightarrow 4)$ for quadratic actions.

This symmetry is broken by the saddle-point solution. From Figure \ref{fig_nonhermitianSYK2} (a), we see that contour $(1,2)$ is disconnected from the contour $(3,4)$. As a result, we expect $G_s^{ab}$ to be block diagonal $G_s^{13}=G_s^{14}=G_s^{23}=G_s^{24}=0$, and satisfying $G_s^{ab}=G_s^{a+2,b+2}$ for $a,b\in \{1,2\}$. Explicitly, on steady states, the solution of \eqref{eq_saddlenonSYK21} and \eqref{eq_saddlenonSYK22} reads
\begin{equation}\label{eq:GBrownian}
G_s^{11}(\omega)=\frac{i\omega}{\omega^2+\Gamma^2/4},\ \ \ \ G_s^{12}(\omega)=-\frac{\Gamma/2}{\omega^2+\Gamma^2/4}.
\end{equation}
Here we have defined $\Gamma=V+J= V_0+V_1+J_0+J_1$.The solution \eqref{eq:GBrownian} is only invariant under the transformation if we have $\gamma_{13}=\gamma_{24}$, and the residue symmetry group is $O(2)$. Since the relative rotation symmetry is broken, there is an associated Goldstone mode living in the coset space $O(2)\otimes O(2)/O(2)=O(2)$. We will use a $\theta\in[0,2\pi)$ field for this Goldstone mode. The effective theory for $\theta$ can be derived by considering fluctuations around the saddle point. The result reads \cite{zhang2021emergent}
\begin{equation}
\label{eq:effective_action_Brownian_theta}
\frac{S_\theta}{N}=\frac{1}{2}\int_{\omega k}\left(\frac{J_1+V_1}{4}k^2+\frac{1}{4V}\omega^2\right)|\theta(\omega,k)|^2.
\end{equation}
This is the 2D XY model, with dynamical exponent $z=1$. It is also a representative conformal field theory, which explains the conformal symmetry of steady states observed in numerical studies. Since this conformal symmetry only emerges after introducing replicas, the authors call it an emergent replica conformal symmetry~\cite{zhang2021emergent}.

The entanglement entropy can be computed using the effective action \eqref{eq:effective_action_Brownian_theta}. The important observation is that the Goldstone mode $\theta$ is exactly the degree of freedom that labels the pairing between contours. For $\theta=0$, contours $(1,2)$ and $(3,4)$ are paired, as shown in \eqref{eq:GBrownian}. This is an analog of $\bm{I}$ in section \ref{subsec_SYKchainunitary}. After performing a relative rotation $\theta$, the Green's functions become $G^{12}=\cos \theta ~G^{12}_s$ and $G^{14}=\sin \theta ~G^{12}_s$. Consequently, $\theta=\pi/2$ corresponds to pairing $(1,4)$ and $(2,3)$. This is an analog of $\bm{\sigma}$ in section \ref{subsec_SYKchainunitary}. Consequently, when computing the entanglement entropy, we should impose $\theta=0$ near the boundary of the contour for sites in subsystem $B$, and $\theta=\pi/2$ for sites in subsystem $A$. This is illustrated in Figure \ref{fig_nonhermitianSYK2} (b). For the XY model, \eqref{eq:effective_action_Brownian_theta}, this boundary condition excites a pair of half-vortexes, with a energy logarithmic in their distance: 
\begin{equation}\label{eq_aaaaaa}
S^{(2)}_A\propto \rho_s \log L_A=\sqrt{\frac{J_1+V_1}{V}}N\log L_A.
\end{equation} 
Here $\rho_s$ is known as the superfluid density \cite{zhai2021ultracold}. \eqref{eq_aaaaaa} has been verified by comparing to numerical results in \cite{zhang2021emergent}.

We finally comment on interaction effects. The absence of interactions is essential for the presence of $O(2)\otimes O(2)$ symmetry, and thus the emergence of replica conformal symmetry. When interactions are introduced, for example an on-site Brownian SYK$_4$ term, the $O(2)\otimes O(2)$ symmetry is broken explicitly back to $Z_4 \otimes Z_4$, where $\theta$ is fixed to be $n \pi/2$, as a combination of $\chi^a\rightarrow -\chi^a$ and permutations between contours. More explanation is given in the next subsection. The Goldstone modes gain mass and a smooth rotation of $\theta$ is no longer allowed. As a result, the boundary condition of $S_A^{(2)}$ will excites a domain wall instead of a half-vortex pair. This leads to a volume law phase similar to section \ref{subsec_SYKchainunitary}. This concludes that no entanglement phase transition shows up in this simple Hamiltonian \eqref{eq:H0non2}.

\subsection{Entanglement phase transitions}
A non-Hermitian SYK chain with entanglement phase transition was proposed in \cite{jian2021measurement}. Comparing to the model in the last subsection, it contains two copy of SYK chains, labeled by $L$ and $R$. The Hamiltonian reads
\begin{equation}
\begin{aligned}
\label{eq:LRhamiltonian}
	H_{\text{R}} =\sum_{x,a} & \Big( \sum_{jk}  i J_{jk}^{x,a}(t) \psi_{x,a,j} \psi_{x+1,a,k} \\&+ \sum_{j_1<...<j_q} i^{q/2} U_{j_1...j_q}^{x,a}(t) \psi_{x,a,j_1}... \psi_{x,a,j_q}  \Big),
	\end{aligned}
\end{equation}
with $a=L,R$, and 
\begin{equation}
H_{\text{I}} =-\frac{\mu}{2} \int dt  \sum_{x,i} i \psi_{x,L,i} \psi_{x,R,i}. 
\end{equation}
The model can be understood as performing continuous forced measurements on the inter-chain fermion parity. We assume $q\geq 4$ and $U=0$ corresponds to the non-interacting limit. 

Following the discussions in the last subsection, we first consider the saddle-point solutions without twist operators. For $U=0$, the system still exhibits the $O(2)\otimes O(2)$ symmetry. The solution of saddle-point equations reads \cite{jian2021measurement}
\begin{equation}
G = \begin{cases} \frac{e^{- \frac{J|t|}2 }}{2} \big( \text{sgn}(t)\sigma^z - \sqrt{1-\tilde\mu^2} i \sigma^y + \tilde \mu \tau^y  \big), \ \ \ \  \tilde \mu < 1  \\
					    \frac{e^{- \frac{\mu |t|}2 }}{2} \left( \text{sgn}(t)\sigma^z +  \tau^y  \right), \qquad \qquad \qquad \quad \tilde \mu \ge 1 \end{cases}
\end{equation}
Here $\tilde \mu=\mu/J$. $\sigma_a$ ($\tau_a$) are the Pauli matrices in the $1,2$
($L$,$R$) space. We find $G^{12}\neq 0$ for $\mu<J$, where the forward evolution pairs with the backward evolution, while $G^{12}> 0$ for $\mu\geq J$, where the pairing is between two copies of the SYK chain. When the pairing between forward and backward evolution presents, the $O(2)\otimes O(2)$ symmetry is broken to $O(2)$, and the Goldstone mode exists. On the other hand, when $G^{12}= 0$, the saddle-point solution preserves the full symmetry, and the system is gapped. Such a transition can be formulated as an effective action \cite{jian2021measurement}:
\begin{equation}
 \frac{S_\phi}{N} = \int dt dx~\left( \frac12 (\partial \vec \phi)^2 + r \vec \phi^2 + \lambda \vec \phi^4  \right),
\end{equation}
Here $\lambda>0$ and $r\propto \frac{\mu-J}{2}$ determines whether the pairing field $\vec \phi\sim(G^{12}+G^{34},G^{14}+G^{23})$ condenses or not. This suggests the entanglement phase transition for non-interacting systems with large local Hilbert space dimension is a large-$N$ $O(2)$ symmetry breaking transition. For $\mu<J$, the Goldstone modes lead to logarithmic entanglement, as in the previous subsection. For $\mu>J$, the system is disordered with respect to boundary conditions. As a result, the insertion of twist operators will not excite any defect, and the entanglement entropy is area law. 

The relation between $G^{12}$ and the scaling of the entanglement entropy can also be understood from the perturbation theory in section \ref{subsec_SYKbathlinear} \cite{liu2021non}. By identifying $G^{21}=iG^>$, the early-time contribution \eqref{eq_Sbathlinear1} always grows linearly. As explained in section \ref{subsec_SYKchainunitary}, this is the area law contribution. In the long-time limit, it becomes larger than the replica wormhole contribution, and the saddle point is dominated by the latter. On the contrary, $G^{12}=0$, the perturbative calculation predicts a saturation value for the area law entanglement. Consequently, it will dominate the entanglement entropy in the long-time limit and lead to an area law entangled phase.

Now we consider small interaction $U/J\ll1$. Similar to \eqref{eq_GSigma}, this introduces an additional term $G^4$ into the action, which contains a contribution $( \phi_1^4 + \phi_2^4)$. Consequently, the revised effective action reads 
\begin{equation}
 \frac{S_\phi}{N} = \int dt dx~\left( \frac12 (\partial \vec \phi)^2 + r \vec \phi^2 + \lambda \vec \phi^4  ,
 + \lambda' ( \phi_1^4 + \phi_2^4)\right),
\end{equation}
with $\lambda' \propto U$. The additional term breaks the $O(2)$ symmetry, and the residue group is $Z_4$. The transition at $r=0$ becomes a large-$N$ $Z_4$ symmetry breaking transition. The disordered phase is similar to the non-interacting case, while the ordered phase exhibits volume law entanglement entropy. In \cite{jian2021quantum}, authors also use effective actions to study the entanglement phase transition from the quantum error correction perspective. The extension to phase transitions of the von Neumann entropy is studied in \cite{jian2021phase}.

\begin{figure}[t]
\center
\includegraphics[width=0.85\linewidth]{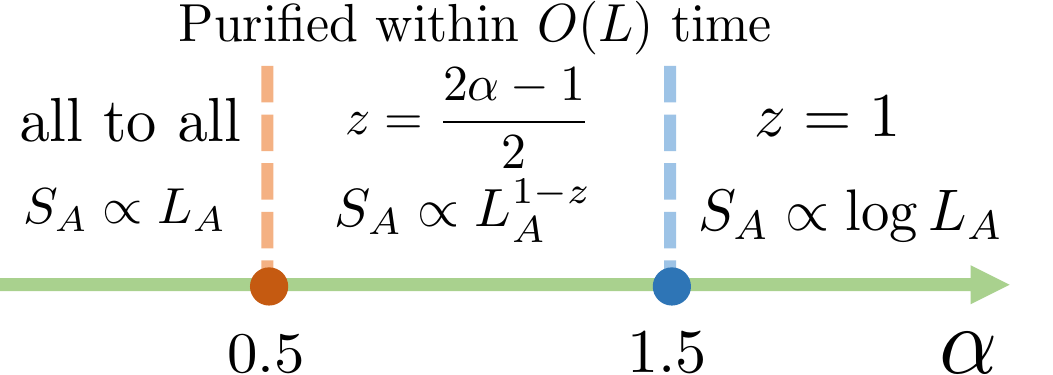}
\caption{ The phase diagram of the non-Hermitian SYK$_2$ chain with long-range couplings. Adapted from \cite{zhang2021universal}.}\label{fig_longrange}
\end{figure}

\subsection{Effects of long-range couplings}

The long-range coupling presents in most of the state-of-the-art experimental platforms for quantum dynamics, including NMR systems and cold atom systems. In this subsection, we review efforts to understand the non-unitary dynamics using SYK chains. We focus on the non-interacting case studied in \cite{zhang2021universal}. The effect of long-range couplings in interacting SYK-like models has been studied in \cite{sahu2021entanglement}.

We consider adding long-range hopping terms to the Hamiltonian \eqref{eq:H0non2} as
\begin{equation}
H_R=i\sum_{x}\Big[\sum_{j,k,r}\frac{J_{jk}^{x,r}(t)}{r^\alpha} \chi_{x,j}\chi_{x+r,k}+\sum_{j<k}\tilde{J}_{jk}^{x}(t) \chi_{x,j}\chi_{x,k}\Big].
\end{equation}
Here $r\in\{1,2,...\}$ labels the range of hopping. For $\alpha\leq 0.5$, the decay of random hopping is too slow, and the system is effectively all-to-all connected. We focus on the $\alpha>0.5$, where the system has a well-defined thermodynamical limit. The solution of the saddle-point equation is still given by \eqref{eq:GBrownian}, with $\Gamma=V_0+V_1+J_0+\sum_r J_1/r^{2\alpha}$. 

Comparing to the short-range hopping case, there is no difference from the symmetry perspective. To determine the scaling of entanglement entropy, we should derive the effective action for Goldstone modes. The $k^2$ term in \eqref{eq:effective_action_Brownian_theta} can be understood as the kinetic energy of a Majorana fermion pair $(V_1+J_1)k^2/2=(V_1+J_1)(1-\cos k)$. After adding long-range coupling, we replace it with
\begin{equation}\label{eq_dispersion}
\epsilon(k)=V_1(1-\cos(k))+\sum_{r=1}^{\infty}\frac{J_1}{r^{2\alpha}}(1-\cos(rk)).
\end{equation}
The low-energy physics is determined by the small $k$ expansion of $\epsilon(k)$, which depends on $\alpha$ as
\begin{equation}\label{eq_epsilon}
  \epsilon(k) \propto
    \begin{cases}
       k^2/2\ \ &\ \ \text{for $\alpha>1.5$,}\\
       k^{2\alpha-1}\ \ &\ \ \text{for $\alpha\leq1.5$.}
    \end{cases}       
\end{equation}
As a result, for $\alpha\geq 1.5$, there is no difference compared to models with short-range hopping. We have dynamical exponent $z=1$ and logarithmic entanglement entropy. For $\alpha<1.5$, the dynamical exponent depends on $\alpha$ as $z=\frac{2\alpha-1}{2}$. Moreover, the entanglement entropy is no longer equal to the energy of a half vortex pair, and a direct calculation using dual fields of $\theta$ shows $S_A^{(2)}\propto L_A^{1-z}$ \cite{zhang2021universal}. This is known as a fractal phase \cite{ippoliti2021fractal}. Implications on the purification dynamics are also discussed in \cite{zhang2021universal}.

\section{Outlook}\label{sec_outlook}
This review summarizes recent studies of the quantum entanglement in the SYK model and its generalizations. The entanglement entropy of the single-site SYK model can be studied using the ED, the ETH, and the path-integral approach. Results from different methods match to good precision. For coupled SYK models, the R\'enyi entanglement entropy generally shows first-order transitions, where the long-time saddle is an analog of replica wormholes. Non-unitary SYK chains can also be used to study measurement induced entanglement phase transitions for the 2-nd R\'enyi entropy, which can be understood as conventional symmetry breaking transitions on replicated Hilbert space. 
\vspace{5pt}

There are several interesting open questions:
\begin{enumerate}[label=\textbf{\arabic*.}]
\item\textbf{Von Neumann Entropy.} In gravity calculations, the von Neumann entropy is much easier to compute compared with R\'enyi entropies \cite{dong2016gravity}, which receive additional backreactions. However, in the SYK model and its generalizations, it's hard to take the limit $n\rightarrow 1$ and derive results for the von Neumann entropy, except for the Brownian models \cite{jian2021phase}, or perturbative calculations \cite{dadras2021perturbative}. It is interesting to develop the methodology for the path-integral approach for the von Neumann entropy for interacting quantum systems.

\item\textbf{General Measurements.} In section \ref{sec_nonunitarySYK}, we focus on forced measurements, under which the evolution is described by non-Hermitian Hamiltonian dynamics. It is interesting to consider more general measurements, in particular, to understand the consequence of random outcomes. It is also inspiring to reveal the gravity picture of the measurement induced phase transitions. 

\item\textbf{General Many-body Systems.} The SYK model and its generalizations are solvable in the large-$N$ limit, and consequently, we can understand their entanglement dynamics using the path-integral representation. However, there is no simple saddle-point description in general quantum systems with strong interactions. It is of vital importance to develop approximation methods for entanglement dynamics using motivations from SYK-like models.
\end{enumerate}

\textit{Acknowledgment.} We acknowledge helpful discussions with Xiao Chen, Yiming Chen, Yingfei Gu, and Chunxiao Liu. We thank Yingfei Gu, Yiming Chen, Chunxiao Liu, and Ning Sun for carefully reading the manuscript and giving valuable suggestions. PZ acknowledges support from the Walter Burke Institute for Theoretical Physics at Caltech. 

\bibliography{ref.bbl}

\begin{thebibliography}{132}%
\makeatletter
\providecommand \@ifxundefined [1]{%
 \@ifx{#1\undefined}
}%
\providecommand \@ifnum [1]{%
 \ifnum #1\expandafter \@firstoftwo
 \else \expandafter \@secondoftwo
 \fi
}%
\providecommand \@ifx [1]{%
 \ifx #1\expandafter \@firstoftwo
 \else \expandafter \@secondoftwo
 \fi
}%
\providecommand \natexlab [1]{#1}%
\providecommand \enquote  [1]{``#1''}%
\providecommand \bibnamefont  [1]{#1}%
\providecommand \bibfnamefont [1]{#1}%
\providecommand \citenamefont [1]{#1}%
\providecommand \href@noop [0]{\@secondoftwo}%
\providecommand \href [0]{\begingroup \@sanitize@url \@href}%
\providecommand \@href[1]{\@@startlink{#1}\@@href}%
\providecommand \@@href[1]{\endgroup#1\@@endlink}%
\providecommand \@sanitize@url [0]{\catcode `\\12\catcode `\$12\catcode
  `\&12\catcode `\#12\catcode `\^12\catcode `\_12\catcode `\%12\relax}%
\providecommand \@@startlink[1]{}%
\providecommand \@@endlink[0]{}%
\providecommand \url  [0]{\begingroup\@sanitize@url \@url }%
\providecommand \@url [1]{\endgroup\@href {#1}{\urlprefix }}%
\providecommand \urlprefix  [0]{URL }%
\providecommand \Eprint [0]{\href }%
\providecommand \doibase [0]{http://dx.doi.org/}%
\providecommand \selectlanguage [0]{\@gobble}%
\providecommand \bibinfo  [0]{\@secondoftwo}%
\providecommand \bibfield  [0]{\@secondoftwo}%
\providecommand \translation [1]{[#1]}%
\providecommand \BibitemOpen [0]{}%
\providecommand \bibitemStop [0]{}%
\providecommand \bibitemNoStop [0]{.\EOS\space}%
\providecommand \EOS [0]{\spacefactor3000\relax}%
\providecommand \BibitemShut  [1]{\csname bibitem#1\endcsname}%
\let\auto@bib@innerbib\@empty
\bibitem [{\citenamefont {Jozsa}(1997)}]{jozsa1997entanglement}%
  \BibitemOpen
  \bibfield  {author} {\bibinfo {author} {\bibfnamefont {R.}~\bibnamefont
  {Jozsa}},\ }\href@noop {} {\bibfield  {journal} {\bibinfo  {journal} {arXiv
  preprint quant-ph/9707034}\ } (\bibinfo {year} {1997})}\BibitemShut {NoStop}%
\bibitem [{\citenamefont {Jozsa}\ and\ \citenamefont
  {Linden}(2003)}]{jozsa2003role}%
  \BibitemOpen
  \bibfield  {author} {\bibinfo {author} {\bibfnamefont {R.}~\bibnamefont
  {Jozsa}}\ and\ \bibinfo {author} {\bibfnamefont {N.}~\bibnamefont {Linden}},\
  }\href@noop {} {\bibfield  {journal} {\bibinfo  {journal} {Proceedings of the
  Royal Society of London. Series A: Mathematical, Physical and Engineering
  Sciences}\ }\textbf {\bibinfo {volume} {459}},\ \bibinfo {pages} {2011}
  (\bibinfo {year} {2003})}\BibitemShut {NoStop}%
\bibitem [{\citenamefont {Ding}\ and\ \citenamefont
  {Jin}(2007)}]{ding2007review}%
  \BibitemOpen
  \bibfield  {author} {\bibinfo {author} {\bibfnamefont {S.}~\bibnamefont
  {Ding}}\ and\ \bibinfo {author} {\bibfnamefont {Z.}~\bibnamefont {Jin}},\
  }\href@noop {} {\bibfield  {journal} {\bibinfo  {journal} {Chinese Science
  Bulletin}\ }\textbf {\bibinfo {volume} {52}},\ \bibinfo {pages} {2161}
  (\bibinfo {year} {2007})}\BibitemShut {NoStop}%
\bibitem [{\citenamefont {Pal}\ and\ \citenamefont {Huse}(2010)}]{pal2010many}%
  \BibitemOpen
  \bibfield  {author} {\bibinfo {author} {\bibfnamefont {A.}~\bibnamefont
  {Pal}}\ and\ \bibinfo {author} {\bibfnamefont {D.~A.}\ \bibnamefont {Huse}},\
  }\href@noop {} {\bibfield  {journal} {\bibinfo  {journal} {Physical review
  b}\ }\textbf {\bibinfo {volume} {82}},\ \bibinfo {pages} {174411} (\bibinfo
  {year} {2010})}\BibitemShut {NoStop}%
\bibitem [{\citenamefont {Nandkishore}\ and\ \citenamefont
  {Huse}(2015)}]{nandkishore2015many}%
  \BibitemOpen
  \bibfield  {author} {\bibinfo {author} {\bibfnamefont {R.}~\bibnamefont
  {Nandkishore}}\ and\ \bibinfo {author} {\bibfnamefont {D.~A.}\ \bibnamefont
  {Huse}},\ }\href@noop {} {\bibfield  {journal} {\bibinfo  {journal} {Annu.
  Rev. Condens. Matter Phys.}\ }\textbf {\bibinfo {volume} {6}},\ \bibinfo
  {pages} {15} (\bibinfo {year} {2015})}\BibitemShut {NoStop}%
\bibitem [{\citenamefont {Abanin}\ \emph {et~al.}(2019)\citenamefont {Abanin},
  \citenamefont {Altman}, \citenamefont {Bloch},\ and\ \citenamefont
  {Serbyn}}]{abanin2019colloquium}%
  \BibitemOpen
  \bibfield  {author} {\bibinfo {author} {\bibfnamefont {D.~A.}\ \bibnamefont
  {Abanin}}, \bibinfo {author} {\bibfnamefont {E.}~\bibnamefont {Altman}},
  \bibinfo {author} {\bibfnamefont {I.}~\bibnamefont {Bloch}}, \ and\ \bibinfo
  {author} {\bibfnamefont {M.}~\bibnamefont {Serbyn}},\ }\href@noop {}
  {\bibfield  {journal} {\bibinfo  {journal} {Reviews of Modern Physics}\
  }\textbf {\bibinfo {volume} {91}},\ \bibinfo {pages} {021001} (\bibinfo
  {year} {2019})}\BibitemShut {NoStop}%
\bibitem [{\citenamefont {Deutsch}(1991)}]{deutsch1991quantum}%
  \BibitemOpen
  \bibfield  {author} {\bibinfo {author} {\bibfnamefont {J.~M.}\ \bibnamefont
  {Deutsch}},\ }\href@noop {} {\bibfield  {journal} {\bibinfo  {journal}
  {Physical review a}\ }\textbf {\bibinfo {volume} {43}},\ \bibinfo {pages}
  {2046} (\bibinfo {year} {1991})}\BibitemShut {NoStop}%
\bibitem [{\citenamefont {Srednicki}(1994)}]{srednicki1994chaos}%
  \BibitemOpen
  \bibfield  {author} {\bibinfo {author} {\bibfnamefont {M.}~\bibnamefont
  {Srednicki}},\ }\href@noop {} {\bibfield  {journal} {\bibinfo  {journal}
  {Physical review e}\ }\textbf {\bibinfo {volume} {50}},\ \bibinfo {pages}
  {888} (\bibinfo {year} {1994})}\BibitemShut {NoStop}%
\bibitem [{\citenamefont {Kitaev}\ and\ \citenamefont
  {Preskill}(2006)}]{kitaev2006topological}%
  \BibitemOpen
  \bibfield  {author} {\bibinfo {author} {\bibfnamefont {A.}~\bibnamefont
  {Kitaev}}\ and\ \bibinfo {author} {\bibfnamefont {J.}~\bibnamefont
  {Preskill}},\ }\href@noop {} {\bibfield  {journal} {\bibinfo  {journal}
  {Physical review letters}\ }\textbf {\bibinfo {volume} {96}},\ \bibinfo
  {pages} {110404} (\bibinfo {year} {2006})}\BibitemShut {NoStop}%
\bibitem [{\citenamefont {Levin}\ and\ \citenamefont
  {Wen}(2006)}]{levin2006detecting}%
  \BibitemOpen
  \bibfield  {author} {\bibinfo {author} {\bibfnamefont {M.}~\bibnamefont
  {Levin}}\ and\ \bibinfo {author} {\bibfnamefont {X.-G.}\ \bibnamefont
  {Wen}},\ }\href@noop {} {\bibfield  {journal} {\bibinfo  {journal} {Physical
  review letters}\ }\textbf {\bibinfo {volume} {96}},\ \bibinfo {pages}
  {110405} (\bibinfo {year} {2006})}\BibitemShut {NoStop}%
\bibitem [{\citenamefont {Ryu}\ and\ \citenamefont
  {Takayanagi}(2006{\natexlab{a}})}]{Ryu:2006bv}%
  \BibitemOpen
  \bibfield  {author} {\bibinfo {author} {\bibfnamefont {S.}~\bibnamefont
  {Ryu}}\ and\ \bibinfo {author} {\bibfnamefont {T.}~\bibnamefont
  {Takayanagi}},\ }\href@noop {} {\bibfield  {journal} {\bibinfo  {journal}
  {Physical review letters}\ }\textbf {\bibinfo {volume} {96}},\ \bibinfo
  {pages} {181602} (\bibinfo {year} {2006}{\natexlab{a}})}\BibitemShut
  {NoStop}%
\bibitem [{\citenamefont {Ryu}\ and\ \citenamefont
  {Takayanagi}(2006{\natexlab{b}})}]{Ryu:2006ef}%
  \BibitemOpen
  \bibfield  {author} {\bibinfo {author} {\bibfnamefont {S.}~\bibnamefont
  {Ryu}}\ and\ \bibinfo {author} {\bibfnamefont {T.}~\bibnamefont
  {Takayanagi}},\ }\href@noop {} {\bibfield  {journal} {\bibinfo  {journal}
  {JHEP}\ }\textbf {\bibinfo {volume} {08}},\ \bibinfo {pages} {045} (\bibinfo
  {year} {2006}{\natexlab{b}})}\BibitemShut {NoStop}%
\bibitem [{\citenamefont {Lewkowycz}\ and\ \citenamefont
  {Maldacena}(2013)}]{Lewkowycz:2013nqa}%
  \BibitemOpen
  \bibfield  {author} {\bibinfo {author} {\bibfnamefont {A.}~\bibnamefont
  {Lewkowycz}}\ and\ \bibinfo {author} {\bibfnamefont {J.}~\bibnamefont
  {Maldacena}},\ }\href@noop {} {\bibfield  {journal} {\bibinfo  {journal}
  {JHEP}\ }\textbf {\bibinfo {volume} {08}},\ \bibinfo {pages} {090} (\bibinfo
  {year} {2013})}\BibitemShut {NoStop}%
\bibitem [{\citenamefont {Hubeny}\ \emph {et~al.}(2007)\citenamefont {Hubeny},
  \citenamefont {Rangamani},\ and\ \citenamefont {Takayanagi}}]{Hubeny:2007xt}%
  \BibitemOpen
  \bibfield  {author} {\bibinfo {author} {\bibfnamefont {V.~E.}\ \bibnamefont
  {Hubeny}}, \bibinfo {author} {\bibfnamefont {M.}~\bibnamefont {Rangamani}}, \
  and\ \bibinfo {author} {\bibfnamefont {T.}~\bibnamefont {Takayanagi}},\
  }\href@noop {} {\bibfield  {journal} {\bibinfo  {journal} {JHEP}\ }\textbf
  {\bibinfo {volume} {07}},\ \bibinfo {pages} {062} (\bibinfo {year}
  {2007})}\BibitemShut {NoStop}%
\bibitem [{\citenamefont {Faulkner}\ \emph {et~al.}(2013)\citenamefont
  {Faulkner}, \citenamefont {Lewkowycz},\ and\ \citenamefont
  {Maldacena}}]{Faulkner:2013ana}%
  \BibitemOpen
  \bibfield  {author} {\bibinfo {author} {\bibfnamefont {T.}~\bibnamefont
  {Faulkner}}, \bibinfo {author} {\bibfnamefont {A.}~\bibnamefont {Lewkowycz}},
  \ and\ \bibinfo {author} {\bibfnamefont {J.}~\bibnamefont {Maldacena}},\
  }\href@noop {} {\bibfield  {journal} {\bibinfo  {journal} {JHEP}\ }\textbf
  {\bibinfo {volume} {11}},\ \bibinfo {pages} {074} (\bibinfo {year}
  {2013})}\BibitemShut {NoStop}%
\bibitem [{\citenamefont {Engelhardt}\ and\ \citenamefont
  {Wall}(2015)}]{Engelhardt:2014gca}%
  \BibitemOpen
  \bibfield  {author} {\bibinfo {author} {\bibfnamefont {N.}~\bibnamefont
  {Engelhardt}}\ and\ \bibinfo {author} {\bibfnamefont {A.~C.}\ \bibnamefont
  {Wall}},\ }\href@noop {} {\bibfield  {journal} {\bibinfo  {journal} {JHEP}\
  }\textbf {\bibinfo {volume} {01}},\ \bibinfo {pages} {073} (\bibinfo {year}
  {2015})}\BibitemShut {NoStop}%
\bibitem [{\citenamefont {Penington}(2020)}]{penington2019entanglement}%
  \BibitemOpen
  \bibfield  {author} {\bibinfo {author} {\bibfnamefont {G.}~\bibnamefont
  {Penington}},\ }\href@noop {} {\bibfield  {journal} {\bibinfo  {journal}
  {Journal of High Energy Physics}\ }\textbf {\bibinfo {volume} {2020}},\
  \bibinfo {pages} {1} (\bibinfo {year} {2020})}\BibitemShut {NoStop}%
\bibitem [{\citenamefont {Almheiri}\ \emph
  {et~al.}(2019{\natexlab{a}})\citenamefont {Almheiri}, \citenamefont
  {Engelhardt}, \citenamefont {Marolf},\ and\ \citenamefont
  {Maxfield}}]{almheiri2019entropy}%
  \BibitemOpen
  \bibfield  {author} {\bibinfo {author} {\bibfnamefont {A.}~\bibnamefont
  {Almheiri}}, \bibinfo {author} {\bibfnamefont {N.}~\bibnamefont
  {Engelhardt}}, \bibinfo {author} {\bibfnamefont {D.}~\bibnamefont {Marolf}},
  \ and\ \bibinfo {author} {\bibfnamefont {H.}~\bibnamefont {Maxfield}},\
  }\href@noop {} {\bibfield  {journal} {\bibinfo  {journal} {Journal of High
  Energy Physics}\ }\textbf {\bibinfo {volume} {2019}},\ \bibinfo {pages} {1}
  (\bibinfo {year} {2019}{\natexlab{a}})}\BibitemShut {NoStop}%
\bibitem [{\citenamefont {Almheiri}\ \emph
  {et~al.}(2019{\natexlab{b}})\citenamefont {Almheiri}, \citenamefont
  {Mahajan}, \citenamefont {Maldacena},\ and\ \citenamefont
  {Zhao}}]{almheiri2019page}%
  \BibitemOpen
  \bibfield  {author} {\bibinfo {author} {\bibfnamefont {A.}~\bibnamefont
  {Almheiri}}, \bibinfo {author} {\bibfnamefont {R.}~\bibnamefont {Mahajan}},
  \bibinfo {author} {\bibfnamefont {J.}~\bibnamefont {Maldacena}}, \ and\
  \bibinfo {author} {\bibfnamefont {Y.}~\bibnamefont {Zhao}},\ }\href@noop {}
  {\bibfield  {journal} {\bibinfo  {journal} {arXiv preprint arXiv:1908.10996}\
  } (\bibinfo {year} {2019}{\natexlab{b}})}\BibitemShut {NoStop}%
\bibitem [{\citenamefont {Almheiri}\ \emph
  {et~al.}(2019{\natexlab{c}})\citenamefont {Almheiri}, \citenamefont
  {Mahajan},\ and\ \citenamefont {Maldacena}}]{almheiri2019islands}%
  \BibitemOpen
  \bibfield  {author} {\bibinfo {author} {\bibfnamefont {A.}~\bibnamefont
  {Almheiri}}, \bibinfo {author} {\bibfnamefont {R.}~\bibnamefont {Mahajan}}, \
  and\ \bibinfo {author} {\bibfnamefont {J.}~\bibnamefont {Maldacena}},\
  }\href@noop {} {\bibfield  {journal} {\bibinfo  {journal} {arXiv preprint
  arXiv:1910.11077}\ } (\bibinfo {year} {2019}{\natexlab{c}})}\BibitemShut
  {NoStop}%
\bibitem [{\citenamefont {Almheiri}\ \emph
  {et~al.}(2019{\natexlab{d}})\citenamefont {Almheiri}, \citenamefont
  {Hartman}, \citenamefont {Maldacena}, \citenamefont {Shaghoulian},\ and\
  \citenamefont {Tajdini}}]{almheiri2019replica}%
  \BibitemOpen
  \bibfield  {author} {\bibinfo {author} {\bibfnamefont {A.}~\bibnamefont
  {Almheiri}}, \bibinfo {author} {\bibfnamefont {T.}~\bibnamefont {Hartman}},
  \bibinfo {author} {\bibfnamefont {J.}~\bibnamefont {Maldacena}}, \bibinfo
  {author} {\bibfnamefont {E.}~\bibnamefont {Shaghoulian}}, \ and\ \bibinfo
  {author} {\bibfnamefont {A.}~\bibnamefont {Tajdini}},\ }\href@noop {}
  {\bibfield  {journal} {\bibinfo  {journal} {arXiv preprint arXiv:1911.12333}\
  } (\bibinfo {year} {2019}{\natexlab{d}})}\BibitemShut {NoStop}%
\bibitem [{\citenamefont {Penington}\ \emph {et~al.}(2019)\citenamefont
  {Penington}, \citenamefont {Shenker}, \citenamefont {Stanford},\ and\
  \citenamefont {Yang}}]{penington2019replica}%
  \BibitemOpen
  \bibfield  {author} {\bibinfo {author} {\bibfnamefont {G.}~\bibnamefont
  {Penington}}, \bibinfo {author} {\bibfnamefont {S.~H.}\ \bibnamefont
  {Shenker}}, \bibinfo {author} {\bibfnamefont {D.}~\bibnamefont {Stanford}}, \
  and\ \bibinfo {author} {\bibfnamefont {Z.}~\bibnamefont {Yang}},\ }\href@noop
  {} {\bibfield  {journal} {\bibinfo  {journal} {arXiv preprint
  arXiv:1911.11977}\ } (\bibinfo {year} {2019})}\BibitemShut {NoStop}%
\bibitem [{\citenamefont {Calabrese}\ and\ \citenamefont
  {Cardy}(2009)}]{calabrese2009entanglement}%
  \BibitemOpen
  \bibfield  {author} {\bibinfo {author} {\bibfnamefont {P.}~\bibnamefont
  {Calabrese}}\ and\ \bibinfo {author} {\bibfnamefont {J.}~\bibnamefont
  {Cardy}},\ }\href@noop {} {\bibfield  {journal} {\bibinfo  {journal} {Journal
  of physics a: mathematical and theoretical}\ }\textbf {\bibinfo {volume}
  {42}},\ \bibinfo {pages} {504005} (\bibinfo {year} {2009})}\BibitemShut
  {NoStop}%
\bibitem [{\citenamefont {Rangamani}\ and\ \citenamefont
  {Takayanagi}(2017)}]{rangamani2017holographic}%
  \BibitemOpen
  \bibfield  {author} {\bibinfo {author} {\bibfnamefont {M.}~\bibnamefont
  {Rangamani}}\ and\ \bibinfo {author} {\bibfnamefont {T.}~\bibnamefont
  {Takayanagi}},\ }in\ \href@noop {} {\emph {\bibinfo {booktitle} {Holographic
  Entanglement Entropy}}}\ (\bibinfo  {publisher} {Springer},\ \bibinfo {year}
  {2017})\ pp.\ \bibinfo {pages} {35--47}\BibitemShut {NoStop}%
\bibitem [{\citenamefont {Metlitski}\ \emph {et~al.}(2009)\citenamefont
  {Metlitski}, \citenamefont {Fuertes},\ and\ \citenamefont
  {Sachdev}}]{metlitski2009entanglement}%
  \BibitemOpen
  \bibfield  {author} {\bibinfo {author} {\bibfnamefont {M.~A.}\ \bibnamefont
  {Metlitski}}, \bibinfo {author} {\bibfnamefont {C.~A.}\ \bibnamefont
  {Fuertes}}, \ and\ \bibinfo {author} {\bibfnamefont {S.}~\bibnamefont
  {Sachdev}},\ }\href@noop {} {\bibfield  {journal} {\bibinfo  {journal}
  {Physical Review B}\ }\textbf {\bibinfo {volume} {80}},\ \bibinfo {pages}
  {115122} (\bibinfo {year} {2009})}\BibitemShut {NoStop}%
\bibitem [{\citenamefont {Whitsitt}\ \emph {et~al.}(2017)\citenamefont
  {Whitsitt}, \citenamefont {Witczak-Krempa},\ and\ \citenamefont
  {Sachdev}}]{whitsitt2017entanglement}%
  \BibitemOpen
  \bibfield  {author} {\bibinfo {author} {\bibfnamefont {S.}~\bibnamefont
  {Whitsitt}}, \bibinfo {author} {\bibfnamefont {W.}~\bibnamefont
  {Witczak-Krempa}}, \ and\ \bibinfo {author} {\bibfnamefont {S.}~\bibnamefont
  {Sachdev}},\ }\href@noop {} {\bibfield  {journal} {\bibinfo  {journal}
  {Physical Review B}\ }\textbf {\bibinfo {volume} {95}},\ \bibinfo {pages}
  {045148} (\bibinfo {year} {2017})}\BibitemShut {NoStop}%
\bibitem [{\citenamefont {Donnelly}\ \emph {et~al.}(2019)\citenamefont
  {Donnelly}, \citenamefont {Timmerman},\ and\ \citenamefont
  {Vald{\'e}s-Meller}}]{donnelly2019entanglement}%
  \BibitemOpen
  \bibfield  {author} {\bibinfo {author} {\bibfnamefont {W.}~\bibnamefont
  {Donnelly}}, \bibinfo {author} {\bibfnamefont {S.}~\bibnamefont {Timmerman}},
  \ and\ \bibinfo {author} {\bibfnamefont {N.}~\bibnamefont
  {Vald{\'e}s-Meller}},\ }\href@noop {} {\bibfield  {journal} {\bibinfo
  {journal} {arXiv preprint arXiv:1911.09302}\ } (\bibinfo {year}
  {2019})}\BibitemShut {NoStop}%
\bibitem [{\citenamefont {Kitaev}(2014)}]{kitaev2014hidden}%
  \BibitemOpen
  \bibfield  {author} {\bibinfo {author} {\bibfnamefont {A.}~\bibnamefont
  {Kitaev}},\ }in\ \href@noop {} {\emph {\bibinfo {booktitle} {Talk given at
  the Fundamental Physics Prize Symposium}}},\ Vol.~\bibinfo {volume} {10}\
  (\bibinfo {year} {2014})\BibitemShut {NoStop}%
\bibitem [{\citenamefont {Sachdev}\ and\ \citenamefont
  {Ye}(1993)}]{sachdev1993gapless}%
  \BibitemOpen
  \bibfield  {author} {\bibinfo {author} {\bibfnamefont {S.}~\bibnamefont
  {Sachdev}}\ and\ \bibinfo {author} {\bibfnamefont {J.}~\bibnamefont {Ye}},\
  }\href@noop {} {\bibfield  {journal} {\bibinfo  {journal} {Physical review
  letters}\ }\textbf {\bibinfo {volume} {70}},\ \bibinfo {pages} {3339}
  (\bibinfo {year} {1993})}\BibitemShut {NoStop}%
\bibitem [{\citenamefont {Maldacena}\ and\ \citenamefont
  {Stanford}(2016)}]{maldacena2016remarks}%
  \BibitemOpen
  \bibfield  {author} {\bibinfo {author} {\bibfnamefont {J.}~\bibnamefont
  {Maldacena}}\ and\ \bibinfo {author} {\bibfnamefont {D.}~\bibnamefont
  {Stanford}},\ }\href@noop {} {\bibfield  {journal} {\bibinfo  {journal}
  {Physical Review D}\ }\textbf {\bibinfo {volume} {94}},\ \bibinfo {pages}
  {106002} (\bibinfo {year} {2016})}\BibitemShut {NoStop}%
\bibitem [{\citenamefont {Maldacena}\ \emph
  {et~al.}(2016{\natexlab{a}})\citenamefont {Maldacena}, \citenamefont
  {Stanford},\ and\ \citenamefont {Yang}}]{maldacena2016conformal}%
  \BibitemOpen
  \bibfield  {author} {\bibinfo {author} {\bibfnamefont {J.}~\bibnamefont
  {Maldacena}}, \bibinfo {author} {\bibfnamefont {D.}~\bibnamefont {Stanford}},
  \ and\ \bibinfo {author} {\bibfnamefont {Z.}~\bibnamefont {Yang}},\
  }\href@noop {} {\bibfield  {journal} {\bibinfo  {journal} {Progress of
  Theoretical and Experimental Physics}\ }\textbf {\bibinfo {volume} {2016}}
  (\bibinfo {year} {2016}{\natexlab{a}})}\BibitemShut {NoStop}%
\bibitem [{\citenamefont {Kitaev}\ and\ \citenamefont
  {Suh}(2018)}]{kitaev2018soft}%
  \BibitemOpen
  \bibfield  {author} {\bibinfo {author} {\bibfnamefont {A.}~\bibnamefont
  {Kitaev}}\ and\ \bibinfo {author} {\bibfnamefont {S.~J.}\ \bibnamefont
  {Suh}},\ }\href@noop {} {\bibfield  {journal} {\bibinfo  {journal} {Journal
  of High Energy Physics}\ }\textbf {\bibinfo {volume} {2018}},\ \bibinfo
  {pages} {1} (\bibinfo {year} {2018})}\BibitemShut {NoStop}%
\bibitem [{\citenamefont {Maldacena}\ \emph
  {et~al.}(2016{\natexlab{b}})\citenamefont {Maldacena}, \citenamefont
  {Shenker},\ and\ \citenamefont {Stanford}}]{maldacena2016bound}%
  \BibitemOpen
  \bibfield  {author} {\bibinfo {author} {\bibfnamefont {J.}~\bibnamefont
  {Maldacena}}, \bibinfo {author} {\bibfnamefont {S.~H.}\ \bibnamefont
  {Shenker}}, \ and\ \bibinfo {author} {\bibfnamefont {D.}~\bibnamefont
  {Stanford}},\ }\href@noop {} {\bibfield  {journal} {\bibinfo  {journal}
  {Journal of High Energy Physics}\ }\textbf {\bibinfo {volume} {2016}},\
  \bibinfo {pages} {1} (\bibinfo {year} {2016}{\natexlab{b}})}\BibitemShut
  {NoStop}%
\bibitem [{\citenamefont {Eberlein}\ \emph {et~al.}(2017)\citenamefont
  {Eberlein}, \citenamefont {Kasper}, \citenamefont {Sachdev},\ and\
  \citenamefont {Steinberg}}]{eberlein2017quantum}%
  \BibitemOpen
  \bibfield  {author} {\bibinfo {author} {\bibfnamefont {A.}~\bibnamefont
  {Eberlein}}, \bibinfo {author} {\bibfnamefont {V.}~\bibnamefont {Kasper}},
  \bibinfo {author} {\bibfnamefont {S.}~\bibnamefont {Sachdev}}, \ and\
  \bibinfo {author} {\bibfnamefont {J.}~\bibnamefont {Steinberg}},\ }\href@noop
  {} {\bibfield  {journal} {\bibinfo  {journal} {Physical Review B}\ }\textbf
  {\bibinfo {volume} {96}},\ \bibinfo {pages} {205123} (\bibinfo {year}
  {2017})}\BibitemShut {NoStop}%
\bibitem [{\citenamefont {Chowdhury}\ \emph {et~al.}(2021)\citenamefont
  {Chowdhury}, \citenamefont {Georges}, \citenamefont {Parcollet},\ and\
  \citenamefont {Sachdev}}]{chowdhury2021sachdev}%
  \BibitemOpen
  \bibfield  {author} {\bibinfo {author} {\bibfnamefont {D.}~\bibnamefont
  {Chowdhury}}, \bibinfo {author} {\bibfnamefont {A.}~\bibnamefont {Georges}},
  \bibinfo {author} {\bibfnamefont {O.}~\bibnamefont {Parcollet}}, \ and\
  \bibinfo {author} {\bibfnamefont {S.}~\bibnamefont {Sachdev}},\ }\href@noop
  {} {\bibfield  {journal} {\bibinfo  {journal} {arXiv preprint
  arXiv:2109.05037}\ } (\bibinfo {year} {2021})}\BibitemShut {NoStop}%
\bibitem [{\citenamefont {Davison}\ \emph {et~al.}(2017)\citenamefont
  {Davison}, \citenamefont {Fu}, \citenamefont {Georges}, \citenamefont {Gu},
  \citenamefont {Jensen},\ and\ \citenamefont
  {Sachdev}}]{davison2017thermoelectric}%
  \BibitemOpen
  \bibfield  {author} {\bibinfo {author} {\bibfnamefont {R.~A.}\ \bibnamefont
  {Davison}}, \bibinfo {author} {\bibfnamefont {W.}~\bibnamefont {Fu}},
  \bibinfo {author} {\bibfnamefont {A.}~\bibnamefont {Georges}}, \bibinfo
  {author} {\bibfnamefont {Y.}~\bibnamefont {Gu}}, \bibinfo {author}
  {\bibfnamefont {K.}~\bibnamefont {Jensen}}, \ and\ \bibinfo {author}
  {\bibfnamefont {S.}~\bibnamefont {Sachdev}},\ }\href@noop {} {\bibfield
  {journal} {\bibinfo  {journal} {Physical Review B}\ }\textbf {\bibinfo
  {volume} {95}},\ \bibinfo {pages} {155131} (\bibinfo {year}
  {2017})}\BibitemShut {NoStop}%
\bibitem [{\citenamefont {Gu}\ \emph {et~al.}(2019)\citenamefont {Gu},
  \citenamefont {Kitaev}, \citenamefont {Sachdev},\ and\ \citenamefont
  {Tarnopolsky}}]{gu2019notes}%
  \BibitemOpen
  \bibfield  {author} {\bibinfo {author} {\bibfnamefont {Y.}~\bibnamefont
  {Gu}}, \bibinfo {author} {\bibfnamefont {A.}~\bibnamefont {Kitaev}}, \bibinfo
  {author} {\bibfnamefont {S.}~\bibnamefont {Sachdev}}, \ and\ \bibinfo
  {author} {\bibfnamefont {G.}~\bibnamefont {Tarnopolsky}},\ }\href@noop {}
  {\bibfield  {journal} {\bibinfo  {journal} {arXiv preprint arXiv:1910.14099}\
  } (\bibinfo {year} {2019})}\BibitemShut {NoStop}%
\bibitem [{\citenamefont {Chaturvedi}\ \emph {et~al.}(2018)\citenamefont
  {Chaturvedi}, \citenamefont {Gu}, \citenamefont {Song},\ and\ \citenamefont
  {Yu}}]{chaturvedi2018note}%
  \BibitemOpen
  \bibfield  {author} {\bibinfo {author} {\bibfnamefont {P.}~\bibnamefont
  {Chaturvedi}}, \bibinfo {author} {\bibfnamefont {Y.}~\bibnamefont {Gu}},
  \bibinfo {author} {\bibfnamefont {W.}~\bibnamefont {Song}}, \ and\ \bibinfo
  {author} {\bibfnamefont {B.}~\bibnamefont {Yu}},\ }\href@noop {} {\bibfield
  {journal} {\bibinfo  {journal} {Journal of High Energy Physics}\ }\textbf
  {\bibinfo {volume} {2018}},\ \bibinfo {pages} {1} (\bibinfo {year}
  {2018})}\BibitemShut {NoStop}%
\bibitem [{\citenamefont {Bulycheva}(2017)}]{bulycheva2017note}%
  \BibitemOpen
  \bibfield  {author} {\bibinfo {author} {\bibfnamefont {K.}~\bibnamefont
  {Bulycheva}},\ }\href@noop {} {\bibfield  {journal} {\bibinfo  {journal}
  {Journal of High Energy Physics}\ }\textbf {\bibinfo {volume} {2017}},\
  \bibinfo {pages} {1} (\bibinfo {year} {2017})}\BibitemShut {NoStop}%
\bibitem [{\citenamefont {Saad}\ \emph {et~al.}(2018)\citenamefont {Saad},
  \citenamefont {Shenker},\ and\ \citenamefont
  {Stanford}}]{saad2018semiclassical}%
  \BibitemOpen
  \bibfield  {author} {\bibinfo {author} {\bibfnamefont {P.}~\bibnamefont
  {Saad}}, \bibinfo {author} {\bibfnamefont {S.~H.}\ \bibnamefont {Shenker}}, \
  and\ \bibinfo {author} {\bibfnamefont {D.}~\bibnamefont {Stanford}},\
  }\href@noop {} {\bibfield  {journal} {\bibinfo  {journal} {arXiv preprint
  arXiv:1806.06840}\ } (\bibinfo {year} {2018})}\BibitemShut {NoStop}%
\bibitem [{\citenamefont {S{\"u}nderhauf}\ \emph {et~al.}(2019)\citenamefont
  {S{\"u}nderhauf}, \citenamefont {Piroli}, \citenamefont {Qi}, \citenamefont
  {Schuch},\ and\ \citenamefont {Cirac}}]{sunderhauf2019quantum}%
  \BibitemOpen
  \bibfield  {author} {\bibinfo {author} {\bibfnamefont {C.}~\bibnamefont
  {S{\"u}nderhauf}}, \bibinfo {author} {\bibfnamefont {L.}~\bibnamefont
  {Piroli}}, \bibinfo {author} {\bibfnamefont {X.-L.}\ \bibnamefont {Qi}},
  \bibinfo {author} {\bibfnamefont {N.}~\bibnamefont {Schuch}}, \ and\ \bibinfo
  {author} {\bibfnamefont {J.~I.}\ \bibnamefont {Cirac}},\ }\href@noop {}
  {\bibfield  {journal} {\bibinfo  {journal} {Journal of High Energy Physics}\
  }\textbf {\bibinfo {volume} {2019}},\ \bibinfo {pages} {1} (\bibinfo {year}
  {2019})}\BibitemShut {NoStop}%
\bibitem [{\citenamefont {Gu}\ \emph {et~al.}(2017{\natexlab{a}})\citenamefont
  {Gu}, \citenamefont {Qi},\ and\ \citenamefont {Stanford}}]{gu2017local}%
  \BibitemOpen
  \bibfield  {author} {\bibinfo {author} {\bibfnamefont {Y.}~\bibnamefont
  {Gu}}, \bibinfo {author} {\bibfnamefont {X.-L.}\ \bibnamefont {Qi}}, \ and\
  \bibinfo {author} {\bibfnamefont {D.}~\bibnamefont {Stanford}},\ }\href@noop
  {} {\bibfield  {journal} {\bibinfo  {journal} {Journal of High Energy
  Physics}\ }\textbf {\bibinfo {volume} {2017}},\ \bibinfo {pages} {1}
  (\bibinfo {year} {2017}{\natexlab{a}})}\BibitemShut {NoStop}%
\bibitem [{\citenamefont {Gu}\ \emph {et~al.}(2017{\natexlab{b}})\citenamefont
  {Gu}, \citenamefont {Lucas},\ and\ \citenamefont {Qi}}]{gu2017energy}%
  \BibitemOpen
  \bibfield  {author} {\bibinfo {author} {\bibfnamefont {Y.}~\bibnamefont
  {Gu}}, \bibinfo {author} {\bibfnamefont {A.}~\bibnamefont {Lucas}}, \ and\
  \bibinfo {author} {\bibfnamefont {X.-L.}\ \bibnamefont {Qi}},\ }\href@noop {}
  {\bibfield  {journal} {\bibinfo  {journal} {SciPost Phys}\ }\textbf {\bibinfo
  {volume} {2}},\ \bibinfo {pages} {018} (\bibinfo {year}
  {2017}{\natexlab{b}})}\BibitemShut {NoStop}%
\bibitem [{\citenamefont {Banerjee}\ and\ \citenamefont
  {Altman}(2017)}]{banerjee2017solvable}%
  \BibitemOpen
  \bibfield  {author} {\bibinfo {author} {\bibfnamefont {S.}~\bibnamefont
  {Banerjee}}\ and\ \bibinfo {author} {\bibfnamefont {E.}~\bibnamefont
  {Altman}},\ }\href@noop {} {\bibfield  {journal} {\bibinfo  {journal}
  {Physical Review B}\ }\textbf {\bibinfo {volume} {95}},\ \bibinfo {pages}
  {134302} (\bibinfo {year} {2017})}\BibitemShut {NoStop}%
\bibitem [{\citenamefont {Chen}\ \emph
  {et~al.}(2017{\natexlab{a}})\citenamefont {Chen}, \citenamefont {Fan},
  \citenamefont {Chen}, \citenamefont {Zhai},\ and\ \citenamefont
  {Zhang}}]{chen2017competition}%
  \BibitemOpen
  \bibfield  {author} {\bibinfo {author} {\bibfnamefont {X.}~\bibnamefont
  {Chen}}, \bibinfo {author} {\bibfnamefont {R.}~\bibnamefont {Fan}}, \bibinfo
  {author} {\bibfnamefont {Y.}~\bibnamefont {Chen}}, \bibinfo {author}
  {\bibfnamefont {H.}~\bibnamefont {Zhai}}, \ and\ \bibinfo {author}
  {\bibfnamefont {P.}~\bibnamefont {Zhang}},\ }\href@noop {} {\bibfield
  {journal} {\bibinfo  {journal} {Physical review letters}\ }\textbf {\bibinfo
  {volume} {119}},\ \bibinfo {pages} {207603} (\bibinfo {year}
  {2017}{\natexlab{a}})}\BibitemShut {NoStop}%
\bibitem [{\citenamefont {Song}\ \emph {et~al.}(2017)\citenamefont {Song},
  \citenamefont {Jian},\ and\ \citenamefont {Balents}}]{song2017strongly}%
  \BibitemOpen
  \bibfield  {author} {\bibinfo {author} {\bibfnamefont {X.-Y.}\ \bibnamefont
  {Song}}, \bibinfo {author} {\bibfnamefont {C.-M.}\ \bibnamefont {Jian}}, \
  and\ \bibinfo {author} {\bibfnamefont {L.}~\bibnamefont {Balents}},\
  }\href@noop {} {\bibfield  {journal} {\bibinfo  {journal} {Physical review
  letters}\ }\textbf {\bibinfo {volume} {119}},\ \bibinfo {pages} {216601}
  (\bibinfo {year} {2017})}\BibitemShut {NoStop}%
\bibitem [{\citenamefont {Jian}\ and\ \citenamefont
  {Yao}(2017)}]{jian2017solvable}%
  \BibitemOpen
  \bibfield  {author} {\bibinfo {author} {\bibfnamefont {S.-K.}\ \bibnamefont
  {Jian}}\ and\ \bibinfo {author} {\bibfnamefont {H.}~\bibnamefont {Yao}},\
  }\href@noop {} {\bibfield  {journal} {\bibinfo  {journal} {Physical review
  letters}\ }\textbf {\bibinfo {volume} {119}},\ \bibinfo {pages} {206602}
  (\bibinfo {year} {2017})}\BibitemShut {NoStop}%
\bibitem [{\citenamefont {Chen}\ \emph
  {et~al.}(2017{\natexlab{b}})\citenamefont {Chen}, \citenamefont {Zhai},\ and\
  \citenamefont {Zhang}}]{chen2017tunable}%
  \BibitemOpen
  \bibfield  {author} {\bibinfo {author} {\bibfnamefont {Y.}~\bibnamefont
  {Chen}}, \bibinfo {author} {\bibfnamefont {H.}~\bibnamefont {Zhai}}, \ and\
  \bibinfo {author} {\bibfnamefont {P.}~\bibnamefont {Zhang}},\ }\href@noop {}
  {\bibfield  {journal} {\bibinfo  {journal} {Journal of High Energy Physics}\
  }\textbf {\bibinfo {volume} {2017}},\ \bibinfo {pages} {1} (\bibinfo {year}
  {2017}{\natexlab{b}})}\BibitemShut {NoStop}%
\bibitem [{\citenamefont {Zhang}(2017)}]{zhang2017dispersive}%
  \BibitemOpen
  \bibfield  {author} {\bibinfo {author} {\bibfnamefont {P.}~\bibnamefont
  {Zhang}},\ }\href@noop {} {\bibfield  {journal} {\bibinfo  {journal}
  {Physical Review B}\ }\textbf {\bibinfo {volume} {96}},\ \bibinfo {pages}
  {205138} (\bibinfo {year} {2017})}\BibitemShut {NoStop}%
\bibitem [{\citenamefont {Bi}\ \emph {et~al.}(2017)\citenamefont {Bi},
  \citenamefont {Jian}, \citenamefont {You}, \citenamefont {Pawlak},\ and\
  \citenamefont {Xu}}]{bi2017instability}%
  \BibitemOpen
  \bibfield  {author} {\bibinfo {author} {\bibfnamefont {Z.}~\bibnamefont
  {Bi}}, \bibinfo {author} {\bibfnamefont {C.-M.}\ \bibnamefont {Jian}},
  \bibinfo {author} {\bibfnamefont {Y.-Z.}\ \bibnamefont {You}}, \bibinfo
  {author} {\bibfnamefont {K.~A.}\ \bibnamefont {Pawlak}}, \ and\ \bibinfo
  {author} {\bibfnamefont {C.}~\bibnamefont {Xu}},\ }\href@noop {} {\bibfield
  {journal} {\bibinfo  {journal} {Physical Review B}\ }\textbf {\bibinfo
  {volume} {95}},\ \bibinfo {pages} {205105} (\bibinfo {year}
  {2017})}\BibitemShut {NoStop}%
\bibitem [{\citenamefont {Narayan}\ and\ \citenamefont
  {Yoon}(2017)}]{narayan2017syk}%
  \BibitemOpen
  \bibfield  {author} {\bibinfo {author} {\bibfnamefont {P.}~\bibnamefont
  {Narayan}}\ and\ \bibinfo {author} {\bibfnamefont {J.}~\bibnamefont {Yoon}},\
  }\href@noop {} {\bibfield  {journal} {\bibinfo  {journal} {Journal of High
  Energy Physics}\ }\textbf {\bibinfo {volume} {2017}},\ \bibinfo {pages} {1}
  (\bibinfo {year} {2017})}\BibitemShut {NoStop}%
\bibitem [{\citenamefont {Liu}\ \emph {et~al.}(2018)\citenamefont {Liu},
  \citenamefont {Chen},\ and\ \citenamefont {Balents}}]{liu2018quantum}%
  \BibitemOpen
  \bibfield  {author} {\bibinfo {author} {\bibfnamefont {C.}~\bibnamefont
  {Liu}}, \bibinfo {author} {\bibfnamefont {X.}~\bibnamefont {Chen}}, \ and\
  \bibinfo {author} {\bibfnamefont {L.}~\bibnamefont {Balents}},\ }\href@noop
  {} {\bibfield  {journal} {\bibinfo  {journal} {Physical Review B}\ }\textbf
  {\bibinfo {volume} {97}},\ \bibinfo {pages} {245126} (\bibinfo {year}
  {2018})}\BibitemShut {NoStop}%
\bibitem [{\citenamefont {Fu}\ and\ \citenamefont
  {Sachdev}(2016)}]{fu2016numerical}%
  \BibitemOpen
  \bibfield  {author} {\bibinfo {author} {\bibfnamefont {W.}~\bibnamefont
  {Fu}}\ and\ \bibinfo {author} {\bibfnamefont {S.}~\bibnamefont {Sachdev}},\
  }\href@noop {} {\bibfield  {journal} {\bibinfo  {journal} {Physical Review
  B}\ }\textbf {\bibinfo {volume} {94}},\ \bibinfo {pages} {035135} (\bibinfo
  {year} {2016})}\BibitemShut {NoStop}%
\bibitem [{\citenamefont {Huang}\ \emph {et~al.}(2019)\citenamefont {Huang},
  \citenamefont {Gu} \emph {et~al.}}]{huang2019eigenstate}%
  \BibitemOpen
  \bibfield  {author} {\bibinfo {author} {\bibfnamefont {Y.}~\bibnamefont
  {Huang}}, \bibinfo {author} {\bibfnamefont {Y.}~\bibnamefont {Gu}},  \emph
  {et~al.},\ }\href@noop {} {\bibfield  {journal} {\bibinfo  {journal}
  {Physical Review D}\ }\textbf {\bibinfo {volume} {100}},\ \bibinfo {pages}
  {041901} (\bibinfo {year} {2019})}\BibitemShut {NoStop}%
\bibitem [{\citenamefont {Zhang}\ \emph
  {et~al.}(2020{\natexlab{a}})\citenamefont {Zhang}, \citenamefont {Liu},\ and\
  \citenamefont {Chen}}]{zhang2020subsystem}%
  \BibitemOpen
  \bibfield  {author} {\bibinfo {author} {\bibfnamefont {P.}~\bibnamefont
  {Zhang}}, \bibinfo {author} {\bibfnamefont {C.}~\bibnamefont {Liu}}, \ and\
  \bibinfo {author} {\bibfnamefont {X.}~\bibnamefont {Chen}},\ }\href@noop {}
  {\bibfield  {journal} {\bibinfo  {journal} {SciPost Physics}\ }\textbf
  {\bibinfo {volume} {8}},\ \bibinfo {pages} {Art} (\bibinfo {year}
  {2020}{\natexlab{a}})}\BibitemShut {NoStop}%
\bibitem [{\citenamefont {Zhang}(2020)}]{zhang2020entanglement}%
  \BibitemOpen
  \bibfield  {author} {\bibinfo {author} {\bibfnamefont {P.}~\bibnamefont
  {Zhang}},\ }\href@noop {} {\bibfield  {journal} {\bibinfo  {journal} {Journal
  of High Energy Physics}\ }\textbf {\bibinfo {volume} {2020}},\ \bibinfo
  {pages} {1} (\bibinfo {year} {2020})}\BibitemShut {NoStop}%
\bibitem [{\citenamefont {Haldar}\ \emph {et~al.}(2020)\citenamefont {Haldar},
  \citenamefont {Bera},\ and\ \citenamefont {Banerjee}}]{haldar2020renyi}%
  \BibitemOpen
  \bibfield  {author} {\bibinfo {author} {\bibfnamefont {A.}~\bibnamefont
  {Haldar}}, \bibinfo {author} {\bibfnamefont {S.}~\bibnamefont {Bera}}, \ and\
  \bibinfo {author} {\bibfnamefont {S.}~\bibnamefont {Banerjee}},\ }\href@noop
  {} {\bibfield  {journal} {\bibinfo  {journal} {Physical Review Research}\
  }\textbf {\bibinfo {volume} {2}},\ \bibinfo {pages} {033505} (\bibinfo {year}
  {2020})}\BibitemShut {NoStop}%
\bibitem [{\citenamefont {Kudler-Flam}\ \emph {et~al.}(2021)\citenamefont
  {Kudler-Flam}, \citenamefont {Sohal},\ and\ \citenamefont
  {Nie}}]{kudler2021information}%
  \BibitemOpen
  \bibfield  {author} {\bibinfo {author} {\bibfnamefont {J.}~\bibnamefont
  {Kudler-Flam}}, \bibinfo {author} {\bibfnamefont {R.}~\bibnamefont {Sohal}},
  \ and\ \bibinfo {author} {\bibfnamefont {L.}~\bibnamefont {Nie}},\
  }\href@noop {} {\bibfield  {journal} {\bibinfo  {journal} {arXiv preprint
  arXiv:2107.04043}\ } (\bibinfo {year} {2021})}\BibitemShut {NoStop}%
\bibitem [{\citenamefont {Gu}\ \emph {et~al.}(2017{\natexlab{c}})\citenamefont
  {Gu}, \citenamefont {Lucas},\ and\ \citenamefont {Qi}}]{gu2017spread}%
  \BibitemOpen
  \bibfield  {author} {\bibinfo {author} {\bibfnamefont {Y.}~\bibnamefont
  {Gu}}, \bibinfo {author} {\bibfnamefont {A.}~\bibnamefont {Lucas}}, \ and\
  \bibinfo {author} {\bibfnamefont {X.-L.}\ \bibnamefont {Qi}},\ }\href@noop {}
  {\bibfield  {journal} {\bibinfo  {journal} {Journal of High Energy Physics}\
  }\textbf {\bibinfo {volume} {2017}},\ \bibinfo {pages} {1} (\bibinfo {year}
  {2017}{\natexlab{c}})}\BibitemShut {NoStop}%
\bibitem [{\citenamefont {Sohal}\ \emph {et~al.}(2022)\citenamefont {Sohal},
  \citenamefont {Nie}, \citenamefont {Sun},\ and\ \citenamefont
  {Fradkin}}]{sohal2022thermalization}%
  \BibitemOpen
  \bibfield  {author} {\bibinfo {author} {\bibfnamefont {R.}~\bibnamefont
  {Sohal}}, \bibinfo {author} {\bibfnamefont {L.}~\bibnamefont {Nie}}, \bibinfo
  {author} {\bibfnamefont {X.-Q.}\ \bibnamefont {Sun}}, \ and\ \bibinfo
  {author} {\bibfnamefont {E.}~\bibnamefont {Fradkin}},\ }\href@noop {}
  {\bibfield  {journal} {\bibinfo  {journal} {Journal of Statistical Mechanics:
  Theory and Experiment}\ }\textbf {\bibinfo {volume} {2022}},\ \bibinfo
  {pages} {013103} (\bibinfo {year} {2022})}\BibitemShut {NoStop}%
\bibitem [{\citenamefont {Chen}\ \emph
  {et~al.}(2020{\natexlab{a}})\citenamefont {Chen}, \citenamefont {Qi},\ and\
  \citenamefont {Zhang}}]{chen2020replica}%
  \BibitemOpen
  \bibfield  {author} {\bibinfo {author} {\bibfnamefont {Y.}~\bibnamefont
  {Chen}}, \bibinfo {author} {\bibfnamefont {X.-L.}\ \bibnamefont {Qi}}, \ and\
  \bibinfo {author} {\bibfnamefont {P.}~\bibnamefont {Zhang}},\ }\href@noop {}
  {\bibfield  {journal} {\bibinfo  {journal} {Journal of High Energy Physics}\
  }\textbf {\bibinfo {volume} {2020}},\ \bibinfo {pages} {1} (\bibinfo {year}
  {2020}{\natexlab{a}})}\BibitemShut {NoStop}%
\bibitem [{\citenamefont {Chen}(2021)}]{chen2021entropy}%
  \BibitemOpen
  \bibfield  {author} {\bibinfo {author} {\bibfnamefont {Y.}~\bibnamefont
  {Chen}},\ }\href@noop {} {\bibfield  {journal} {\bibinfo  {journal} {Journal
  of High Energy Physics}\ }\textbf {\bibinfo {volume} {2021}},\ \bibinfo
  {pages} {1} (\bibinfo {year} {2021})}\BibitemShut {NoStop}%
\bibitem [{\citenamefont {Dadras}\ and\ \citenamefont
  {Kitaev}(2021)}]{dadras2021perturbative}%
  \BibitemOpen
  \bibfield  {author} {\bibinfo {author} {\bibfnamefont {P.}~\bibnamefont
  {Dadras}}\ and\ \bibinfo {author} {\bibfnamefont {A.}~\bibnamefont
  {Kitaev}},\ }\href@noop {} {\bibfield  {journal} {\bibinfo  {journal}
  {Journal of High Energy Physics}\ }\textbf {\bibinfo {volume} {2021}},\
  \bibinfo {pages} {1} (\bibinfo {year} {2021})}\BibitemShut {NoStop}%
\bibitem [{\citenamefont {Jian}\ and\ \citenamefont
  {Swingle}(2021{\natexlab{a}})}]{jian2021note}%
  \BibitemOpen
  \bibfield  {author} {\bibinfo {author} {\bibfnamefont {S.-K.}\ \bibnamefont
  {Jian}}\ and\ \bibinfo {author} {\bibfnamefont {B.}~\bibnamefont {Swingle}},\
  }\href@noop {} {\bibfield  {journal} {\bibinfo  {journal} {Journal of High
  Energy Physics}\ }\textbf {\bibinfo {volume} {2021}},\ \bibinfo {pages} {1}
  (\bibinfo {year} {2021}{\natexlab{a}})}\BibitemShut {NoStop}%
\bibitem [{\citenamefont {Jian}\ and\ \citenamefont
  {Swingle}(2021{\natexlab{b}})}]{jian2021chaos}%
  \BibitemOpen
  \bibfield  {author} {\bibinfo {author} {\bibfnamefont {S.-K.}\ \bibnamefont
  {Jian}}\ and\ \bibinfo {author} {\bibfnamefont {B.}~\bibnamefont {Swingle}},\
  }\href@noop {} {\bibfield  {journal} {\bibinfo  {journal} {arXiv preprint
  arXiv:2109.03825}\ } (\bibinfo {year} {2021}{\natexlab{b}})}\BibitemShut
  {NoStop}%
\bibitem [{\citenamefont {Kaixiang}\ \emph {et~al.}(2021)\citenamefont
  {Kaixiang}, \citenamefont {Zhang},\ and\ \citenamefont
  {Zhai}}]{kaixiang2021page}%
  \BibitemOpen
  \bibfield  {author} {\bibinfo {author} {\bibfnamefont {S.}~\bibnamefont
  {Kaixiang}}, \bibinfo {author} {\bibfnamefont {P.}~\bibnamefont {Zhang}}, \
  and\ \bibinfo {author} {\bibfnamefont {H.}~\bibnamefont {Zhai}},\ }\href@noop
  {} {\bibfield  {journal} {\bibinfo  {journal} {Journal of High Energy
  Physics}\ }\textbf {\bibinfo {volume} {2021}} (\bibinfo {year}
  {2021})}\BibitemShut {NoStop}%
\bibitem [{\citenamefont {Nedel}(2021)}]{nedel2021time}%
  \BibitemOpen
  \bibfield  {author} {\bibinfo {author} {\bibfnamefont {D.~L.}\ \bibnamefont
  {Nedel}},\ }\href@noop {} {\bibfield  {journal} {\bibinfo  {journal} {Physics
  Letters B}\ }\textbf {\bibinfo {volume} {817}},\ \bibinfo {pages} {136340}
  (\bibinfo {year} {2021})}\BibitemShut {NoStop}%
\bibitem [{\citenamefont {Chen}\ and\ \citenamefont
  {Zhang}(2019)}]{chen2019entanglement}%
  \BibitemOpen
  \bibfield  {author} {\bibinfo {author} {\bibfnamefont {Y.}~\bibnamefont
  {Chen}}\ and\ \bibinfo {author} {\bibfnamefont {P.}~\bibnamefont {Zhang}},\
  }\href@noop {} {\bibfield  {journal} {\bibinfo  {journal} {Journal of High
  Energy Physics}\ }\textbf {\bibinfo {volume} {2019}},\ \bibinfo {pages} {1}
  (\bibinfo {year} {2019})}\BibitemShut {NoStop}%
\bibitem [{\citenamefont {Pouria}(2022)}]{pouria2022disentangling}%
  \BibitemOpen
  \bibfield  {author} {\bibinfo {author} {\bibfnamefont {D.}~\bibnamefont
  {Pouria}},\ }\href@noop {} {\bibfield  {journal} {\bibinfo  {journal}
  {Journal of High Energy Physics}\ }\textbf {\bibinfo {volume} {2022}}
  (\bibinfo {year} {2022})}\BibitemShut {NoStop}%
\bibitem [{\citenamefont {Hawking}(1976)}]{hawking1976breakdown}%
  \BibitemOpen
  \bibfield  {author} {\bibinfo {author} {\bibfnamefont {S.~W.}\ \bibnamefont
  {Hawking}},\ }\href@noop {} {\bibfield  {journal} {\bibinfo  {journal}
  {Physical Review D}\ }\textbf {\bibinfo {volume} {14}},\ \bibinfo {pages}
  {2460} (\bibinfo {year} {1976})}\BibitemShut {NoStop}%
\bibitem [{\citenamefont {Nielsen}\ and\ \citenamefont
  {Chuang}(2002)}]{nielsen2002quantum}%
  \BibitemOpen
  \bibfield  {author} {\bibinfo {author} {\bibfnamefont {M.~A.}\ \bibnamefont
  {Nielsen}}\ and\ \bibinfo {author} {\bibfnamefont {I.}~\bibnamefont
  {Chuang}},\ }\href@noop {} {\enquote {\bibinfo {title} {Quantum computation
  and quantum information},}\ } (\bibinfo {year} {2002})\BibitemShut {NoStop}%
\bibitem [{\citenamefont {Li}\ \emph {et~al.}(2018)\citenamefont {Li},
  \citenamefont {Chen},\ and\ \citenamefont {Fisher}}]{li2018quantum}%
  \BibitemOpen
  \bibfield  {author} {\bibinfo {author} {\bibfnamefont {Y.}~\bibnamefont
  {Li}}, \bibinfo {author} {\bibfnamefont {X.}~\bibnamefont {Chen}}, \ and\
  \bibinfo {author} {\bibfnamefont {M.~P.}\ \bibnamefont {Fisher}},\
  }\href@noop {} {\bibfield  {journal} {\bibinfo  {journal} {Physical Review
  B}\ }\textbf {\bibinfo {volume} {98}},\ \bibinfo {pages} {205136} (\bibinfo
  {year} {2018})}\BibitemShut {NoStop}%
\bibitem [{\citenamefont {Cao}\ \emph {et~al.}(2019)\citenamefont {Cao},
  \citenamefont {Tilloy},\ and\ \citenamefont
  {Luca}}]{10.21468/SciPostPhys.7.2.024}%
  \BibitemOpen
  \bibfield  {author} {\bibinfo {author} {\bibfnamefont {X.}~\bibnamefont
  {Cao}}, \bibinfo {author} {\bibfnamefont {A.}~\bibnamefont {Tilloy}}, \ and\
  \bibinfo {author} {\bibfnamefont {A.~D.}\ \bibnamefont {Luca}},\ }\href@noop
  {} {\bibfield  {journal} {\bibinfo  {journal} {SciPost Phys.}\ }\textbf
  {\bibinfo {volume} {7}},\ \bibinfo {pages} {24} (\bibinfo {year}
  {2019})}\BibitemShut {NoStop}%
\bibitem [{\citenamefont {Li}\ \emph {et~al.}(2019)\citenamefont {Li},
  \citenamefont {Chen},\ and\ \citenamefont {Fisher}}]{li2019measurement}%
  \BibitemOpen
  \bibfield  {author} {\bibinfo {author} {\bibfnamefont {Y.}~\bibnamefont
  {Li}}, \bibinfo {author} {\bibfnamefont {X.}~\bibnamefont {Chen}}, \ and\
  \bibinfo {author} {\bibfnamefont {M.~P.}\ \bibnamefont {Fisher}},\
  }\href@noop {} {\bibfield  {journal} {\bibinfo  {journal} {Physical Review
  B}\ }\textbf {\bibinfo {volume} {100}},\ \bibinfo {pages} {134306} (\bibinfo
  {year} {2019})}\BibitemShut {NoStop}%
\bibitem [{\citenamefont {Skinner}\ \emph {et~al.}(2019)\citenamefont
  {Skinner}, \citenamefont {Ruhman},\ and\ \citenamefont
  {Nahum}}]{skinner2019measurement}%
  \BibitemOpen
  \bibfield  {author} {\bibinfo {author} {\bibfnamefont {B.}~\bibnamefont
  {Skinner}}, \bibinfo {author} {\bibfnamefont {J.}~\bibnamefont {Ruhman}}, \
  and\ \bibinfo {author} {\bibfnamefont {A.}~\bibnamefont {Nahum}},\
  }\href@noop {} {\bibfield  {journal} {\bibinfo  {journal} {Physical Review
  X}\ }\textbf {\bibinfo {volume} {9}},\ \bibinfo {pages} {031009} (\bibinfo
  {year} {2019})}\BibitemShut {NoStop}%
\bibitem [{\citenamefont {Chan}\ \emph {et~al.}(2019)\citenamefont {Chan},
  \citenamefont {Nandkishore}, \citenamefont {Pretko},\ and\ \citenamefont
  {Smith}}]{chan2019unitary}%
  \BibitemOpen
  \bibfield  {author} {\bibinfo {author} {\bibfnamefont {A.}~\bibnamefont
  {Chan}}, \bibinfo {author} {\bibfnamefont {R.~M.}\ \bibnamefont
  {Nandkishore}}, \bibinfo {author} {\bibfnamefont {M.}~\bibnamefont {Pretko}},
  \ and\ \bibinfo {author} {\bibfnamefont {G.}~\bibnamefont {Smith}},\
  }\href@noop {} {\bibfield  {journal} {\bibinfo  {journal} {Physical Review
  B}\ }\textbf {\bibinfo {volume} {99}},\ \bibinfo {pages} {224307} (\bibinfo
  {year} {2019})}\BibitemShut {NoStop}%
\bibitem [{\citenamefont {Bao}\ \emph {et~al.}(2020)\citenamefont {Bao},
  \citenamefont {Choi},\ and\ \citenamefont {Altman}}]{bao2020theory}%
  \BibitemOpen
  \bibfield  {author} {\bibinfo {author} {\bibfnamefont {Y.}~\bibnamefont
  {Bao}}, \bibinfo {author} {\bibfnamefont {S.}~\bibnamefont {Choi}}, \ and\
  \bibinfo {author} {\bibfnamefont {E.}~\bibnamefont {Altman}},\ }\href@noop {}
  {\bibfield  {journal} {\bibinfo  {journal} {Physical Review B}\ }\textbf
  {\bibinfo {volume} {101}},\ \bibinfo {pages} {104301} (\bibinfo {year}
  {2020})}\BibitemShut {NoStop}%
\bibitem [{\citenamefont {Choi}\ \emph {et~al.}(2020)\citenamefont {Choi},
  \citenamefont {Bao}, \citenamefont {Qi},\ and\ \citenamefont
  {Altman}}]{choi2020quantum}%
  \BibitemOpen
  \bibfield  {author} {\bibinfo {author} {\bibfnamefont {S.}~\bibnamefont
  {Choi}}, \bibinfo {author} {\bibfnamefont {Y.}~\bibnamefont {Bao}}, \bibinfo
  {author} {\bibfnamefont {X.-L.}\ \bibnamefont {Qi}}, \ and\ \bibinfo {author}
  {\bibfnamefont {E.}~\bibnamefont {Altman}},\ }\href@noop {} {\bibfield
  {journal} {\bibinfo  {journal} {Physical Review Letters}\ }\textbf {\bibinfo
  {volume} {125}},\ \bibinfo {pages} {030505} (\bibinfo {year}
  {2020})}\BibitemShut {NoStop}%
\bibitem [{\citenamefont {Gullans}\ and\ \citenamefont
  {Huse}(2020{\natexlab{a}})}]{gullans2020dynamical}%
  \BibitemOpen
  \bibfield  {author} {\bibinfo {author} {\bibfnamefont {M.~J.}\ \bibnamefont
  {Gullans}}\ and\ \bibinfo {author} {\bibfnamefont {D.~A.}\ \bibnamefont
  {Huse}},\ }\href@noop {} {\bibfield  {journal} {\bibinfo  {journal} {Physical
  Review X}\ }\textbf {\bibinfo {volume} {10}},\ \bibinfo {pages} {041020}
  (\bibinfo {year} {2020}{\natexlab{a}})}\BibitemShut {NoStop}%
\bibitem [{\citenamefont {Gullans}\ and\ \citenamefont
  {Huse}(2020{\natexlab{b}})}]{gullans2020scalable}%
  \BibitemOpen
  \bibfield  {author} {\bibinfo {author} {\bibfnamefont {M.~J.}\ \bibnamefont
  {Gullans}}\ and\ \bibinfo {author} {\bibfnamefont {D.~A.}\ \bibnamefont
  {Huse}},\ }\href@noop {} {\bibfield  {journal} {\bibinfo  {journal} {Physical
  review letters}\ }\textbf {\bibinfo {volume} {125}},\ \bibinfo {pages}
  {070606} (\bibinfo {year} {2020}{\natexlab{b}})}\BibitemShut {NoStop}%
\bibitem [{\citenamefont {Jian}\ \emph
  {et~al.}(2020{\natexlab{a}})\citenamefont {Jian}, \citenamefont {You},
  \citenamefont {Vasseur},\ and\ \citenamefont {Ludwig}}]{jian2020measurement}%
  \BibitemOpen
  \bibfield  {author} {\bibinfo {author} {\bibfnamefont {C.-M.}\ \bibnamefont
  {Jian}}, \bibinfo {author} {\bibfnamefont {Y.-Z.}\ \bibnamefont {You}},
  \bibinfo {author} {\bibfnamefont {R.}~\bibnamefont {Vasseur}}, \ and\
  \bibinfo {author} {\bibfnamefont {A.~W.}\ \bibnamefont {Ludwig}},\
  }\href@noop {} {\bibfield  {journal} {\bibinfo  {journal} {Physical Review
  B}\ }\textbf {\bibinfo {volume} {101}},\ \bibinfo {pages} {104302} (\bibinfo
  {year} {2020}{\natexlab{a}})}\BibitemShut {NoStop}%
\bibitem [{\citenamefont {Szyniszewski}\ \emph {et~al.}(2019)\citenamefont
  {Szyniszewski}, \citenamefont {Romito},\ and\ \citenamefont
  {Schomerus}}]{szyniszewski2019entanglement}%
  \BibitemOpen
  \bibfield  {author} {\bibinfo {author} {\bibfnamefont {M.}~\bibnamefont
  {Szyniszewski}}, \bibinfo {author} {\bibfnamefont {A.}~\bibnamefont
  {Romito}}, \ and\ \bibinfo {author} {\bibfnamefont {H.}~\bibnamefont
  {Schomerus}},\ }\href@noop {} {\bibfield  {journal} {\bibinfo  {journal}
  {Physical Review B}\ }\textbf {\bibinfo {volume} {100}},\ \bibinfo {pages}
  {064204} (\bibinfo {year} {2019})}\BibitemShut {NoStop}%
\bibitem [{\citenamefont {Zabalo}\ \emph {et~al.}(2020)\citenamefont {Zabalo},
  \citenamefont {Gullans}, \citenamefont {Wilson}, \citenamefont
  {Gopalakrishnan}, \citenamefont {Huse},\ and\ \citenamefont
  {Pixley}}]{zabalo2020critical}%
  \BibitemOpen
  \bibfield  {author} {\bibinfo {author} {\bibfnamefont {A.}~\bibnamefont
  {Zabalo}}, \bibinfo {author} {\bibfnamefont {M.~J.}\ \bibnamefont {Gullans}},
  \bibinfo {author} {\bibfnamefont {J.~H.}\ \bibnamefont {Wilson}}, \bibinfo
  {author} {\bibfnamefont {S.}~\bibnamefont {Gopalakrishnan}}, \bibinfo
  {author} {\bibfnamefont {D.~A.}\ \bibnamefont {Huse}}, \ and\ \bibinfo
  {author} {\bibfnamefont {J.}~\bibnamefont {Pixley}},\ }\href@noop {}
  {\bibfield  {journal} {\bibinfo  {journal} {Physical Review B}\ }\textbf
  {\bibinfo {volume} {101}},\ \bibinfo {pages} {060301} (\bibinfo {year}
  {2020})}\BibitemShut {NoStop}%
\bibitem [{\citenamefont {Tang}\ and\ \citenamefont
  {Zhu}(2020)}]{tang2020measurement}%
  \BibitemOpen
  \bibfield  {author} {\bibinfo {author} {\bibfnamefont {Q.}~\bibnamefont
  {Tang}}\ and\ \bibinfo {author} {\bibfnamefont {W.}~\bibnamefont {Zhu}},\
  }\href@noop {} {\bibfield  {journal} {\bibinfo  {journal} {Physical Review
  Research}\ }\textbf {\bibinfo {volume} {2}},\ \bibinfo {pages} {013022}
  (\bibinfo {year} {2020})}\BibitemShut {NoStop}%
\bibitem [{\citenamefont {Zhang}\ \emph
  {et~al.}(2020{\natexlab{b}})\citenamefont {Zhang}, \citenamefont {Reyes},
  \citenamefont {Kourtis}, \citenamefont {Chamon}, \citenamefont {Mucciolo},\
  and\ \citenamefont {Ruckenstein}}]{zhang2020nonuniversal}%
  \BibitemOpen
  \bibfield  {author} {\bibinfo {author} {\bibfnamefont {L.}~\bibnamefont
  {Zhang}}, \bibinfo {author} {\bibfnamefont {J.~A.}\ \bibnamefont {Reyes}},
  \bibinfo {author} {\bibfnamefont {S.}~\bibnamefont {Kourtis}}, \bibinfo
  {author} {\bibfnamefont {C.}~\bibnamefont {Chamon}}, \bibinfo {author}
  {\bibfnamefont {E.~R.}\ \bibnamefont {Mucciolo}}, \ and\ \bibinfo {author}
  {\bibfnamefont {A.~E.}\ \bibnamefont {Ruckenstein}},\ }\href@noop {}
  {\bibfield  {journal} {\bibinfo  {journal} {Physical Review B}\ }\textbf
  {\bibinfo {volume} {101}},\ \bibinfo {pages} {235104} (\bibinfo {year}
  {2020}{\natexlab{b}})}\BibitemShut {NoStop}%
\bibitem [{\citenamefont {Goto}\ and\ \citenamefont
  {Danshita}(2020)}]{goto2020measurement}%
  \BibitemOpen
  \bibfield  {author} {\bibinfo {author} {\bibfnamefont {S.}~\bibnamefont
  {Goto}}\ and\ \bibinfo {author} {\bibfnamefont {I.}~\bibnamefont
  {Danshita}},\ }\href@noop {} {\bibfield  {journal} {\bibinfo  {journal}
  {Physical Review A}\ }\textbf {\bibinfo {volume} {102}},\ \bibinfo {pages}
  {033316} (\bibinfo {year} {2020})}\BibitemShut {NoStop}%
\bibitem [{\citenamefont {Jian}\ \emph
  {et~al.}(2020{\natexlab{b}})\citenamefont {Jian}, \citenamefont {Bauer},
  \citenamefont {Keselman},\ and\ \citenamefont
  {Ludwig}}]{jian2020criticality}%
  \BibitemOpen
  \bibfield  {author} {\bibinfo {author} {\bibfnamefont {C.-M.}\ \bibnamefont
  {Jian}}, \bibinfo {author} {\bibfnamefont {B.}~\bibnamefont {Bauer}},
  \bibinfo {author} {\bibfnamefont {A.}~\bibnamefont {Keselman}}, \ and\
  \bibinfo {author} {\bibfnamefont {A.~W.}\ \bibnamefont {Ludwig}},\
  }\href@noop {} {\bibfield  {journal} {\bibinfo  {journal} {arXiv preprint
  arXiv:2012.04666}\ } (\bibinfo {year} {2020}{\natexlab{b}})}\BibitemShut
  {NoStop}%
\bibitem [{\citenamefont {Bao}\ \emph {et~al.}(2021)\citenamefont {Bao},
  \citenamefont {Choi},\ and\ \citenamefont {Altman}}]{bao2021symmetry}%
  \BibitemOpen
  \bibfield  {author} {\bibinfo {author} {\bibfnamefont {Y.}~\bibnamefont
  {Bao}}, \bibinfo {author} {\bibfnamefont {S.}~\bibnamefont {Choi}}, \ and\
  \bibinfo {author} {\bibfnamefont {E.}~\bibnamefont {Altman}},\ }\href@noop {}
  {\bibfield  {journal} {\bibinfo  {journal} {arXiv preprint arXiv:2102.09164}\
  } (\bibinfo {year} {2021})}\BibitemShut {NoStop}%
\bibitem [{\citenamefont {Alberton}\ \emph {et~al.}(2020)\citenamefont
  {Alberton}, \citenamefont {Buchhold},\ and\ \citenamefont
  {Diehl}}]{alberton2020trajectory}%
  \BibitemOpen
  \bibfield  {author} {\bibinfo {author} {\bibfnamefont {O.}~\bibnamefont
  {Alberton}}, \bibinfo {author} {\bibfnamefont {M.}~\bibnamefont {Buchhold}},
  \ and\ \bibinfo {author} {\bibfnamefont {S.}~\bibnamefont {Diehl}},\
  }\href@noop {} {\bibfield  {journal} {\bibinfo  {journal} {arXiv preprint
  arXiv:2005.09722}\ } (\bibinfo {year} {2020})}\BibitemShut {NoStop}%
\bibitem [{\citenamefont {Chen}\ \emph
  {et~al.}(2020{\natexlab{b}})\citenamefont {Chen}, \citenamefont {Li},
  \citenamefont {Fisher},\ and\ \citenamefont {Lucas}}]{Chen_2020}%
  \BibitemOpen
  \bibfield  {author} {\bibinfo {author} {\bibfnamefont {X.}~\bibnamefont
  {Chen}}, \bibinfo {author} {\bibfnamefont {Y.}~\bibnamefont {Li}}, \bibinfo
  {author} {\bibfnamefont {M.~P.~A.}\ \bibnamefont {Fisher}}, \ and\ \bibinfo
  {author} {\bibfnamefont {A.}~\bibnamefont {Lucas}},\ }\href@noop {}
  {\bibfield  {journal} {\bibinfo  {journal} {Physical Review Research}\
  }\textbf {\bibinfo {volume} {2}} (\bibinfo {year}
  {2020}{\natexlab{b}})}\BibitemShut {NoStop}%
\bibitem [{\citenamefont {Nahum}\ and\ \citenamefont
  {Skinner}(2020)}]{Nahum_2020}%
  \BibitemOpen
  \bibfield  {author} {\bibinfo {author} {\bibfnamefont {A.}~\bibnamefont
  {Nahum}}\ and\ \bibinfo {author} {\bibfnamefont {B.}~\bibnamefont
  {Skinner}},\ }\href@noop {} {\bibfield  {journal} {\bibinfo  {journal}
  {Physical Review Research}\ }\textbf {\bibinfo {volume} {2}} (\bibinfo {year}
  {2020})}\BibitemShut {NoStop}%
\bibitem [{\citenamefont {Liu}\ \emph {et~al.}(2021)\citenamefont {Liu},
  \citenamefont {Zhang},\ and\ \citenamefont {Chen}}]{liu2021non}%
  \BibitemOpen
  \bibfield  {author} {\bibinfo {author} {\bibfnamefont {C.}~\bibnamefont
  {Liu}}, \bibinfo {author} {\bibfnamefont {P.}~\bibnamefont {Zhang}}, \ and\
  \bibinfo {author} {\bibfnamefont {X.}~\bibnamefont {Chen}},\ }\href@noop {}
  {\bibfield  {journal} {\bibinfo  {journal} {SciPost Physics}\ }\textbf
  {\bibinfo {volume} {10}},\ \bibinfo {pages} {Art} (\bibinfo {year}
  {2021})}\BibitemShut {NoStop}%
\bibitem [{\citenamefont {Zhang}\ \emph
  {et~al.}(2021{\natexlab{a}})\citenamefont {Zhang}, \citenamefont {Jian},
  \citenamefont {Liu},\ and\ \citenamefont {Chen}}]{zhang2021emergent}%
  \BibitemOpen
  \bibfield  {author} {\bibinfo {author} {\bibfnamefont {P.}~\bibnamefont
  {Zhang}}, \bibinfo {author} {\bibfnamefont {S.-K.}\ \bibnamefont {Jian}},
  \bibinfo {author} {\bibfnamefont {C.}~\bibnamefont {Liu}}, \ and\ \bibinfo
  {author} {\bibfnamefont {X.}~\bibnamefont {Chen}},\ }\href@noop {} {\bibfield
   {journal} {\bibinfo  {journal} {Quantum}\ }\textbf {\bibinfo {volume} {5}},\
  \bibinfo {pages} {579} (\bibinfo {year} {2021}{\natexlab{a}})}\BibitemShut
  {NoStop}%
\bibitem [{\citenamefont {Jian}\ \emph
  {et~al.}(2021{\natexlab{a}})\citenamefont {Jian}, \citenamefont {Liu},
  \citenamefont {Chen}, \citenamefont {Swingle},\ and\ \citenamefont
  {Zhang}}]{jian2021measurement}%
  \BibitemOpen
  \bibfield  {author} {\bibinfo {author} {\bibfnamefont {S.-K.}\ \bibnamefont
  {Jian}}, \bibinfo {author} {\bibfnamefont {C.}~\bibnamefont {Liu}}, \bibinfo
  {author} {\bibfnamefont {X.}~\bibnamefont {Chen}}, \bibinfo {author}
  {\bibfnamefont {B.}~\bibnamefont {Swingle}}, \ and\ \bibinfo {author}
  {\bibfnamefont {P.}~\bibnamefont {Zhang}},\ }\href@noop {} {\bibfield
  {journal} {\bibinfo  {journal} {Physical Review Letters}\ }\textbf {\bibinfo
  {volume} {127}},\ \bibinfo {pages} {140601} (\bibinfo {year}
  {2021}{\natexlab{a}})}\BibitemShut {NoStop}%
\bibitem [{\citenamefont {Zhang}\ \emph
  {et~al.}(2021{\natexlab{b}})\citenamefont {Zhang}, \citenamefont {Liu},
  \citenamefont {Jian},\ and\ \citenamefont {Chen}}]{zhang2021universal}%
  \BibitemOpen
  \bibfield  {author} {\bibinfo {author} {\bibfnamefont {P.}~\bibnamefont
  {Zhang}}, \bibinfo {author} {\bibfnamefont {C.}~\bibnamefont {Liu}}, \bibinfo
  {author} {\bibfnamefont {S.-K.}\ \bibnamefont {Jian}}, \ and\ \bibinfo
  {author} {\bibfnamefont {X.}~\bibnamefont {Chen}},\ }\href@noop {} {\bibfield
   {journal} {\bibinfo  {journal} {arXiv preprint arXiv:2105.08895}\ }
  (\bibinfo {year} {2021}{\natexlab{b}})}\BibitemShut {NoStop}%
\bibitem [{\citenamefont {Sahu}\ \emph {et~al.}(2021)\citenamefont {Sahu},
  \citenamefont {Jian}, \citenamefont {Bentsen},\ and\ \citenamefont
  {Swingle}}]{sahu2021entanglement}%
  \BibitemOpen
  \bibfield  {author} {\bibinfo {author} {\bibfnamefont {S.}~\bibnamefont
  {Sahu}}, \bibinfo {author} {\bibfnamefont {S.-K.}\ \bibnamefont {Jian}},
  \bibinfo {author} {\bibfnamefont {G.}~\bibnamefont {Bentsen}}, \ and\
  \bibinfo {author} {\bibfnamefont {B.}~\bibnamefont {Swingle}},\ }\href@noop
  {} {\bibfield  {journal} {\bibinfo  {journal} {arXiv preprint
  arXiv:2109.00013}\ } (\bibinfo {year} {2021})}\BibitemShut {NoStop}%
\bibitem [{\citenamefont {Jian}\ \emph
  {et~al.}(2021{\natexlab{b}})\citenamefont {Jian}, \citenamefont {Liu},
  \citenamefont {Chen}, \citenamefont {Swingle},\ and\ \citenamefont
  {Zhang}}]{jian2021quantum}%
  \BibitemOpen
  \bibfield  {author} {\bibinfo {author} {\bibfnamefont {S.-K.}\ \bibnamefont
  {Jian}}, \bibinfo {author} {\bibfnamefont {C.}~\bibnamefont {Liu}}, \bibinfo
  {author} {\bibfnamefont {X.}~\bibnamefont {Chen}}, \bibinfo {author}
  {\bibfnamefont {B.}~\bibnamefont {Swingle}}, \ and\ \bibinfo {author}
  {\bibfnamefont {P.}~\bibnamefont {Zhang}},\ }\href@noop {} {\bibfield
  {journal} {\bibinfo  {journal} {arXiv preprint arXiv:2106.09635}\ } (\bibinfo
  {year} {2021}{\natexlab{b}})}\BibitemShut {NoStop}%
\bibitem [{\citenamefont {Jian}\ and\ \citenamefont
  {Swingle}(2021{\natexlab{c}})}]{jian2021phase}%
  \BibitemOpen
  \bibfield  {author} {\bibinfo {author} {\bibfnamefont {S.-K.}\ \bibnamefont
  {Jian}}\ and\ \bibinfo {author} {\bibfnamefont {B.}~\bibnamefont {Swingle}},\
  }\href@noop {} {\bibfield  {journal} {\bibinfo  {journal} {arXiv preprint
  arXiv:2108.11973}\ } (\bibinfo {year} {2021}{\natexlab{c}})}\BibitemShut
  {NoStop}%
\bibitem [{\citenamefont {Coleman}(1988)}]{coleman1988aspects}%
  \BibitemOpen
  \bibfield  {author} {\bibinfo {author} {\bibfnamefont {S.}~\bibnamefont
  {Coleman}},\ }\href@noop {} {\emph {\bibinfo {title} {Aspects of symmetry:
  selected Erice lectures}}}\ (\bibinfo  {publisher} {Cambridge University
  Press},\ \bibinfo {year} {1988})\BibitemShut {NoStop}%
\bibitem [{\citenamefont {Altland}\ and\ \citenamefont
  {Simons}(2010)}]{altland2010condensed}%
  \BibitemOpen
  \bibfield  {author} {\bibinfo {author} {\bibfnamefont {A.}~\bibnamefont
  {Altland}}\ and\ \bibinfo {author} {\bibfnamefont {B.~D.}\ \bibnamefont
  {Simons}},\ }\href@noop {} {\emph {\bibinfo {title} {Condensed matter field
  theory}}}\ (\bibinfo  {publisher} {Cambridge university press},\ \bibinfo
  {year} {2010})\BibitemShut {NoStop}%
\bibitem [{\citenamefont {Grinstead}\ and\ \citenamefont
  {Snell}(1997)}]{grinstead1997introduction}%
  \BibitemOpen
  \bibfield  {author} {\bibinfo {author} {\bibfnamefont {C.~M.}\ \bibnamefont
  {Grinstead}}\ and\ \bibinfo {author} {\bibfnamefont {J.~L.}\ \bibnamefont
  {Snell}},\ }\href@noop {} {\emph {\bibinfo {title} {Introduction to
  probability}}}\ (\bibinfo  {publisher} {American Mathematical Soc.},\
  \bibinfo {year} {1997})\BibitemShut {NoStop}%
\bibitem [{\citenamefont {Georges}\ \emph {et~al.}(2001)\citenamefont
  {Georges}, \citenamefont {Parcollet},\ and\ \citenamefont
  {Sachdev}}]{georges2001quantum}%
  \BibitemOpen
  \bibfield  {author} {\bibinfo {author} {\bibfnamefont {A.}~\bibnamefont
  {Georges}}, \bibinfo {author} {\bibfnamefont {O.}~\bibnamefont {Parcollet}},
  \ and\ \bibinfo {author} {\bibfnamefont {S.}~\bibnamefont {Sachdev}},\
  }\href@noop {} {\bibfield  {journal} {\bibinfo  {journal} {Physical Review
  B}\ }\textbf {\bibinfo {volume} {63}},\ \bibinfo {pages} {134406} (\bibinfo
  {year} {2001})}\BibitemShut {NoStop}%
\bibitem [{\citenamefont {Garc{\'\i}a-Garc{\'\i}a}\ and\ \citenamefont
  {Verbaarschot}(2017)}]{garcia2017analytical}%
  \BibitemOpen
  \bibfield  {author} {\bibinfo {author} {\bibfnamefont {A.~M.}\ \bibnamefont
  {Garc{\'\i}a-Garc{\'\i}a}}\ and\ \bibinfo {author} {\bibfnamefont {J.~J.}\
  \bibnamefont {Verbaarschot}},\ }\href@noop {} {\bibfield  {journal} {\bibinfo
   {journal} {Physical Review D}\ }\textbf {\bibinfo {volume} {96}},\ \bibinfo
  {pages} {066012} (\bibinfo {year} {2017})}\BibitemShut {NoStop}%
\bibitem [{\citenamefont {Bagrets}\ \emph {et~al.}(2016)\citenamefont
  {Bagrets}, \citenamefont {Altland},\ and\ \citenamefont
  {Kamenev}}]{bagrets2016sachdev}%
  \BibitemOpen
  \bibfield  {author} {\bibinfo {author} {\bibfnamefont {D.}~\bibnamefont
  {Bagrets}}, \bibinfo {author} {\bibfnamefont {A.}~\bibnamefont {Altland}}, \
  and\ \bibinfo {author} {\bibfnamefont {A.}~\bibnamefont {Kamenev}},\
  }\href@noop {} {\bibfield  {journal} {\bibinfo  {journal} {Nuclear Physics
  B}\ }\textbf {\bibinfo {volume} {911}},\ \bibinfo {pages} {191} (\bibinfo
  {year} {2016})}\BibitemShut {NoStop}%
\bibitem [{\citenamefont {Stanford}\ and\ \citenamefont
  {Witten}(2017)}]{stanford2017fermionic}%
  \BibitemOpen
  \bibfield  {author} {\bibinfo {author} {\bibfnamefont {D.}~\bibnamefont
  {Stanford}}\ and\ \bibinfo {author} {\bibfnamefont {E.}~\bibnamefont
  {Witten}},\ }\href@noop {} {\bibfield  {journal} {\bibinfo  {journal}
  {Journal of High Energy Physics}\ }\textbf {\bibinfo {volume} {2017}},\
  \bibinfo {pages} {1} (\bibinfo {year} {2017})}\BibitemShut {NoStop}%
\bibitem [{\citenamefont {Mertens}\ \emph {et~al.}(2017)\citenamefont
  {Mertens}, \citenamefont {Turiaci},\ and\ \citenamefont
  {Verlinde}}]{mertens2017solving}%
  \BibitemOpen
  \bibfield  {author} {\bibinfo {author} {\bibfnamefont {T.~G.}\ \bibnamefont
  {Mertens}}, \bibinfo {author} {\bibfnamefont {G.~J.}\ \bibnamefont
  {Turiaci}}, \ and\ \bibinfo {author} {\bibfnamefont {H.~L.}\ \bibnamefont
  {Verlinde}},\ }\href@noop {} {\bibfield  {journal} {\bibinfo  {journal}
  {Journal of High Energy Physics}\ }\textbf {\bibinfo {volume} {2017}},\
  \bibinfo {pages} {1} (\bibinfo {year} {2017})}\BibitemShut {NoStop}%
\bibitem [{\citenamefont {Yang}(2019)}]{yang2019quantum}%
  \BibitemOpen
  \bibfield  {author} {\bibinfo {author} {\bibfnamefont {Z.}~\bibnamefont
  {Yang}},\ }\href@noop {} {\bibfield  {journal} {\bibinfo  {journal} {Journal
  of High Energy Physics}\ }\textbf {\bibinfo {volume} {2019}},\ \bibinfo
  {pages} {1} (\bibinfo {year} {2019})}\BibitemShut {NoStop}%
\bibitem [{\citenamefont {Kamenev}(2011)}]{kamenev2011field}%
  \BibitemOpen
  \bibfield  {author} {\bibinfo {author} {\bibfnamefont {A.}~\bibnamefont
  {Kamenev}},\ }\href@noop {} {\emph {\bibinfo {title} {Field theory of
  non-equilibrium systems}}}\ (\bibinfo  {publisher} {Cambridge University
  Press},\ \bibinfo {year} {2011})\BibitemShut {NoStop}%
\bibitem [{\citenamefont {Zhang}\ \emph
  {et~al.}(2021{\natexlab{c}})\citenamefont {Zhang}, \citenamefont {Gu},\ and\
  \citenamefont {Kitaev}}]{zhang2021obstacle}%
  \BibitemOpen
  \bibfield  {author} {\bibinfo {author} {\bibfnamefont {P.}~\bibnamefont
  {Zhang}}, \bibinfo {author} {\bibfnamefont {Y.}~\bibnamefont {Gu}}, \ and\
  \bibinfo {author} {\bibfnamefont {A.}~\bibnamefont {Kitaev}},\ }\href@noop {}
  {\bibfield  {journal} {\bibinfo  {journal} {Journal of High Energy Physics}\
  }\textbf {\bibinfo {volume} {2021}},\ \bibinfo {pages} {1} (\bibinfo {year}
  {2021}{\natexlab{c}})}\BibitemShut {NoStop}%
\bibitem [{\citenamefont {Page}(1993)}]{page1993average}%
  \BibitemOpen
  \bibfield  {author} {\bibinfo {author} {\bibfnamefont {D.~N.}\ \bibnamefont
  {Page}},\ }\href@noop {} {\bibfield  {journal} {\bibinfo  {journal} {Physical
  review letters}\ }\textbf {\bibinfo {volume} {71}},\ \bibinfo {pages} {1291}
  (\bibinfo {year} {1993})}\BibitemShut {NoStop}%
\bibitem [{\citenamefont {Kourkoulou}\ and\ \citenamefont
  {Maldacena}(2017)}]{kourkoulou2017pure}%
  \BibitemOpen
  \bibfield  {author} {\bibinfo {author} {\bibfnamefont {I.}~\bibnamefont
  {Kourkoulou}}\ and\ \bibinfo {author} {\bibfnamefont {J.}~\bibnamefont
  {Maldacena}},\ }\href@noop {} {\bibfield  {journal} {\bibinfo  {journal}
  {arXiv preprint arXiv:1707.02325}\ } (\bibinfo {year} {2017})}\BibitemShut
  {NoStop}%
\bibitem [{\citenamefont {Israel}(1976)}]{israel1976thermo}%
  \BibitemOpen
  \bibfield  {author} {\bibinfo {author} {\bibfnamefont {W.}~\bibnamefont
  {Israel}},\ }\href@noop {} {\bibfield  {journal} {\bibinfo  {journal}
  {Physics Letters A}\ }\textbf {\bibinfo {volume} {57}},\ \bibinfo {pages}
  {107} (\bibinfo {year} {1976})}\BibitemShut {NoStop}%
\bibitem [{\citenamefont {{\L}yd{\.z}ba}\ \emph {et~al.}(2020)\citenamefont
  {{\L}yd{\.z}ba}, \citenamefont {Rigol},\ and\ \citenamefont
  {Vidmar}}]{lydzba2020eigenstate}%
  \BibitemOpen
  \bibfield  {author} {\bibinfo {author} {\bibfnamefont {P.}~\bibnamefont
  {{\L}yd{\.z}ba}}, \bibinfo {author} {\bibfnamefont {M.}~\bibnamefont
  {Rigol}}, \ and\ \bibinfo {author} {\bibfnamefont {L.}~\bibnamefont
  {Vidmar}},\ }\href@noop {} {\bibfield  {journal} {\bibinfo  {journal}
  {Physical review letters}\ }\textbf {\bibinfo {volume} {125}},\ \bibinfo
  {pages} {180604} (\bibinfo {year} {2020})}\BibitemShut {NoStop}%
\bibitem [{\citenamefont {Casini}\ and\ \citenamefont
  {Huerta}(2009)}]{casini2009entanglement}%
  \BibitemOpen
  \bibfield  {author} {\bibinfo {author} {\bibfnamefont {H.}~\bibnamefont
  {Casini}}\ and\ \bibinfo {author} {\bibfnamefont {M.}~\bibnamefont
  {Huerta}},\ }\href@noop {} {\bibfield  {journal} {\bibinfo  {journal}
  {Journal of Physics A: Mathematical and Theoretical}\ }\textbf {\bibinfo
  {volume} {42}},\ \bibinfo {pages} {504007} (\bibinfo {year}
  {2009})}\BibitemShut {NoStop}%
\bibitem [{\citenamefont {Forrester}(2006)}]{forrester2006quantum}%
  \BibitemOpen
  \bibfield  {author} {\bibinfo {author} {\bibfnamefont {P.~J.}\ \bibnamefont
  {Forrester}},\ }\href@noop {} {\bibfield  {journal} {\bibinfo  {journal}
  {Journal of Physics A: Mathematical and General}\ }\textbf {\bibinfo {volume}
  {39}},\ \bibinfo {pages} {6861} (\bibinfo {year} {2006})}\BibitemShut
  {NoStop}%
\bibitem [{\citenamefont {Forrester}(2010)}]{forrester2010log}%
  \BibitemOpen
  \bibfield  {author} {\bibinfo {author} {\bibfnamefont {P.~J.}\ \bibnamefont
  {Forrester}},\ }\href@noop {} {\emph {\bibinfo {title} {Log-gases and random
  matrices (LMS-34)}}}\ (\bibinfo  {publisher} {Princeton University Press},\
  \bibinfo {year} {2010})\BibitemShut {NoStop}%
\bibitem [{\citenamefont {Bianchi}\ \emph {et~al.}(2021)\citenamefont
  {Bianchi}, \citenamefont {Hackl},\ and\ \citenamefont
  {Kieburg}}]{bianchi2021page}%
  \BibitemOpen
  \bibfield  {author} {\bibinfo {author} {\bibfnamefont {E.}~\bibnamefont
  {Bianchi}}, \bibinfo {author} {\bibfnamefont {L.}~\bibnamefont {Hackl}}, \
  and\ \bibinfo {author} {\bibfnamefont {M.}~\bibnamefont {Kieburg}},\
  }\href@noop {} {\bibfield  {journal} {\bibinfo  {journal} {Physical Review
  B}\ }\textbf {\bibinfo {volume} {103}},\ \bibinfo {pages} {L241118} (\bibinfo
  {year} {2021})}\BibitemShut {NoStop}%
\bibitem [{Note1()}]{Note1}%
  \BibitemOpen
  \bibinfo {note} {There are also studies on subsystem entropy of systems
  prepared in thermal ensembles and coupled to a bath \cite
  {kaixiang2021page}.}\BibitemShut {Stop}%
\bibitem [{\citenamefont {Zhang}(2019)}]{zhang2019evaporation}%
  \BibitemOpen
  \bibfield  {author} {\bibinfo {author} {\bibfnamefont {P.}~\bibnamefont
  {Zhang}},\ }\href@noop {} {\bibfield  {journal} {\bibinfo  {journal}
  {Physical Review B}\ }\textbf {\bibinfo {volume} {100}},\ \bibinfo {pages}
  {245104} (\bibinfo {year} {2019})}\BibitemShut {NoStop}%
\bibitem [{\citenamefont {Almheiri}\ \emph
  {et~al.}(2019{\natexlab{e}})\citenamefont {Almheiri}, \citenamefont
  {Milekhin},\ and\ \citenamefont {Swingle}}]{almheiri2019universal}%
  \BibitemOpen
  \bibfield  {author} {\bibinfo {author} {\bibfnamefont {A.}~\bibnamefont
  {Almheiri}}, \bibinfo {author} {\bibfnamefont {A.}~\bibnamefont {Milekhin}},
  \ and\ \bibinfo {author} {\bibfnamefont {B.}~\bibnamefont {Swingle}},\
  }\href@noop {} {\bibfield  {journal} {\bibinfo  {journal} {arXiv preprint
  arXiv:1912.04912}\ } (\bibinfo {year} {2019}{\natexlab{e}})}\BibitemShut
  {NoStop}%
\bibitem [{Note2()}]{Note2}%
  \BibitemOpen
  \bibinfo {note} {Similar calculations has been carried out in \cite
  {gu2017spread} for SYK chains.}\BibitemShut {Stop}%
\bibitem [{\citenamefont {Do~Carmo}(2016)}]{do2016differential}%
  \BibitemOpen
  \bibfield  {author} {\bibinfo {author} {\bibfnamefont {M.~P.}\ \bibnamefont
  {Do~Carmo}},\ }\href@noop {} {\emph {\bibinfo {title} {Differential geometry
  of curves and surfaces: revised and updated second edition}}}\ (\bibinfo
  {publisher} {Courier Dover Publications},\ \bibinfo {year}
  {2016})\BibitemShut {NoStop}%
\bibitem [{Note3()}]{Note3}%
  \BibitemOpen
  \bibinfo {note} {We thank Yingfei Gu for explaining this
  example.}\BibitemShut {Stop}%
\bibitem [{\citenamefont {Dong}\ \emph {et~al.}(2016)\citenamefont {Dong},
  \citenamefont {Harlow},\ and\ \citenamefont {Wall}}]{dong2016reconstruction}%
  \BibitemOpen
  \bibfield  {author} {\bibinfo {author} {\bibfnamefont {X.}~\bibnamefont
  {Dong}}, \bibinfo {author} {\bibfnamefont {D.}~\bibnamefont {Harlow}}, \ and\
  \bibinfo {author} {\bibfnamefont {A.~C.}\ \bibnamefont {Wall}},\ }\href@noop
  {} {\bibfield  {journal} {\bibinfo  {journal} {Physical review letters}\
  }\textbf {\bibinfo {volume} {117}},\ \bibinfo {pages} {021601} (\bibinfo
  {year} {2016})}\BibitemShut {NoStop}%
\bibitem [{\citenamefont {Chen}(2020)}]{chen2020pulling}%
  \BibitemOpen
  \bibfield  {author} {\bibinfo {author} {\bibfnamefont {Y.}~\bibnamefont
  {Chen}},\ }\href@noop {} {\bibfield  {journal} {\bibinfo  {journal} {Journal
  of High Energy Physics}\ }\textbf {\bibinfo {volume} {2020}},\ \bibinfo
  {pages} {1} (\bibinfo {year} {2020})}\BibitemShut {NoStop}%
\bibitem [{\citenamefont {Hayden}\ \emph {et~al.}(2016)\citenamefont {Hayden},
  \citenamefont {Nezami}, \citenamefont {Qi}, \citenamefont {Thomas},
  \citenamefont {Walter},\ and\ \citenamefont {Yang}}]{hayden2016holographic}%
  \BibitemOpen
  \bibfield  {author} {\bibinfo {author} {\bibfnamefont {P.}~\bibnamefont
  {Hayden}}, \bibinfo {author} {\bibfnamefont {S.}~\bibnamefont {Nezami}},
  \bibinfo {author} {\bibfnamefont {X.-L.}\ \bibnamefont {Qi}}, \bibinfo
  {author} {\bibfnamefont {N.}~\bibnamefont {Thomas}}, \bibinfo {author}
  {\bibfnamefont {M.}~\bibnamefont {Walter}}, \ and\ \bibinfo {author}
  {\bibfnamefont {Z.}~\bibnamefont {Yang}},\ }\href@noop {} {\bibfield
  {journal} {\bibinfo  {journal} {Journal of High Energy Physics}\ }\textbf
  {\bibinfo {volume} {2016}},\ \bibinfo {pages} {1} (\bibinfo {year}
  {2016})}\BibitemShut {NoStop}%
\bibitem [{\citenamefont {Nahum}\ \emph {et~al.}(2018)\citenamefont {Nahum},
  \citenamefont {Vijay},\ and\ \citenamefont {Haah}}]{nahum2018operator}%
  \BibitemOpen
  \bibfield  {author} {\bibinfo {author} {\bibfnamefont {A.}~\bibnamefont
  {Nahum}}, \bibinfo {author} {\bibfnamefont {S.}~\bibnamefont {Vijay}}, \ and\
  \bibinfo {author} {\bibfnamefont {J.}~\bibnamefont {Haah}},\ }\href@noop {}
  {\bibfield  {journal} {\bibinfo  {journal} {Physical Review X}\ }\textbf
  {\bibinfo {volume} {8}},\ \bibinfo {pages} {021014} (\bibinfo {year}
  {2018})}\BibitemShut {NoStop}%
\bibitem [{\citenamefont {Von~Keyserlingk}\ \emph {et~al.}(2018)\citenamefont
  {Von~Keyserlingk}, \citenamefont {Rakovszky}, \citenamefont {Pollmann},\ and\
  \citenamefont {Sondhi}}]{von2018operator}%
  \BibitemOpen
  \bibfield  {author} {\bibinfo {author} {\bibfnamefont {C.}~\bibnamefont
  {Von~Keyserlingk}}, \bibinfo {author} {\bibfnamefont {T.}~\bibnamefont
  {Rakovszky}}, \bibinfo {author} {\bibfnamefont {F.}~\bibnamefont {Pollmann}},
  \ and\ \bibinfo {author} {\bibfnamefont {S.~L.}\ \bibnamefont {Sondhi}},\
  }\href@noop {} {\bibfield  {journal} {\bibinfo  {journal} {Physical Review
  X}\ }\textbf {\bibinfo {volume} {8}},\ \bibinfo {pages} {021013} (\bibinfo
  {year} {2018})}\BibitemShut {NoStop}%
\bibitem [{Note4()}]{Note4}%
  \BibitemOpen
  \bibinfo {note} {Here $H_I$ is not positive semidefinite. However, we can
  always make it positive semidefinite by shifting a large enough
  constant.}\BibitemShut {Stop}%
\bibitem [{\citenamefont {Zhai}(2021)}]{zhai2021ultracold}%
  \BibitemOpen
  \bibfield  {author} {\bibinfo {author} {\bibfnamefont {H.}~\bibnamefont
  {Zhai}},\ }\href@noop {} {\emph {\bibinfo {title} {Ultracold Atomic
  Physics}}}\ (\bibinfo  {publisher} {Cambridge University Press},\ \bibinfo
  {year} {2021})\BibitemShut {NoStop}%
\bibitem [{\citenamefont {Ippoliti}\ \emph {et~al.}(2021)\citenamefont
  {Ippoliti}, \citenamefont {Rakovszky},\ and\ \citenamefont
  {Khemani}}]{ippoliti2021fractal}%
  \BibitemOpen
  \bibfield  {author} {\bibinfo {author} {\bibfnamefont {M.}~\bibnamefont
  {Ippoliti}}, \bibinfo {author} {\bibfnamefont {T.}~\bibnamefont {Rakovszky}},
  \ and\ \bibinfo {author} {\bibfnamefont {V.}~\bibnamefont {Khemani}},\
  }\href@noop {} {\bibfield  {journal} {\bibinfo  {journal} {arXiv preprint
  arXiv:2103.06873}\ } (\bibinfo {year} {2021})}\BibitemShut {NoStop}%
\bibitem [{\citenamefont {Dong}(2016)}]{dong2016gravity}%
  \BibitemOpen
  \bibfield  {author} {\bibinfo {author} {\bibfnamefont {X.}~\bibnamefont
  {Dong}},\ }\href@noop {} {\bibfield  {journal} {\bibinfo  {journal} {Nature
  communications}\ }\textbf {\bibinfo {volume} {7}},\ \bibinfo {pages} {1}
  (\bibinfo {year} {2016})}\BibitemShut {NoStop}%
\end{thebibliography}%
	
\end{document}